\documentclass[fleqn,usenatbib]{mnras}

\usepackage{newtxtext,newtxmath}
 
\usepackage[T1]{fontenc}

\DeclareRobustCommand{\VAN}[3]{#2}
\let\VANthebibliography\thebibliography
\def\thebibliography{\DeclareRobustCommand{\VAN}[3]{##3}\VANthebibliography}

\usepackage{graphicx}	
\usepackage{amsmath}	
\usepackage{amssymb}	
\usepackage{algorithm} 
\usepackage{algpseudocode} 
\usepackage{bm}
\usepackage{comment}

\newcommand{\fr}[1]{\textcolor{black}{ #1}}

\title[BANG-MaNGA]{\fr{Decomposing} galaxies with \textsc{bang}: an automated morpho-kinematic decomposition of the SDSS-DR17 MaNGA survey}

\author[F. Rigamonti et al.]{
Fabio Rigamonti,$^{1,2,3}$\thanks{E-mail: frigamonti@uninsubria.it}
Massimo Dotti,$^{4,2,3}$
Stefano Covino,$^{2,1}$
Francesco Haardt,$^{1,2,3}$
Luca Cortese,$^{5,6}$
\newauthor{
Marco Landoni,$^{2,1}$
Ludovica Varisco, $^{3,4}$
}
\\
$^{1}$DiSAT, Universit\`a degli Studi dell'Insubria, via Valleggio 11, I-22100 Como, Italy\\
$^{2}$INAF, Osservatorio Astronomico di Brera, Via E. Bianchi 46, I-23807 Merate, Italy\\
$^{3}$INFN, Sezione di Milano-Bicocca, Piazza della Scienza 3, I-20126 Milano, Italy\\
$^{4}$Dipartimento di Fisica G. Occhialini, Universit\`a di Milano-Bicocca, Piazza della Scienza 3, I-20126 Milano, Italy\\
$^{5}$ International Centre for Radio Astronomy Research, The University of Western Australia, Crawley, WA 6009, Australia\\
$^{6}$ ARC Centre of Excellence for All Sky Astrophysics in 3 Dimensions (ASTRO 3D), Australia\\
}

\date{Accepted . Received ; in original form}

\pubyear{2015}
\newcommand{\rowsep}[0]{\vspace{0.2cm}}
\begin{document}
\label{firstpage}
\pagerange{\pageref{firstpage}--\pageref{lastpage}}
\maketitle

\begin{abstract}
From a purely photometric perspective galaxies are generally decomposed into a bulge+disc system, with bulges being dispersion-dominated and discs rotationally-supported. However, recent observations have demonstrated that such a framework oversimplifies complexity, especially if one considers galaxy kinematics.
To address this issue we introduced with the GPU-based code \textsc{bang} a novel approach that employs analytical potential-density pairs as galactic components, allowing for a computationally fast, still reliable fit of the morphological and kinematic properties of galaxies. Here we apply \textsc{bang} to the SDSS-MaNGA survey, estimating key parameters such as mass, radial extensions, and dynamics, for both bulges and discs of +10,000 objects. We test our methodology against a smaller subsample of galaxies independently analysed with an orbit-based algorithm, finding agreement in the recovered total stellar mass. We also manage to reproduce well-established scaling relations, demonstrating how proper dynamical modelling can result in tighter correlations and provide corrections to standard approaches. Finally, we propose a more general way of decomposing galaxies into "hot" and "cold" components, showing a correlation with orbit-based approaches and visually determined morphological type. Unexpected tails in the "hot-to-total" mass-ratio distribution are present for galaxies of all morphologies, possibly due to visual morphology misclassifications.


\end{abstract}


\section{Introduction}
Galaxies have been long studied using the "bulge+disc" decomposition framework \citep[see, e.g.,][]{Mendel_2014}, with bulges thought to be spheroidal components with little net rotation, possibly originated via merger events and well described by the de Vaucouleurs surface brightness profile \citep[][]{de-Vaucouleurs1948}, while discs are seen as the result of accretion and cooling of angular momentum conserving gas, typically described by a flat exponential profile supported by ordered rotation \citep[][]{Freeman1970,White_1978}. 

Recent high-resolution observations have challenged such two-component decomposition. For instance, bulges are commonly divided into "classical bulges" and "pseudobulges", with the latter frequently showing disc-like features \citep[see][for a review]{Laurikainen_2016}. To model this structural complexity, purely photometric decomposition methods and codes have been developed, initially working on 1D surface brightness profiles \citep[e.g.,][]{Gavazzi_2000_pV}, then applied to 2D images \citep[e.g., GIM2D, by][]{simard_2002} and including discs, bars \citep[e.g., BUDDA, by][]{budda_Gadotti}, spiral arms, rings and warps \citep[e.g., GALFIT, by][]{Peng_2002}.


In recent years, the development of integral field unit (IFU) spectroscopy has allowed for an unprecedentedly detailed description of the velocity field of galaxies, forcing  
the "bulge+disc" decomposition to take into account also the observed kinematics; this has mainly been done by combining a photometry-assumed decomposition of the galaxy images with the extraction of each component kinematics from the spectra \citep[][]{Oh_2016,Tabor_2017}.

These methods have been successfully applied on sub-samples of surveys such as SAMI \citep[][]{Oh_2016}, MaNGA \citep[][]{Tabor_2019} and CALIFA \citep[][]{Pak_2021}, helping to disentangle and characterise the kinematic properties of bulges and discs. Relying and/or assuming a first decomposition based only on photometric data, these methods can inevitably introduce bias in any eventual extraction of kinematic information without guaranteeing any self-consistency between the density and the underlying gravitational potential.    


An alternative approach is the orbit superposition method \citep[][]{Schwarzschild1979}, in which both the surface brightness and the line-of-sight kinematic observables are reproduced as the superposition of purposely weighted orbital families. In this method, the structural components of the galaxies are estimated ex-post the fit, 
associating orbits to bulge and disc components depending on their angular momentum and spatial extension \citep[][]{Zhu_2018a,Zhu_2018b}.
Despite the strength and the generalization capabilities of these methods, their applicability is limited by the huge computational power required for orbit integration and weighting. 

In a former paper \citep[][hereinafter Paper I.]{Rigamonti_2022}, we proposed a new approach to this problem that lies between the purely photometric decomposition methods and the orbit-based ones. To this aim, we developed and publicly released \textsc{bang}\footnote{\url{https://pypi.org/project/BANGal/}}, a GPU-optimized code aimed at a fully Bayesian estimation of the global structure of disc galaxies. \textsc{bang} performs a nested sampling analysis 
\citep[][]{Skilling2004}, 
assuming that each galaxy is composed of simple "classical" components, such as a bulge and one or more discs, parameterized in terms of analytic density distributions and characterized by well-defined dynamics. In Paper I, as a case study, we applied \textsc{bang} to NGC 7683, an S0 galaxy carefully selected for its regular velocity pattern and disc-like geometry, finding substantial agreement with the orbit-based code \textsc{dynamite} \citep[][]{DYNAMITE} at a fraction of the computational cost despite the fact that the number of models explored by \textsc{bang} far exceeds the orbit-based galaxy realizations. 

Motivated by the first success of our effort, we decided to apply an updated version of \textsc{bang} to a large sample of galaxies, namely, the full SDSS-DR17 MaNGA survey \citep[][]{Abdurro'uf_2022}, consisting in $+10,000$ local ($0.01\lesssim z \lesssim 0.1$) galaxies with IFU data. The present paper is the first of a series describing our study and findings, and it is structured as follows. In \S2 we briefly set out the sample we analysed, in \S3 we summarize the methodology upon which \textsc{bang} is based. \S4 is devoted to the analysis of a single galaxy as a showcase, while in \S5 we perform a comparison of \textsc{bang} results for a selected subsample to orbit superposition methods. Our novel methodology is then applied to the whole MaNGA sample in \S6 highlighting some preliminary results. In \S7 we show scaling relations in comparison to similar existing studies, while \S8 is devoted to summary and discussion about follow-up studies. 

\section{Data Sample}

This study is based on a morpho-kinematic dataset built upon the SDSS imaging survey \fr{\citep[][]{Strauss_2002}} and the final release of the MaNGA survey from the Seventeenth Data Release \citep[DR17,][]{Abdurro'uf_2022} of the fourth phase of the Sloan Digital Sky Survey \citep[SDSS-IV,][]{Blanton_2017}. In the following, we briefly describe the most relevant features of these surveys and their data reduction pipelines.

\subsection{Kinematics: The MaNGA survey}\label{sec:mangasurvey}
The Mapping Nearby Galaxies at Apache Point Observatory (MaNGA) survey \citep{Bundy_2015,Drory_2015} has collected integral field spectroscopic measurements for $+10,000$ local galaxies selected to have a roughly flat stellar mass distribution above $10^{9}~\rm M_{\sun}$ in the redshift range $0.01 < \rm z <0.15$. The integral field units (IFU) observations have been carried out with different setups ranging from 19 IFU-fiber bundles ($\simeq 12''$ diameter) to 127 IFU-fiber bundles ($\simeq 32''$ diameter). The adopted configuration changes depending on different galaxy properties, we refer to \citealt{Wake_2017} for more details about the selection criteria and the distribution of the IFU sizes. The MaNGA survey is divided into a primary sample ($\simeq 67\%$) consisting of galaxies observed up to $1.5 ~\rm R_e$ and a secondary sample ($\simeq 33\%$) with spatial coverage of $2.5 ~\rm R_e$. The raw data are collected after a $\simeq 3$ hour-long dithered exposure and reduced by the data reduction pipeline \citep[][]{Law_2016,Yan_2016}. The reduced data cubes contain a spectrum at each spatial element with a resolution $\rm R \simeq 2000$ (i.e. $\sigma_{\rm inst}\simeq70\rm km/s$) in the wavelength range 3600\r{A} - 10000\r{A}. During each exposure, the hexagonal plate of the IFU is subject to dithered movements resulting in a \fr{datacube binned to a $0.5''$ grid in order to optimally sample the $2.5''$ full width at half maximum (FWHM) reconstructed point spread function (PSF)} \citep[][]{Westfall_2019}.  
\fr{In addition to each data cube, the MaNGA collaboration provides a data analysis pipeline \citep[\textsc{dap}]{Belfiore_2019,Law_2021} comprising a set of spatially resolved \fr{maps} in different configurations depending on the spatial binning and on the stellar templates adopted in the data-cube spectral fitting process (see \citealt{Westfall_2019} for further details).} In this work, we \fr{collected, through the \textsc{dap} web interface \textsc{marvin} \citep[][]{Cherinka_2019}}, line of sight (l.o.s.) stellar velocity and velocity dispersion estimates and their associated errors assuming the same spatial binning as for the data cube. For each galaxy, we discarded all the pixels with S/N<4 (<6) for the  velocity (velocity dispersion) to avoid spurious data in the outskirt of galaxies. On top of this, we applied the isolation forest algorithm \citep{Liu_2012} to each of the two maps to further mask possible outliers still present especially in the outskirt of velocity dispersion data.

\subsection{Photometry: The SDSS survey}
The observed MaNGA galaxies are a small subsample drawn from the SDSS imaging survey for which a wealth of ancillary information is available. For each MaNGA galaxy, we downloaded the original \fr{$i$ and $g$ band} ($10'\times13'$) SDSS \fr{\citep[DR17,][]{Abdurro'uf_2022}{}{}} brightness maps and cropped them to the same spatial extension  and resolution of the MaNGA kinematics squared cut-outs\footnote{Note that the Manga square cut-outs always exceed the hexagonal region for which IFU data are available. Since it is not obvious how to limit the region within which  brightness measurements are considered for large galaxy samples, we chose to never exceed the MaNGA squared cut-outs and to mask all the pixels outside an ellipse containing more than 60 noisy (i.e. NaN) measurements.} using the procedure discussed in \citep{Bertin_2002}.
We then computed the associated error by considering a Poissonian statistics on the number of detected photo-electrons and the contribution of additional sources of noise (i.e. read-noise and noise in the dark current)\footnote{See \url{https://dr12.sdss.org/datamodel/files/BOSS_PHOTOOBJ/frames/RERUN/RUN/CAMCOL/frame.html} for further details about the computation of the photometric errors.}. We then converted, by knowing the galaxy redshift, the photometric maps to astrophysical units ($\rm L_{\sun}/\rm kpc^2$) and masked the possible presence of stars using DAOStarFinder, a \textsc{python} package based on the \citet{Stetson_1987} implementation, combined with the isolation forest algorithm mentioned above.
This procedure has been automated and applied to all the MaNGA galaxies in both the $i$ and $g$ bands allowing us to compute estimates of the stellar mass-to-light ratio ($\rm M/L$) from standard color-$\rm M/L$ relation \citep{Zibetti_2009}. The associated error, computed by propagation, is generally large enough to account for the scatter of the linear relation and for any assumption on the initial mass function. $\rm M/L$ ratio data are masked following the same procedure as done in the case of the velocity dispersion. 
In the end, we obtained a dataset comprising spatially resolved information of brightness, $\rm M/L$, l.o.s. velocity and velocity dispersion of 10,005 galaxies \fr{morphologically characterized according to the classification presented in the MaNGA Visual Morphology Catalogue \citep[VMC-VAC,][]{Vazquez_2022}{}{}}.

\section{Methodology}
\label{sec:methodology}
For each galaxy, we fit simultaneously the logarithm of the $i$-band brightness, the stellar $\rm M/L$, the  velocity ($\rm v_{los}$) and the l.o.s. velocity dispersion ($\rm \sigma_{los}$) using an upgraded version of \textsc{bang} \citep{Rigamonti_2022_software}. In the following, we will briefly summarize the main assumption underlying our dynamical model, the assumed priors and the adopted criteria for selecting among different models. To perform this study we extensively used high-performance computing resources dividing our samples into groups of 20 galaxies and iteratively analysing them with a multi-GPU parallelization strategy; the average duration of each GPU job was $\simeq 12~\rm hours$ ($\simeq 30~$ minutes per galaxy) implying a total amount of $\simeq 6000~ \rm GPU/h$ per sample run. 

\subsection{Model}
Each galaxy is the superposition of a spherical bulge, two exponential razor-thin discs and a dark-matter halo.
Each model is axially symmetric and has a well-defined centre, that corresponds to both the photometric and dynamical centre shared by all the components. 

\begin{itemize}
    \item The bulge is assumed to be spherically symmetric with an isotropic velocity distribution described, differently from what is assumed in Paper I, by a Dehnen profile \citep{Dehnen1993}. The l.o.s. projected quantities, namely the surface density ($\Sigma_b$) and the velocity dispersion ($\sigma_b$) at any location in the plane of the sky $R$, are:
    \begin{equation}
        \label{eq:surface_density_dehnen}
        \Sigma_b(R) = \frac{M_b R_b(3-\gamma)}{2\pi}\int_{R}^{\infty}\frac{r^{1-\gamma}}{(r+R_b)^{4-\gamma}}\frac{dr}{\sqrt{r^2-R^2}},
    \end{equation}
    \begin{equation}
        \label{eq:dispersion_dehnen}
        \sigma_b^2(R) = \frac{(3-\gamma)GM_b^2R^2R_b}{4\pi\Sigma(R)} \int_R^\infty  \frac{r^{1-2\gamma}}{(r+R_b)^{7-2\gamma}}\frac{dr}{\sqrt{r^2-R^2}} ,
    \end{equation}
    where $M_b$ and $R_b$ are the mass and the scale radius of the bulge while $\gamma$ is the inner slope of the density profile.  

    \item \fr{The inner and the outer discs are modelled as exponential razor-thin discs whose intrinsic velocity dispersion and tangential velocity are approximated assuming that a fraction $k_j$ (with the convention of $j=1$ for the inner disc and $j=2$ for the outer disc) of the total kinetic energy is in ordered bulk motion and the rest goes into an isotropic velocity dispersion component. The "kinematic decomposition parameter" $k_j$ is a free parameter of the model. Its value ranges from $k_j=0$ for a dispersion-supported regime to $k_j=1$ for a rotation-supported regime.} \fr{More specifically, the bulk l.o.s. velocity component for each disc is}
    \begin{equation}
        \label{eq:disc_projected_velocity}
        v_{d,j}(R) = -k_j\, \sin{i}\, \cos{\phi}\, v_{{\rm c}}(r),
    \end{equation}
    \fr{where $r$ is the deprojected distance from the centre of the galaxy of the $R$ position, $\phi$ is the azimuthal angle evaluated in the plane of the disc from the major axis of the projected disc and $ v_{{\rm c}}$ is the total circular velocity.}
    
    \fr{The l.o.s. velocity dispersion of the disc is instead}
    \begin{equation}
        \label{eq:disc_projected_dispersion}
        \sigma^2_{d,j}(R) = (1-k_j^2)\frac{v^2_{c}(r)}{3},
    \end{equation}
    \fr{where the factor 3 at the denominator is due to the assumption of an isotropic velocity dispersion tensor. As discussed in Paper I, the assumption of razor-thin discs is inconsistent with having a finite velocity dispersion; still, the effect on the fit is small as long as the discs have aspect ratio $\lesssim0.1$. }

    \item The dark matter halo is described by a Navarro-Frenk-White profile \citep{Navarro1996} with a cut-off radius $R_{200}$ equals to the distance at which the mean density corresponds to 200 times the cosmological critical density. The circular velocity associated to the dark matter contribution to the total potential reads:
    \begin{equation}
        \label{eq:NFW_circular_velocity}
        v_h^2(r) = \frac{GM_h}{r} \frac{\log{(1+r/a)}-r/(r+a)}{\log{(1+c)}-c/(1+c)},
    \end{equation}
    where $M_h$ is the halo mass, $c$ is the concentration and $a$ is the halo scale radius computed as $R_{200}/c$. Within our fitting strategy, we parameterised the halo mass in terms of the logarithm of the ratio between the dark and the visible mass $\log_{10}{f_{\star}} $. 
\end{itemize}

SDSS i-band image data and MaNGA kinematics come with different PSFs, with an average of about $1.2''$ and $2.5''$, respectively. To avoid any loss of information in the data, we accounted for different PSF convolutions between the photometry and the kinematic, considering for each galaxy the actual PSF FWHM as provided by the SDSS ancillary data and the MaNGA \textsc{dap} catalogue. For this analysis, \fr{we  successfully fitted $10,005$ galaxies with} three possible components configurations, namely a bulge+disc+disc ("$\rm B+D_1+D_2$") model, a bulge+disc ("$\rm B+D_1$") model, and a disc+disc ("$\rm D_1+D_2$") model, whose parameters are summarised in Tab.~\ref{tab:parameters}. \fr{For each galaxy we selected the best-fit model as the one with the highest evidence, resulting in $\simeq 86\%$ of the sample preferring the "$\rm B + D_1 + D_2$" model, $\simeq 6\%$ the "$\rm B+D_1$" model, and the remaining $\simeq \rm 8\%$ the "$\rm D_1+D_2$" model}.

\begin{table*}
	\centering
	\caption{Summary of the model parameters in the three different configurations, namely a bulge+disc+disc ($\rm"B+D_1+D_2"$), a bulge+disc ($\rm"B+D_1"$) and a disc+disc ($\rm"D_1+D_2"$) model. The first two columns describe each parameter and its reference name used in this work. Columns three, four and five visually represent the inclusion of the parameters in the different models, while the last column describes the adopted prior ranges. When not differently specified positions, masses, angles and mass-to-light ratios are respectively normalized to $\rm kpc$, $\rm M_{\sun}$, degrees and $\rm M_{\sun}/L_{\sun}$. We refer to Sec.~\ref{sec:methodology_priors} for a more detailed discussion about the assumed priors.}
	\label{tab:parameters}
	\begin{tabular}{llcccc} 
		\hline
		description & name & "$\rm B+D_1+D_2$" & "$\rm B+D_1$" & "$\rm D_1+D_2$" & prior range\\
		\hline
		horizontal/vertical position of the center & $x_0$ ; $y_0$ & \checkmark & \checkmark & \checkmark & [-2,2]\\
		position angle & $\mathrm{P.A.}$ & \checkmark & \checkmark & \checkmark & [$-180$,$180$]\\
		inclination angle & $i$ & \checkmark & \checkmark & \checkmark & [$3$,$87$] \\
		mass, scale radius and inner slope of the bulge & $\log_{10}{M_{b}}$ ; $\log_{10}{R_{b}}$; $\gamma$ & \checkmark & \checkmark & $\times$ & [7.75,13.0], [-2,2], [0,2]\\
		mass and scale radius of the inner disc & $\log_{10}{M_{d,1}}$ ; $\log_{10}{R_{d,1}}$ & \checkmark & \checkmark & \checkmark & [7.75,13], [-2,2] \\
		mass and scale radius of the outer disc & $\log_{10}{M_{d,2}}$ ; $\log_{10}{R_{d,2}}$ & \checkmark & $\times$ & \checkmark & [7.75,13], [-2,2] \\
		halo-to-stellar mass fraction and concentration & $\log_{10}{f_{\star}}$ ; $c$  & \checkmark & \checkmark & \checkmark & [0.1,2.7], [1,20] \\
		mass-to-light ratio of the bulge & $(M/L)_{b}$ & \checkmark & \checkmark & $\times$ & [0.1,20]\\
		mass-to-light ratio of the inner disc & $(M/L)_{d,1}$ & \checkmark & \checkmark & \checkmark & [0.1,20]\\
		 mass-to-light ratio of the outer disc & $(M/L)_{d,2}$ & \checkmark & $\times$ & \checkmark & [0.1,20]\\
		kinematic decomposition parameter of the inner disc & $k_1$ & \checkmark & \checkmark & \checkmark & [0.01,1]\\
		kinematic decomposition parameter of the outer disc & $k_2$ &\checkmark &$\times$ & \checkmark & [0.01,1]\\
		\hline
	\end{tabular}
\end{table*}

\subsection{Priors}
\label{sec:methodology_priors}
Our Bayesian approach requires the inclusion of any \textit{a priori} knowledge of the probability distribution of each model parameter. In our case, since we are dealing with a statistically large sample of galaxies, it is mandatory to assume well-motivated representative priors. Given the complex nature of the problem at hand, we assumed different levels of priors combining information from both simulations and theoretical predictions. \fr{We emphasize that our modelling scheme significantly deviates from any other approach employed in the study of galactic dynamic decomposition, particularly when it comes to analyzing such a large sample of objects while also considering kinematic information. This is the reason why we have not taken into account observational priors, which are typically derived solely from photometric decomposition and thus not aligned with our methodology. We also stress that in the majority of cases, where the data provide sufficient information, our results are in fact independent of the assumed priors.}

\begin{itemize}

    \item \textbf{Strong Prior}\\
    Using the output from a z=0 snapshot of the TNG50 simulation \citep[][]{TNG50_a,TNG50_b} we estimated a relation between the total stellar mass ($M_{\star}$) and the halo mass ($M_{h}$) (see the next bullet point for a detailed description on how these mass estimates have been obtained). Moreover, following previous works in the literature \citep[][]{Cappellari_2013,Zhu_2018,Jin_2020} we assumed a relation between the halo mass and its concentration following estimates from \citet{Dutton}.
    For both the halo mass and its concentration parameter we opted for two different fitting set-ups. In both cases, we still assumed 2D Gaussian priors respectively centred around the assumed relations ($M_{\star}-M_{h}$ and $M_{h}-c$), but in a second, more conservative approach we do not allow for halo parameter exploration outside 1-$\sigma$ of the assumed relation. We refer to the two cases respectively as "halo-free" and "halo-fixed" configurations, we anticipate that generally, the two approaches lead to similar results in terms of the estimated stellar parameters.

    \item\textbf{Mild-informative Prior}\\
    We assumed prior relations among the following pairs of model parameters: $(M_b,R_{b-half})$, $(M_{d1},R_{d1-half})$, $(M_{d2},R_{d2-half})$, where the \textit{half} subscript refers to the half-mass radius of that component. Starting from a z=0 snapshot of the TNG50 simulation \citep[][]{TNG50_a,TNG50_b}, we collected masses and half-mass radii for bulges, discs and halos of a representative sample of simulated galaxies analysed with \textsc{mordor}\footnote{\url{https://github.com/thanatom/mordor}} \citep{Zana_2022}, a publicly available code which decomposes galaxies in their structural components depending on  the energy and angular momentum of the stellar distribution.
    Note that this methodology, being phase-space based, allow us to include prior information in the most consistent way with \textsc{bang}. For each galaxy component analysed by \textsc{mordor} we estimated scaling relations for the above-mentioned parameter pairs by computing the average and standard deviation in different mass bins. 
    We stress that, in our fitting approach, we do not force those parameters to strictly obey the simulated-obtained relations, instead, we included that additional information as a Gaussian prior on the considered radii whose dispersion does depend on the mass of the component such that any parameter pair is still free to converge possibly far away from the assumed relation.    

    \item \textbf{Uninformative Prior}\\
    We assumed uniform or log-uniform prior for most of the parameters over a range broad enough to account for the large variability present in our sample. Similarly to what was done in Paper I, we adopted a prior proportional to the sine of the inclination angle for disc-like galaxies (S and S0) while a flat distribution is assumed in the case of ellipticals. We summarise the adopted choices in the $6^{\rm th}$ column of Tab.~\ref{tab:parameters}

\end{itemize}

\subsection{Caveats}
Before any further analysis, we discuss some caveats regarding our methodology. 

In $\simeq 15\%$ of the cases, the spherical bulge is almost not resolved, with a half-mass radius, even if generally affected by large errors, well below the pixel resolution. It is possible that in some circumstances this indicates the presence of a sub-nuclear structure, even though, more probably, it is an effect arising either from some limitations of our modelling scheme or from spurious features in the central regions of the data. This happens only in models made up of three visible components and it roughly anti-correlates with the model evidence, meaning that for small bulges, generally the evidence difference between the three visible components model and the two visible components one generally decreases. As we will discuss again in Sec.~\ref{sec:Preliminary analysis on the proposed decomposition} this should not affect our analysis in terms of the proposed decomposition.

The brightness spatial extension is typically larger than the field of view covered by the kinematics. It is not obvious where to stop including brightness measurements, especially if the process must be automated for many galaxies. Our choice is never to exceed the squared MaNGA cut-outs and to mask all the pixels outside an ellipse containing more than 60 noisy (i.e. NaN) measurements. This approach generally prevents our fits to be affected by brightness noise-dominated regions, even though, in some cases, our estimates converge towards unrealistically extended discs (for "$\rm D_2$" only); similarly to what is discussed for the bulge, this happens only for a three visible components model. As a conservative approach, in these cases, when the disc scale radius ($R_{d,2}$) is larger than $2.5 R_e$ \footnote{Which roughly corresponds to the maximum region probed by the MaNGA kinematics.} we replace the total mass of the external disc with the mass within $2.5 R_e$. For the forthcoming analysis we adopt $R_e$ as provided by the MaNGA DAP \citep[][]{Blanton_2011}{}{}.

\fr{Before moving to discuss our results, it is important to stress the main focus of our work. Galaxy automated decomposition can sometimes lead to incorrect classification of the components. In fact, some confusion can arise between classical bulges, pseudo-bulges and bars. Similarly, degenerations can arise between extended bars and galactic thin and thick discs \citep[][]{Kormendy_2004,Erwin_2015,Kruk_2018}{}{}. At present, our model is not complex enough to account for such a detailed description. It must be said that it is not clear whether the current spatial and spectral resolution of the largest IFS surveys (i.e., MaNGA) is sufficient to break the degeneracy between these components. Nonetheless, we argue that including kinematics in the decomposition, as we do in the present work, leads to more robust results than purely photometric approaches. We are not focusing on any simple "bulge+disc" decomposition here, instead, we are to propose a more general approach, as detailed starting from Sec.~\ref{sec:individual_galaxy}).} 

\fr{It is indeed possible for structures not directly modelled by \textsc{bang} (such as  bars, thick discs, etc.) to be mimicked by some combination of the structures actually included in the modelling (i.e., the bulge, the inner and outer discs). As an example, thick discs could possibly be modelled as the superposition of components "$\rm D_1$" and "$\rm D_2$" weighted with appropriate values of the kinematic decomposition parameters $k_1$ and $k_2$, while strongly non-axisymmetric components such as bars can only be modelled in their azimuth-averaged properties. We emphasize that even when the inner and outer discs are modelled as razor-thin exponential components, caution must be payed in interpreting them as dynamically cold thin discs. A physically motivated discussion cannot neglect considering their kinematic state.}

\section{Analysis of an individual galaxy}
\label{sec:individual_galaxy}
In this section, we show the results of our analysis focusing on a specific galaxy. For this purpose we choose \fr{1-256009} (MaNGA-ID) an S0a galaxy\footnote{According to the MaNGA Visual Morphology Catalogue (VMC-VAC) \citep[][]{Vazquez_2022}} exhibiting a regular rotation pattern at redshift  $z=0.04935$.

\begin{table}
	\centering
	\caption{Results of the fit over the 1-256009 galaxy. The first column is the name of the parameter while the second and the third columns refer the best-fit value and the errors, estimated as the $50^{th}$, the $95^{th}$ and the $5^{th}$ percentiles of its marginalised posterior distribution, respectively in the "halo-fixed" and "halo-free" configurations. }
	\label{tab:best_fit_parameters}
	\begin{tabular}{lcr} 
		\hline
		Parameter & halo-fixed & halo-free \\
		\hline \rowsep
		$x_0 [\mathrm{kpc}] $ &  $0.227^{+0.002}_{-0.003}$ & $0.225^{+0.002}_{-0.002}$ \\ \rowsep
		$y_0 [\mathrm{kpc}]$ &  $0.244^{+0.002}_{-0.002}$ & $0.245^{+0.002}_{-0.002}$\\  \rowsep
		$\rm P.A. [\mathrm{deg}]$ &  $22.6^{+0.2}_{-0.2}$ & $22.7^{+0.2}_{-0.2}$ \\ \rowsep
		$i$ &  $50.7^{+0.3}_{-0.3}$ & $50.7^{+0.3}_{-0.3}$\\ \rowsep
		$\log_{10}(M_{\rm b}/\rm M_{\sun}) $ &  $10.69^{+0.01}_{-0.02}$ & $10.73^{+0.01}_{-0.01}$\\ \rowsep
		$\log_{10}(R_{\rm b}/\mathrm{kpc}) $ &  $-1.05^{+0.04}_{-0.04}$ & $-0.96^{+0.03}_{-0.03}$ \\ \rowsep
		$\gamma $ &  $0.06^{+0.14}_{-0.06}$ & $0.06^{+0.14}_{-0.06}$ \\ \rowsep
		$\log_{10}(M_{\rm d,1}/\rm M_{\sun})$ &  $10.89^{+0.01}_{-0.01}$ & $10.87^{0.01}_{0.01}$ \\  \rowsep
		$\log_{10}(R_{\rm d,1}/\mathrm{kpc}) $ &  $0.20^{+0.01}_{-0.01}$ & $0.26^{+0.01}_{-0.01}$ \\ \rowsep
		$\log_{10}(M_{\rm d,2}/\rm M_{\sun})$ &  $10.79^{+0.02}_{-0.02}$ & $10.76^{+0.03}_{0.04}$ \\ \rowsep
		$\log_{10}(R_{\rm d,2}/\mathrm{kpc}) $ &  $0.95^{+0.01}_{-0.01}$ & $1.11^{+0.02}_{-0.02}$ \\ \rowsep
		$\log_{10}(f)$ & $1.94^{+0.01}_{-0.01}$ & $1.8^{+0.2}_{-0.1}$ \\ \rowsep
		$c$ &  $7.81^{+0.02}_{-0.03}$ & $10.0^{+1}_{-1.3}$ \\ \rowsep
		$\log_{10}{M_{\rm h,5}/\rm M_{\sun}}$ & $12.15^{+0.02}_{-0.02}$ & $12.38^{+0.06}_{-0.05}$ \\ \rowsep
		$(M/L)_{\rm b} [\rm{M_{\sun}/L_{\sun}}]$ &  $2.8^{+0.2}_{-0.2}$ & $3.2^{+0.1}_{-0.1}$ \\ \rowsep
		$(M/L)_{\rm d,1} [\mathrm{M_{\sun}/L_{\sun}}]$ &  $3.78^{+0.08}_{-0.09}$ & $3.3^{+0.1}_{-0.1}$ \\ \rowsep
		$(M/L)_{\rm d,2} [\mathrm{M_{\sun}/L_{\sun}}]$ &  $1.17^{+0.06}_{-0.06}$ & $0.81^{+0.1}_{-0.1}$ \\ \rowsep
		$k_1$ &  $0.42^{+0.01}_{-0.01}$ & $0.44^{0.01}_{0.01}$ \\ \rowsep
		$k_2$ &  $0.53^{+0.01}_{-0.01}$ & $0.53^{-0.01}_{+0.01}$ \\
		\hline
	\end{tabular}
\end{table}

We fitted this galaxy, as we did for the whole sample, with three different models ("$\rm B+D_1+D_2$","$\rm B+D_1$","$\rm D_1+D_2$") choosing the most statistically favoured according to the highest evidence criterion, "$\rm B+D_1+D_2$" in this case. We report in Tab.~\ref{tab:best_fit_parameters} the best-fit parameters and their upper and lower errors \footnote{As already pointed out in \citealt{Rigamonti_2022} most of the statistical errors are often smaller than a few percent; this is a quite common situation when using nested sampling algorithms, related to the simplified nature of the model not taking into account the whole underlying physics governing the problem at hand thus implicitly reducing possible degeneracies between parameters.} computed as the $50^{th}$, the $95^{th}$ and the $5^{th}$ percentiles of the marginalised posterior distribution, both in the "halo-fixed" and "halo-free" configurations. The two models generally agree with a small scatter between most of the relevant parameters, especially regarding the stellar components. Given the limited spatial extension of the MaNGA kinematics, it is not clear whether and in which cases the halo parameters can be really constrained by the data, this is why, from now on, following a conservative approach, we will focus our analysis on the "halo-fixed" configuration.\footnote{Please refer to table~\ref{tab:parameters} and appendix~\ref{app:Scaling_halo_fixed} for the discussion about the statistical consistency of the results obtained using  the two different fitting procedures.}

\begin{figure*}
    \centering
    \includegraphics[scale=0.385]{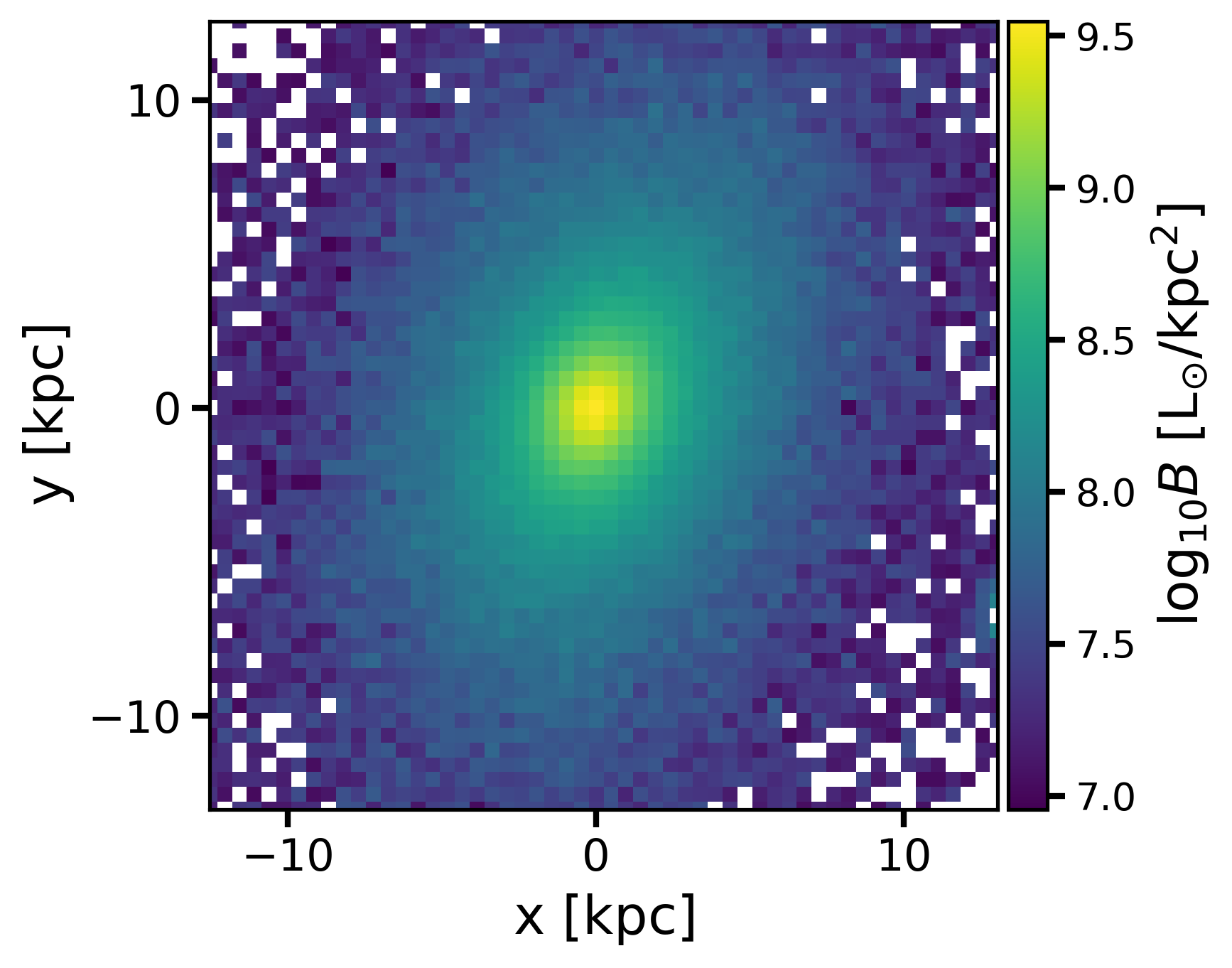}\hspace{0.015\textwidth}
    \includegraphics[scale=0.385]{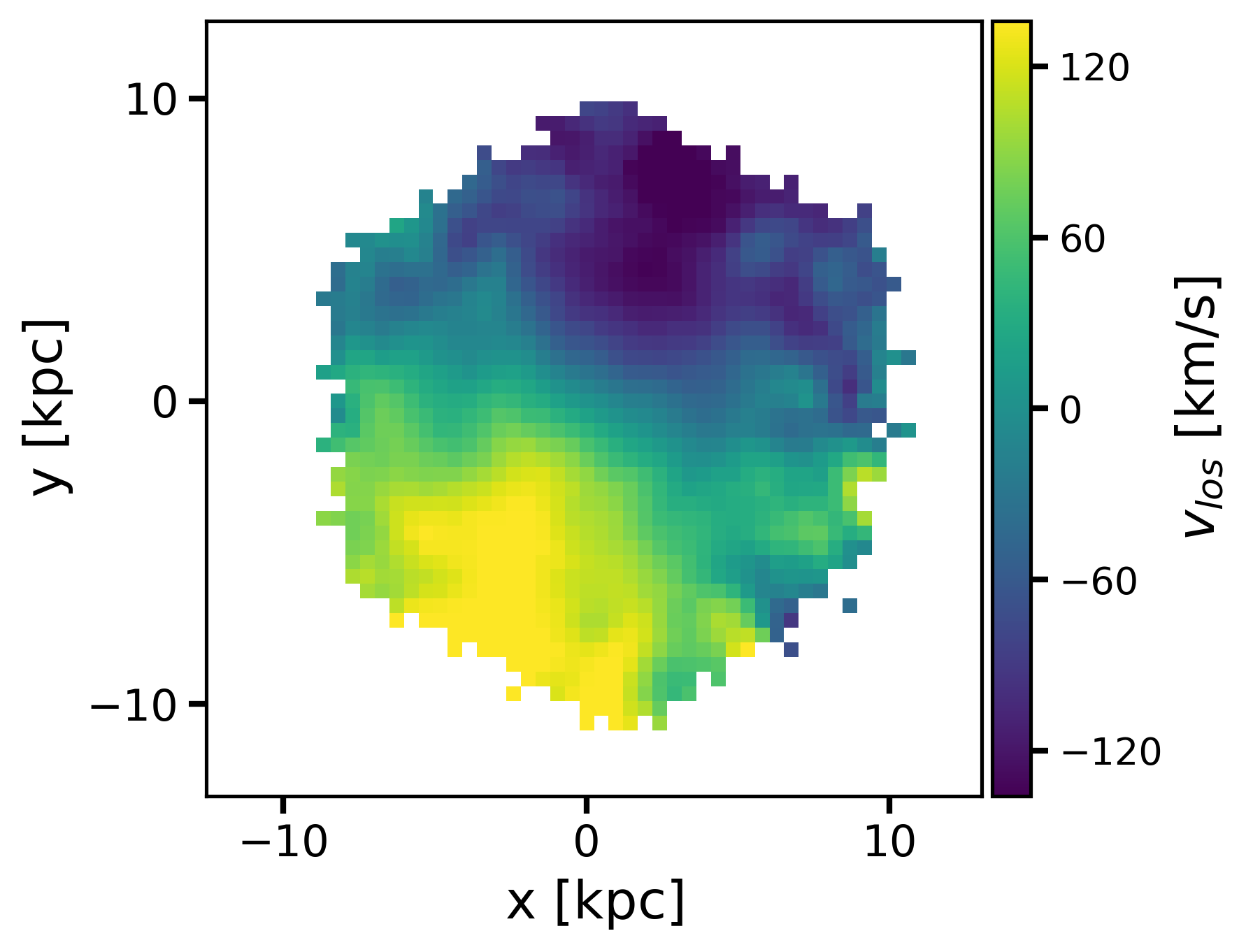}\hspace{0.015\textwidth}
    \includegraphics[scale=0.385]{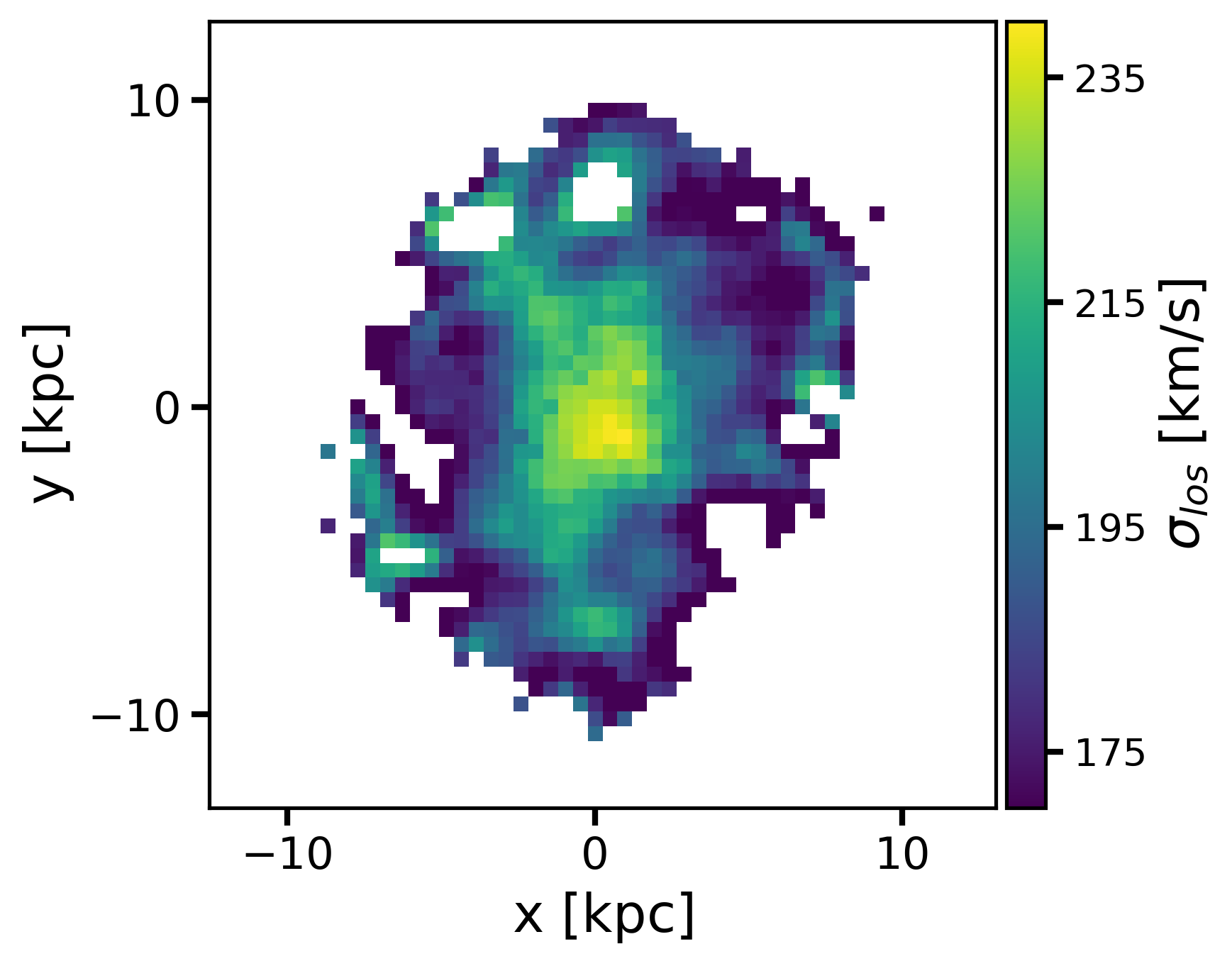}\vspace{0.0075\textheight}

    \includegraphics[scale=0.385]{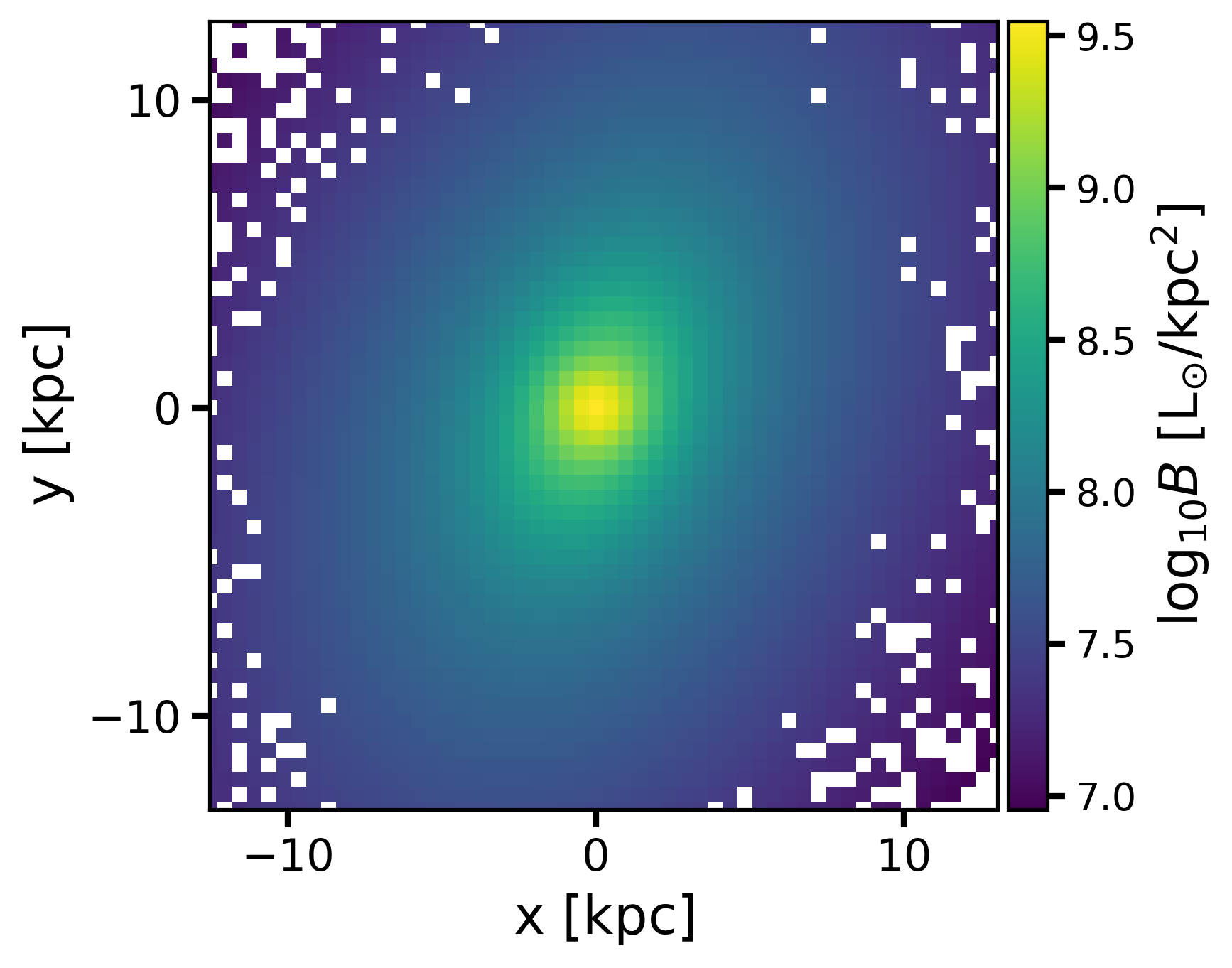}\hspace{0.015\textwidth}
    \includegraphics[scale=0.385]{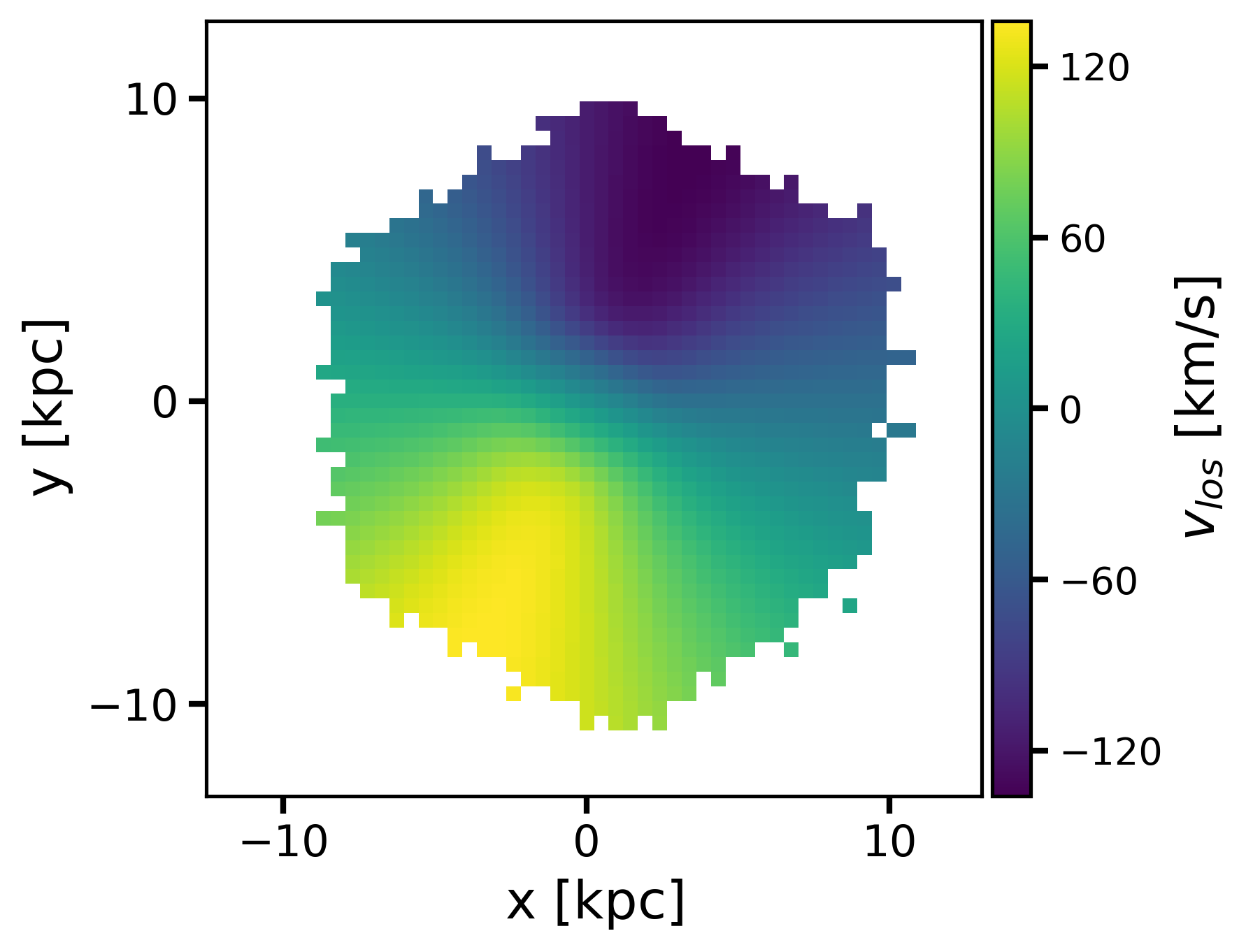}\hspace{0.015\textwidth}
    \includegraphics[scale=0.385]{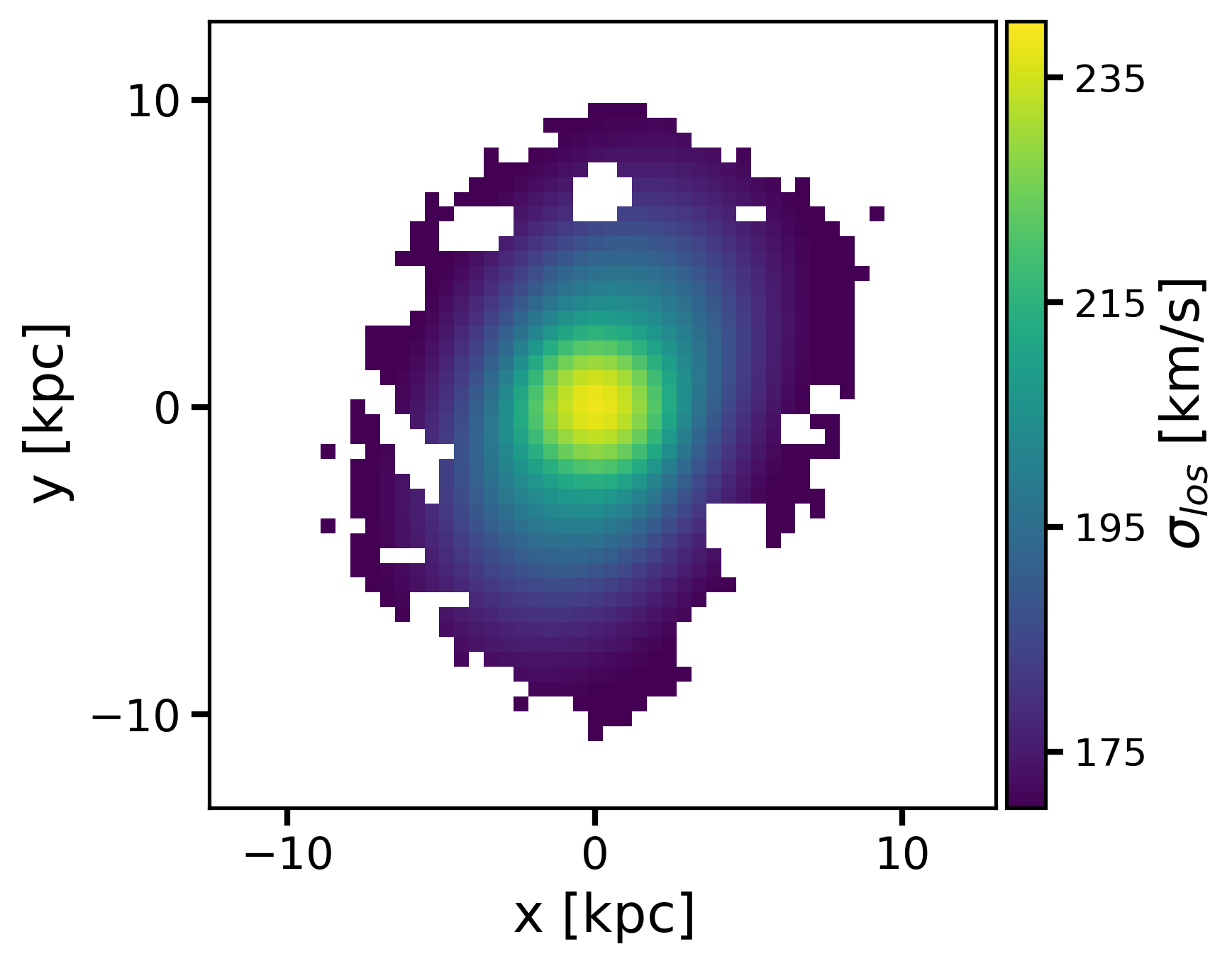}\vspace{0.0075\textheight}
    
    \includegraphics[scale=0.385]{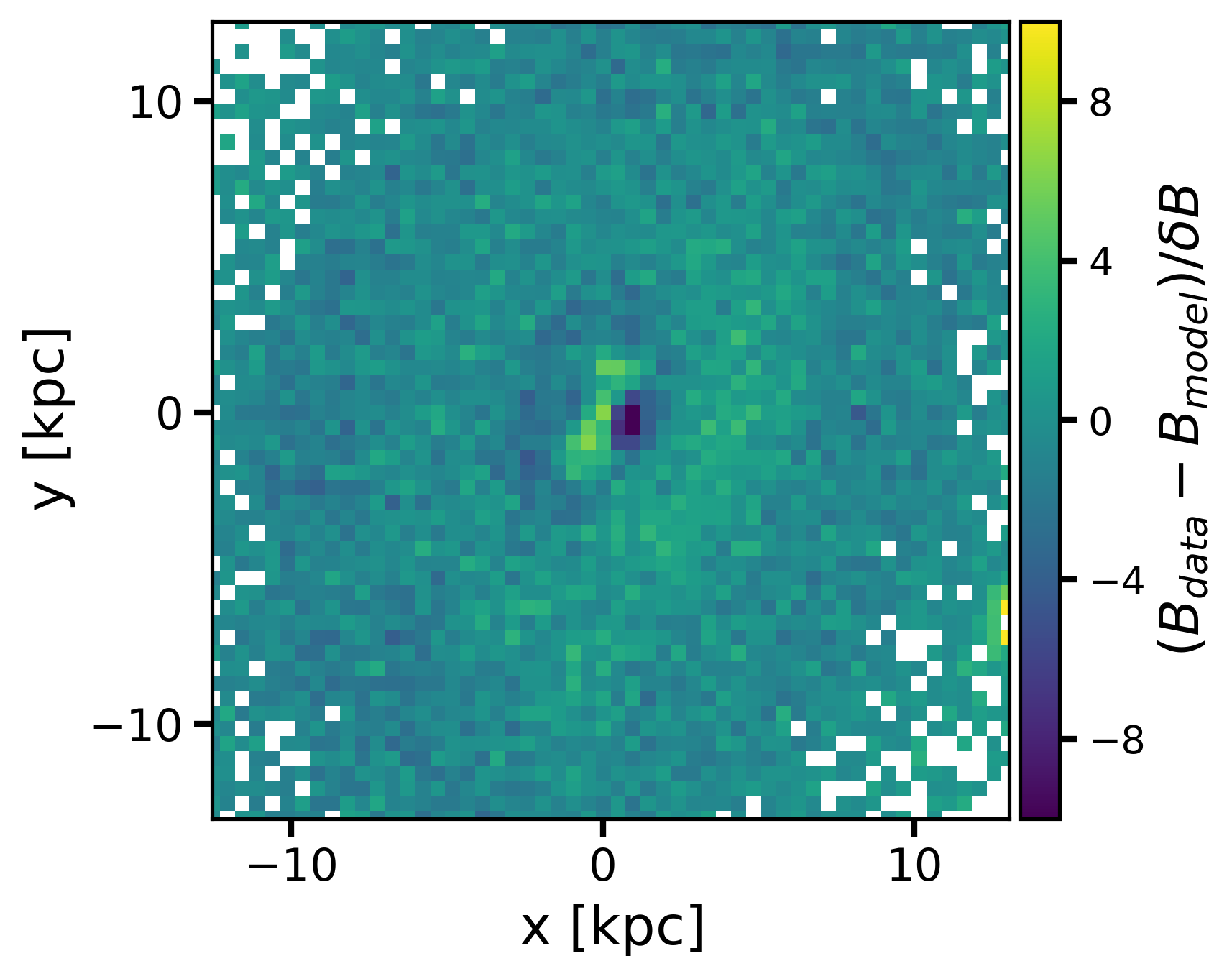}\hspace{0.015\textwidth}
    \includegraphics[scale=0.385]{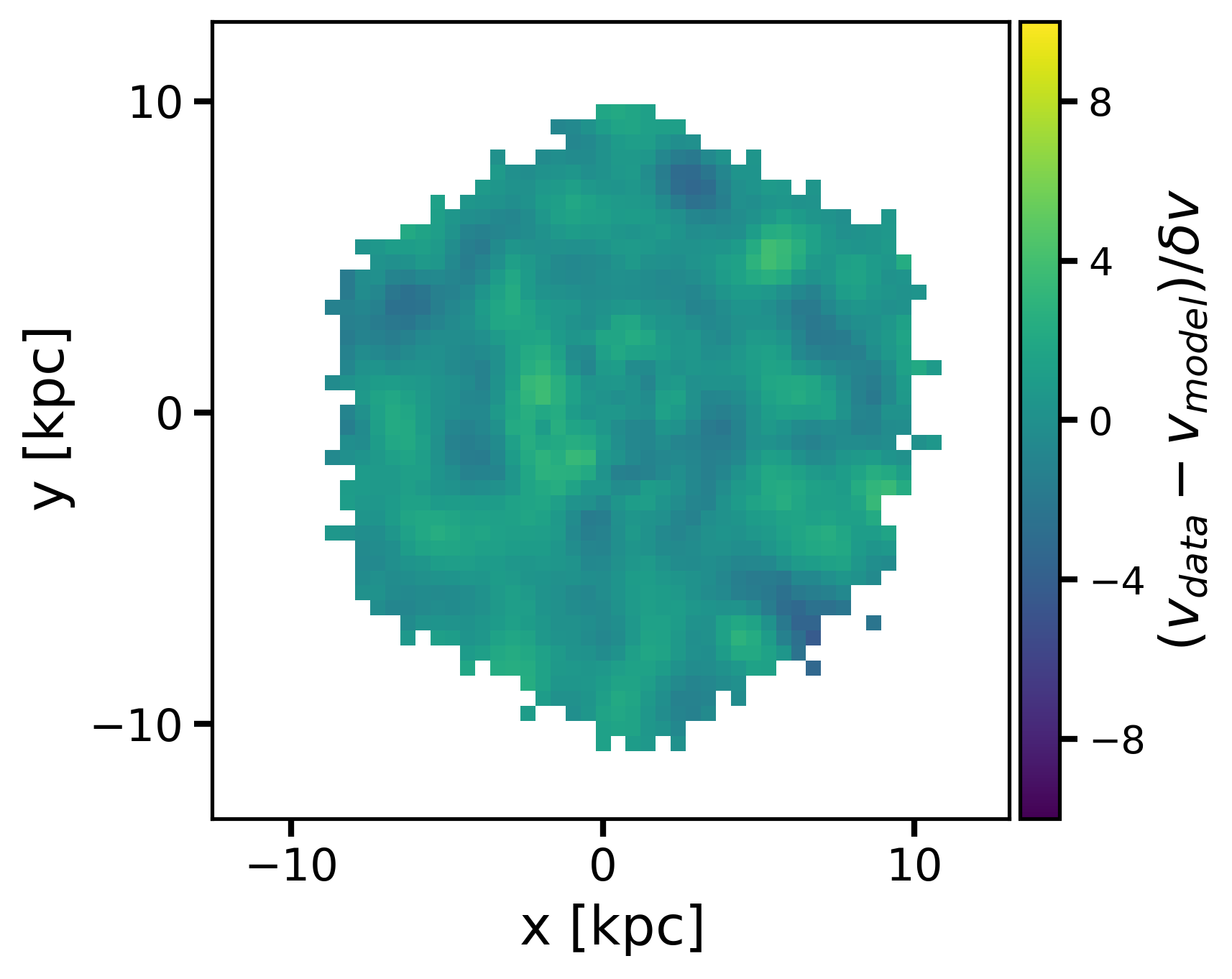}\hspace{0.015\textwidth}
    \includegraphics[scale=0.385]{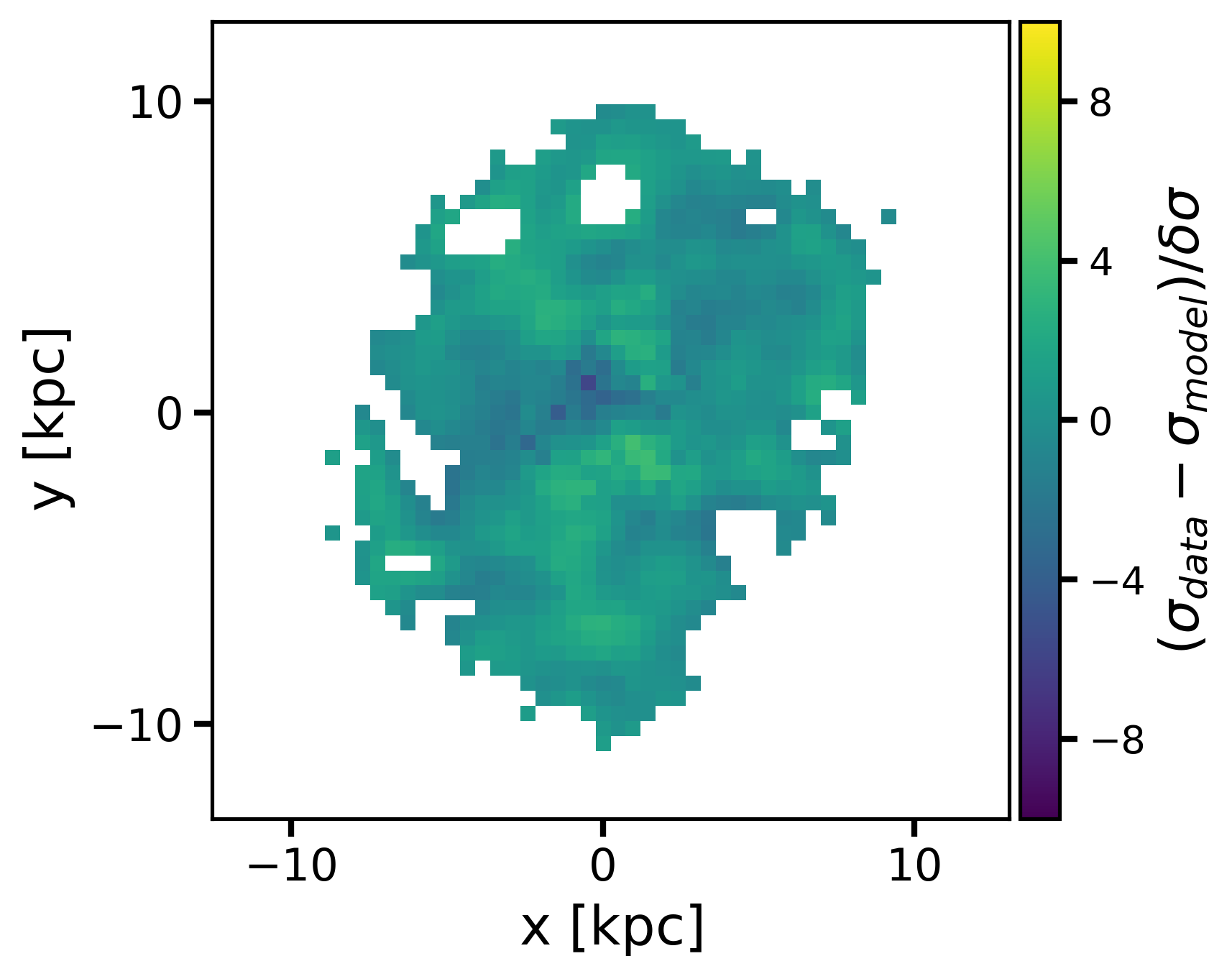}
    
    \caption{Best fit model of the 8258-6102 galaxy. The first, second and third columns refer to the surface brightness, the line of sight velocity and the line of sight velocity dispersion, respectively. From top to bottom we report the observational data, the best fit model and the residual maps. The residuals are computed as differences between the data and the model divided by the data errors $\delta B$ $\delta v$, $\delta \sigma$.}
    \label{fig:8258-6102_nine_2d_maps}
\end{figure*}

\begin{figure*}
    \centering
    \includegraphics[scale=0.37]{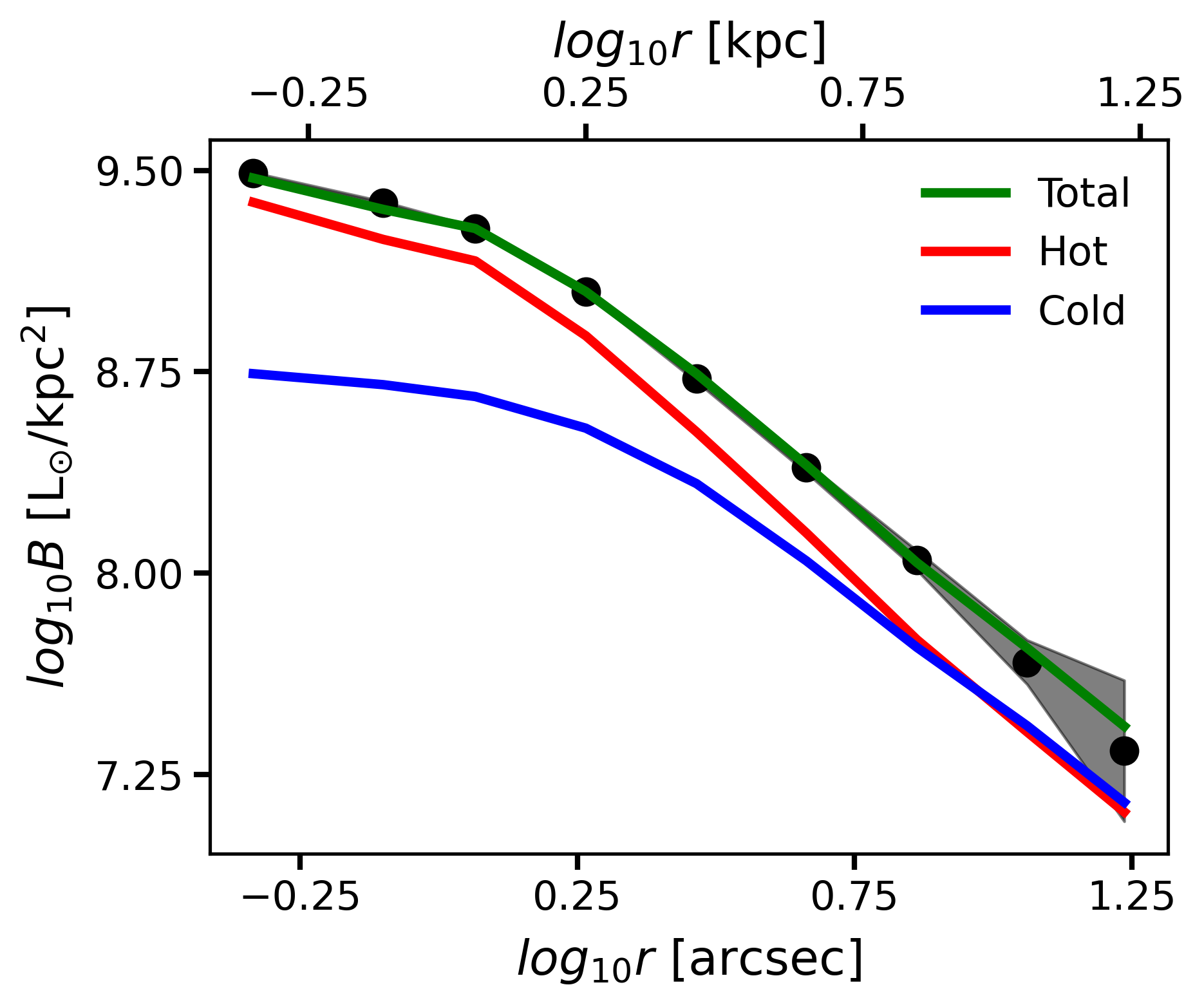}\hspace{0.0095\textwidth}
    \includegraphics[scale=0.37]{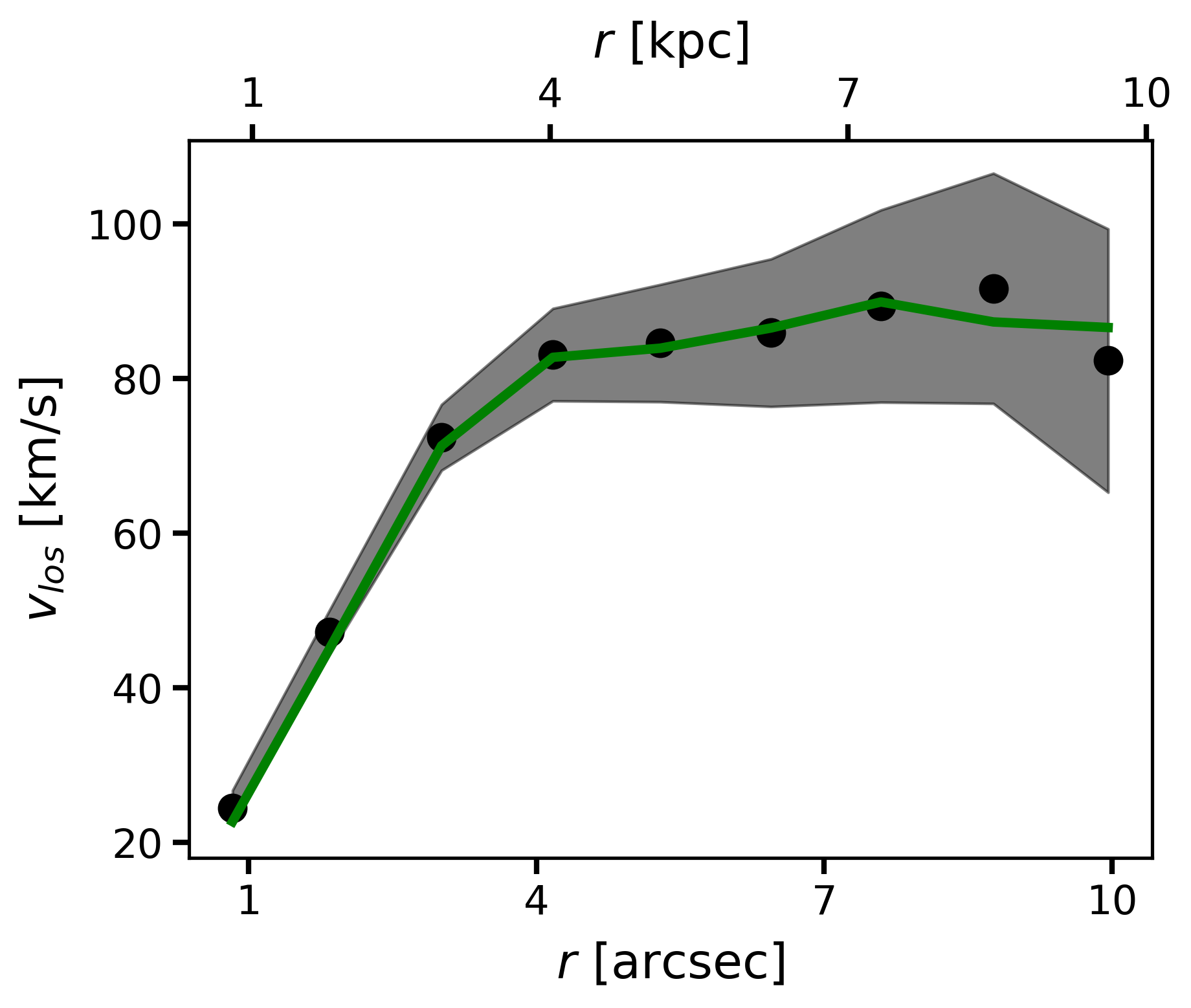}\hspace{0.0095\textwidth}
    \includegraphics[scale=0.37]{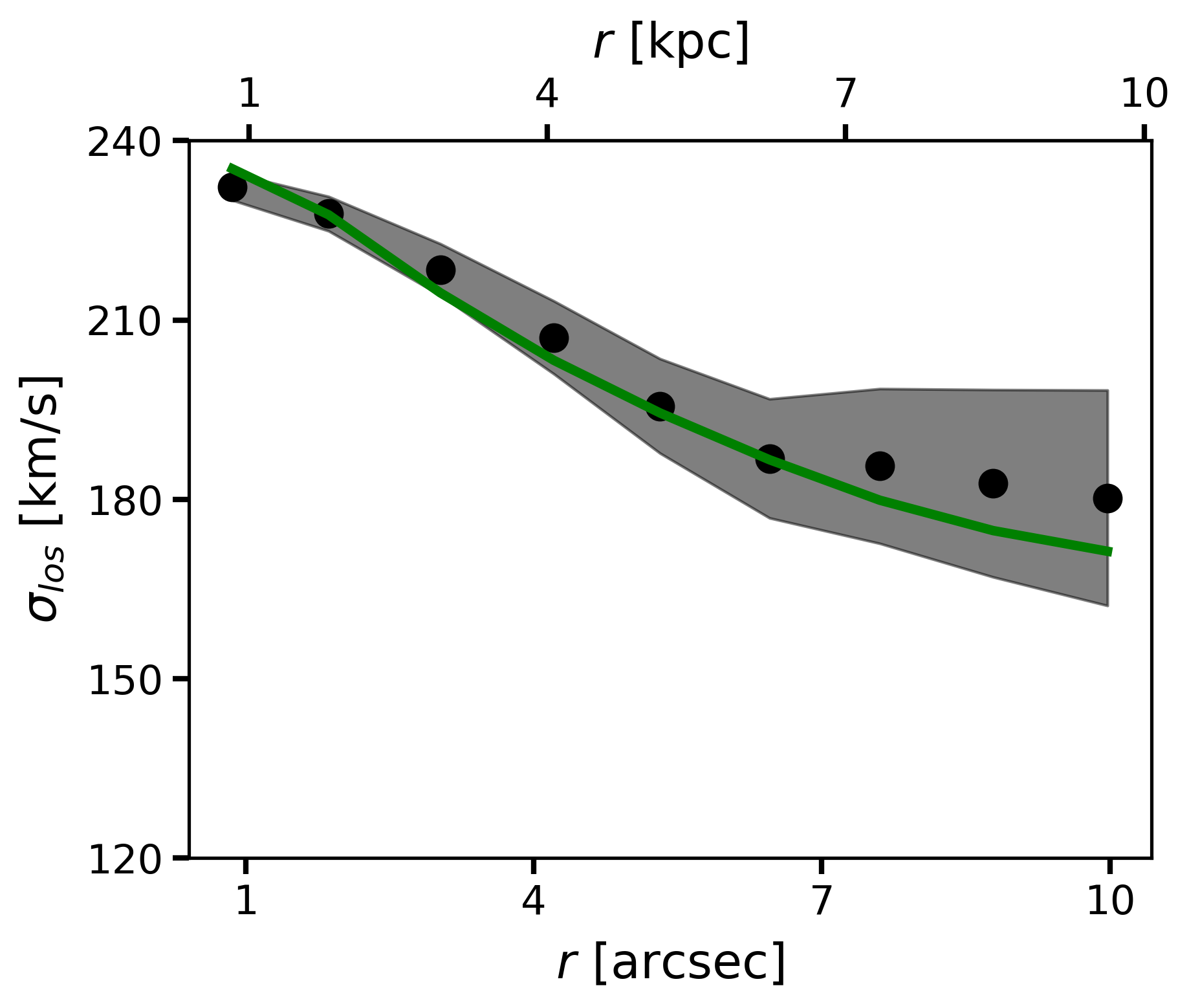}
    \caption{Average radial profiles of the logarithm of the surface brightness (left-hand panel), of the velocity (middle panel) and of the line of sight velocity dispersion (right-hand panel). The data are reported as black dots, with the errors, computed by an azimuthal average, shown as a grey-shaded area. The green lines refer to the best-fit model, while the red and blue lines refer to a kinematic decomposition of the galaxy in its dispersion-dominated and rotation-dominated components (we refer to Sec.~\ref{sec:individual_galaxy} for further details on this tentative decomposition). }
    \label{fig:radial_profile_8258-6102_nine}
\end{figure*}

Fig.~\ref{fig:8258-6102_nine_2d_maps} illustrates the best-fit model obtained from our fit for the logarithm of the surface brightness (left column), the line of sight velocity (middle column), and the line of sight velocity dispersion (right column). The first and the second rows refer respectively to the data and the best-fit model, while in the last row, we show the residuals respectively divided by the data errors: $\delta B$ (for the brightness), $\delta v$ (for the velocity) and $\delta \sigma$ (for the velocity dispersion). Comparing our model predictions with the data, we generally find a quite good agreement between all three quantities, as can be shown from the residual maps in the third row. In this case, the brightness differences between our model and the data (left panel, third row) highlight the presence of a nuclear structure poorly fitted by our smooth and axially symmetric model.

In Fig. \ref{fig:radial_profile_8258-6102_nine}, we show the azimuthally average profiles of the brightness (left), velocity (middle) and velocity dispersion (centre) for the full model (green line) showing a good agreement between our best-fit model and the data (black dots).


We decided to decompose the galaxy into two more general components,  dynamically hot and  cold respectively. Starting from the parameter of Tab.~\ref{tab:best_fit_parameters} we use the information about the originally assumed (bulge and discs) components to define the brightness profiles of the  hot and cold constituents as:    
    \begin{equation}
        \label{eq:hot_brightness}
        B_{\rm{hot}}(r) = B_{b}(r)+(1-k_1) B_{d,1}(r)+(1-k_2)  B_{d,2}(r),
    \end{equation}
and similarly for the cold component:
    \begin{equation}
        \label{eq:cold_brightness}
        B_{\rm{cold}}(r) = k_1 B_{d,1}(r)+ k_2  B_{d,2}(r).
    \end{equation}

We propose here a decomposition which accounts both for the geometrical structure of the galaxy constituents and for their kinematics state; keeping this in mind we include in the "hot" component the bulge, which is entirely supported by dispersion, and a fraction of each disc in a way proportional to their degree of random motion \footnote{Remember that the larger is $0<k_{1;2}<1$ the more is the rotational motion of the disc, thus implying that $1-k_{1;2}$ is proportional to the dispersion support.}. As opposed, we define the "cold" component as the disc brightness proportional to the rotation and hence proportional to $k_1$ and $k_2$.

Our decomposition can be visualised in the left panel of Fig.~\ref{fig:radial_profile_8258-6102_nine} where we show the "hot" and "cold" brightness profiles, respectively in red and blue colours, as a function of the distance from the galaxy centre. The "hot" component almost dominates in the whole physical region being, as expected, more important in the centre; we note that the two discs, given their mild rotation ($k_1 = 0.42$ and $k_2 = 0.53$), add a relevant contribution to the "hot" component.  

We report in Tab.~\ref{tab:best_fit_parameters_decoposition} masses and half-mass radii of the "hot" and "cold" decomposition together with their fraction for the whole galaxy and within the effective radius ($R_e$). We refer to Sec.~\ref{sec:Comparison with previous dynamical modelling estimates} for further discussion about any biases and limitations about our definition of "hot" and "cold" components.


\begin{table}
	\centering
	\caption{Table summary of decomposition adopted in par.~\ref{sec:individual_galaxy} for the 1-256009 galaxy reporting total stellar masses and half-mass radii of each component together with their light and mass fraction in respect to the whole galaxy and within the $R_e$. The first two columns are the description of the parameter together with the associated name while the third column refers to the best-fit value and the errors, estimated as in Tab.~\ref{tab:best_fit_parameters}.}
	\label{tab:best_fit_parameters_decoposition}
	\begin{tabular}{lcr} 
		\hline
		description & name & value \\
		\hline \rowsep
		 total stellar mass & $\log_{10}{(M_{\star}/M_{\odot})} $ & $11.27^{+0.01}_{-0.01}$ \\ \rowsep
		 total hot mass & $ \log_{10}{(M_{\rm hot}/M_{\odot})}$ & $11.09^{+0.01}_{-0.01}$\\  \rowsep
		total cold mass &  $ \log_{10}{(M_{\rm cold}/M_{\odot})}$ & $10.81^{+0.01}_{-0.01}$\\ \rowsep
		half-mass radius &  $ \log_{10}{(R_{\rm half,\star}/\rm kpc)}$ & $0.50^{+0.01}_{-0.01}$\\ \rowsep
		 half-mass radius hot component &  $ \log_{10}{(R_{\rm half,\rm hot}/\rm kpc)}$ & $0.33^{+0.01}_{-0.01}$\\ \rowsep
		 half-mass radius cold component &  $ \log_{10}{(R_{\rm half,\rm cold}/\rm kpc)}$ & $0.75^{+0.02}_{-0.02}$ \\ \rowsep
		 hot ratio (mass) &  $ M_{\rm {hot}}/M_{\star}$ & $0.65^{+0.01}_{-0.01}$ \\ \rowsep
		 cold ratio (mass) &  $M_{\rm {cold}}/M_{\star}$ & $0.35^{+0.01}_{-0.01}$ \\ \rowsep
		 hot ratio (light) &  $L_{\rm {hot}}/L_{\star}$ & $0.60^{+0.01}_{-0.01}$ \\ \rowsep
		 cold ratio (light) &  $ L_{\rm {cold}}/L_{\star}$ & $0.40^{+0.01}_{-0.01}$ \\
		 \rowsep
         hot ratio within $R_e$ (mass) &  $ M_{\rm {hot}}(R_e)/M_{\star}(R_e)$ & $0.73^{+0.01}_{-0.01}$ \\ \rowsep
		 cold ratio within $R_e$ (mass) &  $M_{\rm {cold}}(R_e)/M_{\star}(R_e)$ & $0.27^{+0.01}_{-0.01}$ \\ \rowsep
		 hot ratio within $R_e$ (light) &  $ L_{\rm {hot}}(R_e)/L_{\star}(R_e)$ & $0.71^{+0.01}_{-0.01}$ \\ \rowsep
		 cold ratio within $R_e$ (light) &  $ L_{\rm{cold}}(R_e)/L_{\star}(R_e)$ & $0.29^{+0.01}_{-0.01}$ \\
		\hline
	\end{tabular}
\end{table}

\section{Comparison with previous dynamical modelling estimates}
\label{sec:Comparison with previous dynamical modelling estimates}

\begin{figure*}
    \centering
    \includegraphics[scale=0.37]{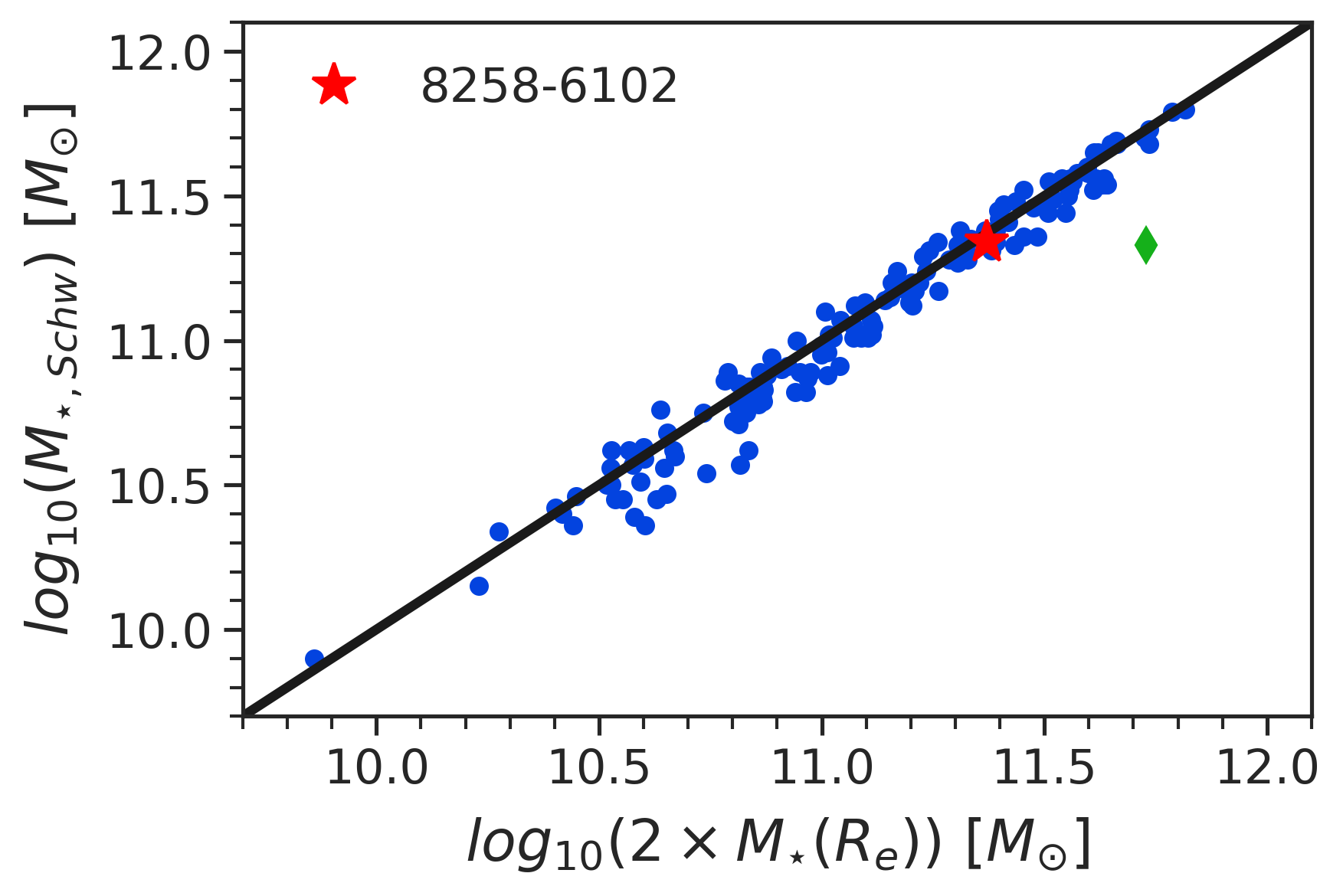}\hspace{0.0095\textwidth}
    \includegraphics[scale=0.37]{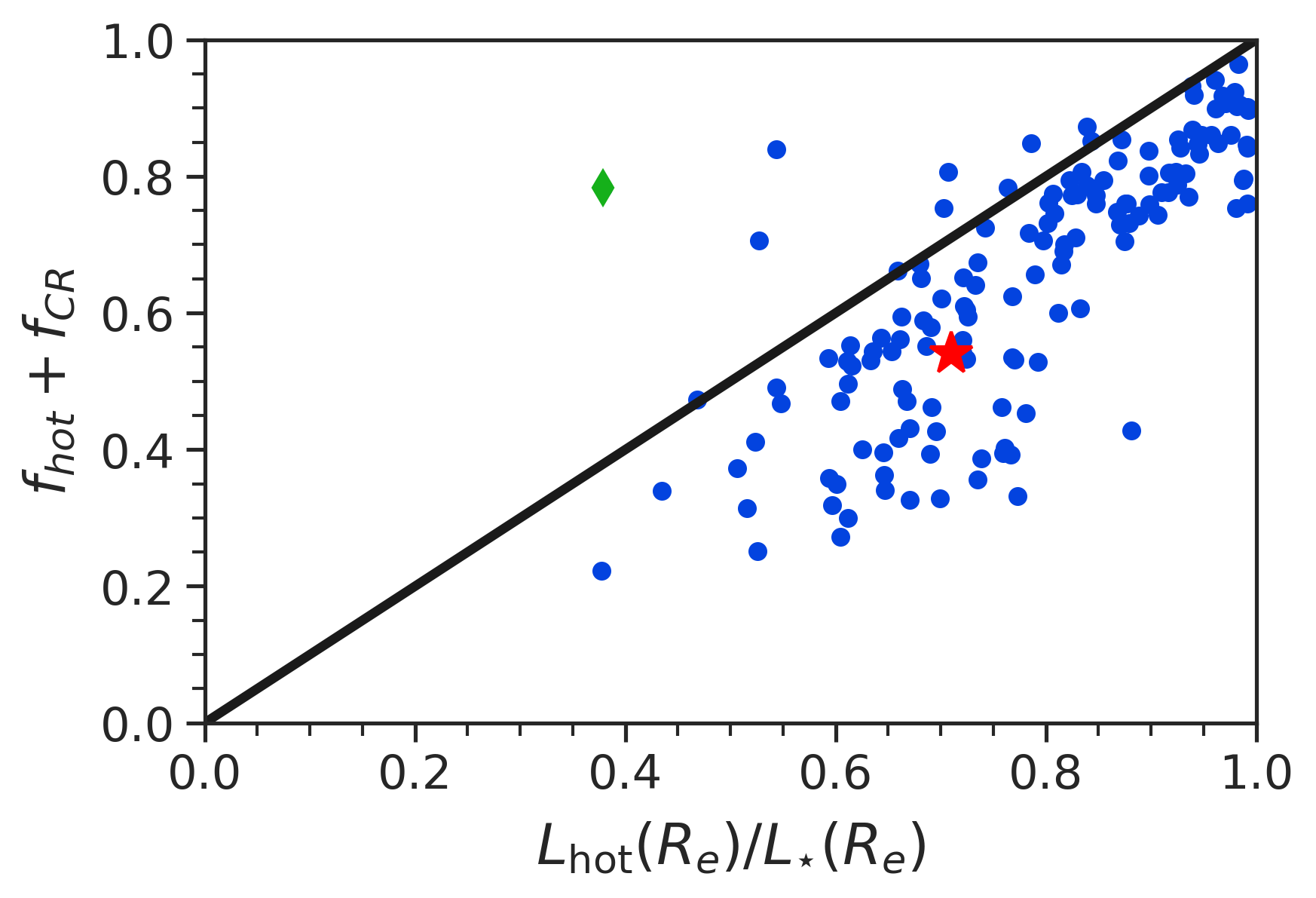}\hspace{0.0095\textwidth}
    \includegraphics[scale=0.37]{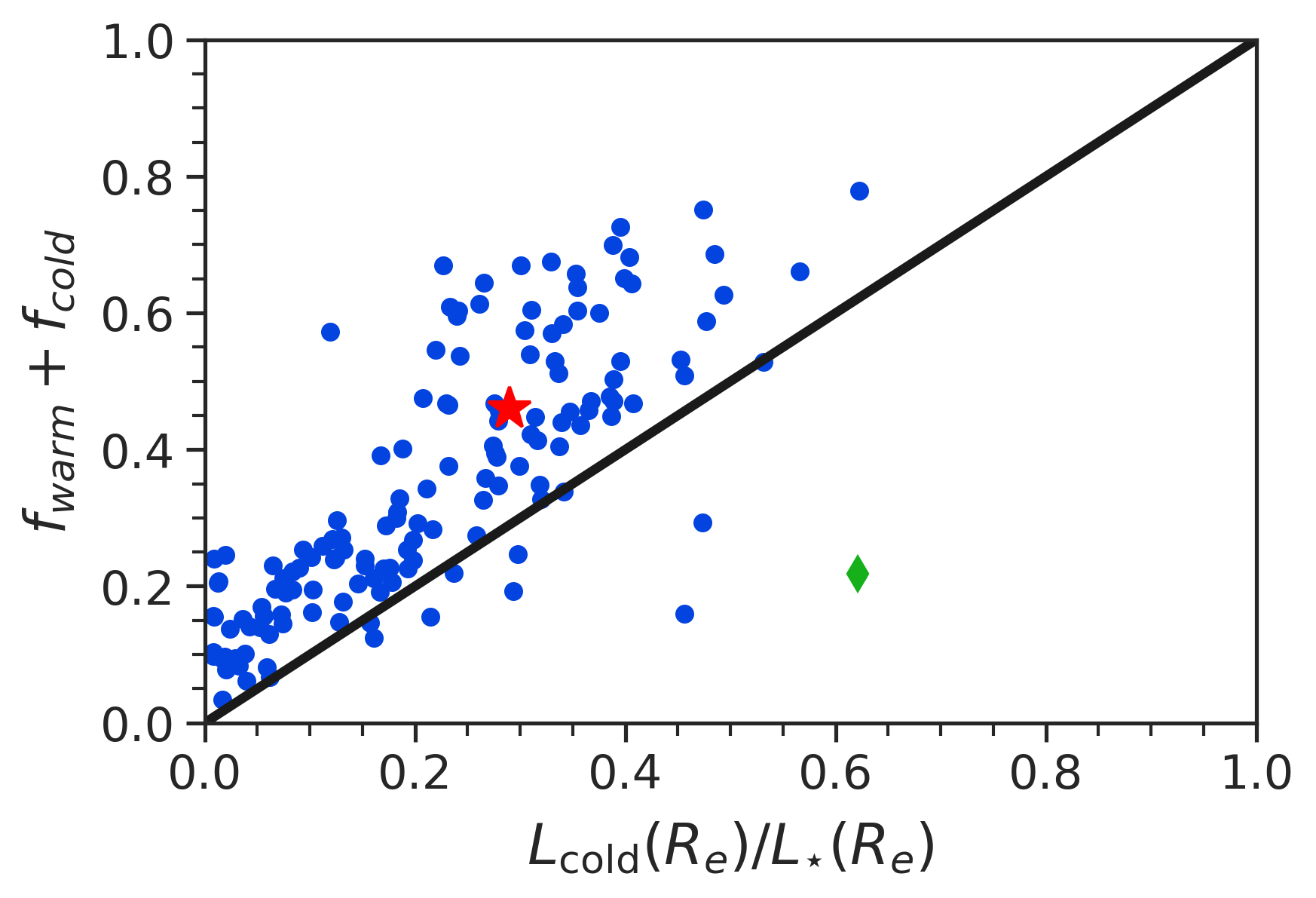}
    \caption{Comparison between our estimates and predictions from the Schwarzschild orbit superposition method on the 149 galaxies analysed in \citep{Jin_2020}. From left to right we compare our estimates (on the x-axis) of stellar mass, hot and cold luminosity weighted fractions respectively with the stellar mass, "hot"+"counter-rotating" orbital fractions and "warm" + "cold" orbital fractions presented in \citealt{Jin_2020}. The stellar masses are estimated as twice the mass at the effective radius ($R_e$) while our hot and cold fractions are estimated from eqs.~\ref{eq:hot_luminosity_Re},~\ref{eq:cold_luminosity_Re} and ~ \ref{eq:tot_luminosity_Re}. The red star refers to the galaxy presented in \ref{sec:individual_galaxy} (i.e.  8258-6102), the blue dots refer to the other 148 galaxies while the black line shows the one-to-one relation.}
    \label{fig:compare_with_dynamite}
\end{figure*}

\fr{In this section, we compare our results for the same sub-sample of 149 early-type galaxies (ETGs) analysed by \citet{Jin_2020} with a Schwarzschild orbit superposition method \citep[][]{Schwarzschild1979,van_den_Bosch_2008,DYNAMITE}}\footnote{Note that the work presented in \cite{Jin_2020} has been completed before the discovery of a minor and statistically irrelevant bug in the orbit mirroring scheme present in their implementation \citep{Thater_2022}.}.

Orbit-based methods represent the state-of-the-art dynamical modelling techniques allowing deep investigation of the internal structure of galaxies, fully characterizing their phase-space distribution and thus offering an optimal benchmark for comparison. 

Despite the robustness of these modelling techniques, their applicability, mainly due to the large computational cost, has been limited to a few hundred galaxies.
\textsc{bang} can offer an alternative and also a complementary approach to the problem of galactic dynamical modelling, especially in cases of large data samples. We, therefore, test our approach against such reliable modelling techniques to prove and check the robustness of our modeling.

In the left panel of Fig.~\ref{fig:compare_with_dynamite}, we compare stellar mass predictions, again as 
twice the mass within the effective radius. The agreement is good in the whole range of masses with an average scatter, computed as the standard deviation of the differences, of $\simeq 0.07~\rm dex$, corresponding to a mean relative error of $\simeq 6 \%$. Only one outlier is evident, identified in 8263-9102 (green diamond on the plots), an elliptical galaxy clearly showing a twisting of the rotation pattern implying a geometrical structure far from axial symmetry and, hence, not consistent with \textsc{bang}'s modelling assumptions. Given the rarity of those examples in the MaNGA dataset, we expect our analysis not to suffer from any biases arising from those peculiar cases \citep[][]{Jin_2016,Feng_2022}{}{}.

In the middle and right panels of Fig.~\ref{fig:compare_with_dynamite} we show a comparison between our tentative "hot" vs "cold" decomposition (horizontal axis) and the orbital analysis extrapolated from Schwarzschild modelling \citep[][vertical axis]{Jin_2020}. Finding a way to compare the two analyses 
is not straightforward. \cite{Jin_2020} orbital decomposition is luminosity-weighted and limited to $R_e$. We therefore computed our "hot", "cold" and "total" luminosity as:

\begin{equation}
    \label{eq:hot_luminosity_Re}
    L_{\rm{hot}}(R_e) = \frac{M_{b}(R_e)}{(M/L)_b} + (1-k_1)\frac{M_{d,1}(R_e)}{(M/L)_{d,1}} + (1-k_2)\frac{M_{d,2}(R_e)}{(M/L)_{d,2}},\\
\end{equation}
\begin{equation}
    \label{eq:cold_luminosity_Re}
    L_{\rm{cold}}(R_e) =  k_1\frac{M_{d,1}(R_e)}{(M/L)_{d,1}} + k_2\frac{M_{d,2}(R_e)}{(M/L)_{d,2}},
\end{equation}

\begin{equation}
    \label{eq:tot_luminosity_Re}
    L_{\star}(R_e) =  \frac{M_{b}(R_e)}{(M/L)_b} + \frac{M_{d,1}(R_e)}{(M/L)_{d,1}} + \frac{M_{d,2}(R_e)}{(M/L)_{d,2}},
\end{equation}
where $M_j(R_e)$ with ($j=\{b;d,1; d,2\}$) is the enclosed mass of the j-th component within $R_e$.
Moreover, we compared our "hot" fraction with the sum of "hot" and "counter-rotating" orbits since those are the ones expected to have the largest contribution to dispersion; similarly, most of the rotational support (our "cold" fraction) should reside in both "warm" and "cold" orbits. We highlight that the counter-rotating fraction decreases only by a slight amount the scatter in the middle panel of Fig.~\ref{fig:compare_with_dynamite}, acting more as a systematic shift and indicating almost no correlation with $L_{\rm{hot}}(R_e)/L_{\star}(R_e)$. Similarly, in these galaxies, $f_{\rm{cold}}$ is generally small, making it difficult to quantify whether $L_{\rm{cold}}(R_e)/L_{\star}(R_e)$ is more a proxy of the "warm" or "cold" orbital families.  

The two pictures show a correlation in the "hot" and the "cold" cases\footnote{Note that our definitions of "hot" and "cold" components are complementary to each other so that a relationship seen with one of the two quantities is present also with the other by construction.} even though the scatter
is quite relevant ($\simeq 0.12$) given the limited dynamic range. A systematic offset ($\simeq 0.12$) is present in both the "hot" and "cold" fractions mainly due to intrinsic differences in the two approaches; 
we account for a decomposition proportional to $k_1$ and $k_2$, meaning that even when they are large enough (i.e. $k_{1,2}\gtrsim0.8$) to suggest the presence of a dynamically cold disc, still a non-negligible amount of the galaxy mass will be added to the hot component possibly resulting in an overestimation of its contribution. Differently, in Schwarzschild-based approaches, orbits are divided into families depending on constant circularity thresholds thus assigning, in the example of a dynamically cold disc, all of the mass to the cold component. \fr{Note that the offset between the hot (cold) fraction predicted by \textsc{dynamite} analysis on simulated galaxies shown in \citet{Jin_2019} is consistent (but opposite in sign) to the offset of our decomposition with respect to \textsc{dynamite} shown in the middle (right) panel of Fig.~\ref{fig:compare_with_dynamite}. This may suggest that the main source of disagreement between \textsc{dynamite} and \textsc{bang} could be due to an underestimate of hot orbits in \textsc{dynamite} and not necessarily to an overestimate in \textsc{bang}. A more quantitative analysis of the differences between the two approaches is beyond the scope of this paper.}


\section{Preliminary analysis on the proposed decomposition}
\label{sec:Preliminary analysis on the proposed decomposition}
In this section, we present some of the results obtained from our analysis focusing mainly on the hot-cold decomposition proposed in the previous paragraphs and characterizing it in terms of visual morphology and kinematics. \fr{We show here the results of our analysis performed over the whole MaNGA sample comprising, as reported in Sec.~\ref{sec:methodology}, 10,005 galaxies roughly composed by $ 62 \%$ of spirals, $21\%$ of lenticulars and $17\%$ of ellipticals\footnote{According to the VMC-VAC \citealt{Vazquez_2022}} with a best fit corresponding to a "$\rm B + D_1 + D_2$", "$\rm B+D_1$" a "$\rm D_1+D_2$" model respectively in $\simeq 86\%$, $\simeq 6\%$ and $\simeq \rm 8\%$ of the cases. Unless otherwise specified, all the forthcoming analysis is performed on such sample, which we refer to as the "main sample". We moreover refer to Appendix~\ref{app:Analysis_golden_sample} for the same analysis on a selected subsample.} 

\begin{figure}
    \centering
    \includegraphics[scale=0.4]{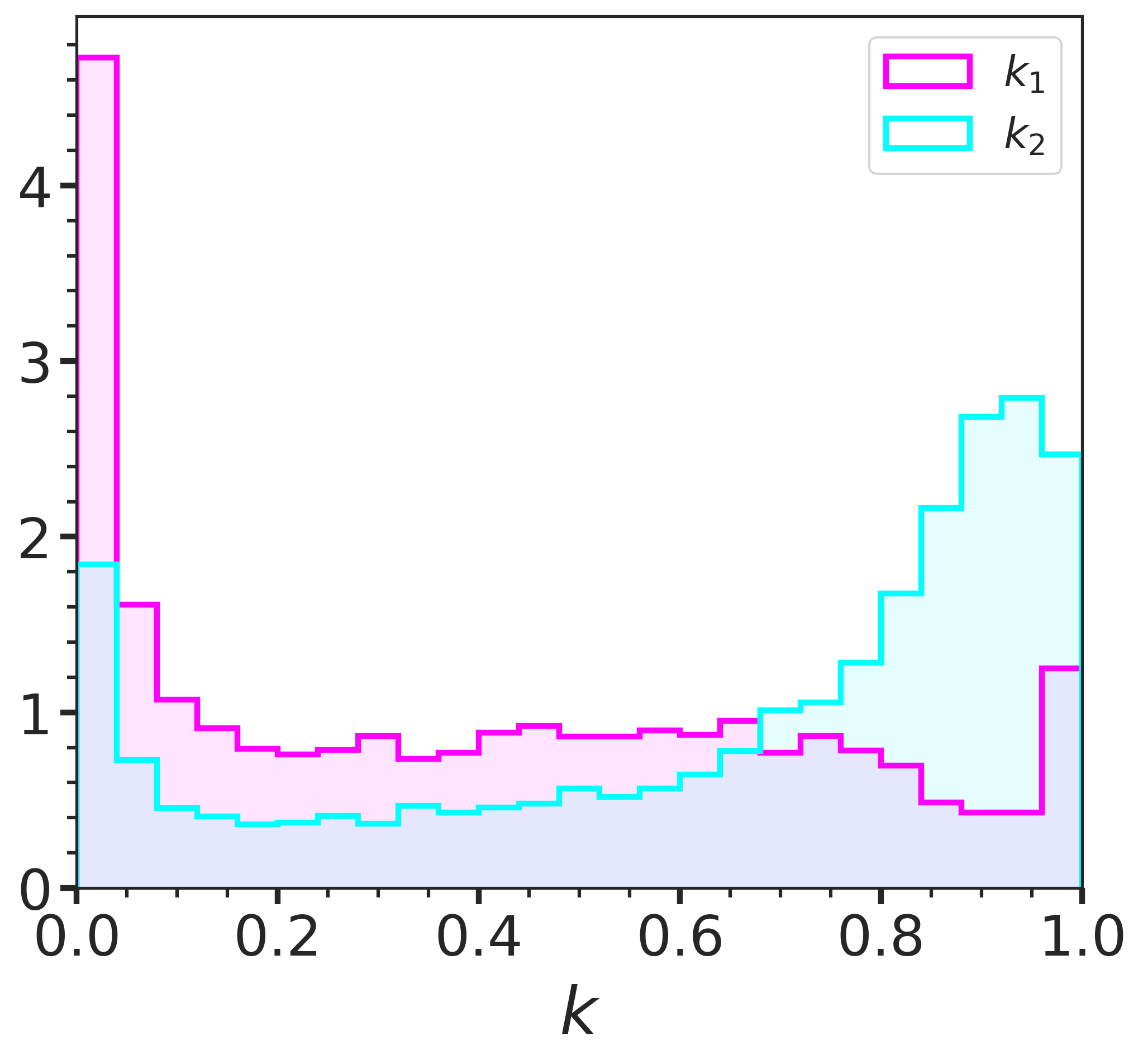}
    \caption{Distribution of the kinematic decomposition parameters of the inner ($k_1$) and outer ($k_2$) discs respectively in magenta and cyan. Each histogram is normalized such that the total area sums to one. }
    \label{fig:k1_k2_dist}
\end{figure}

Fig.~\ref{fig:k1_k2_dist} shows the distributions of the kinematic decomposition parameters $k_1$ and $k_2$ for the inner and the outer discs. The inner disc has a significant peak for low values of $k_1$, indicating the presence of dispersion-supported exponential structures in galaxy internal regions;  we argue that those discs account for corrections to the classical bulge, assumed to be perfectly spherical in our modelling. A relevant number of galaxies show intermediate rotational support spanning a wide range of $k_1$, probably indicating the presence of a variety of more or less rotating structures in galaxy centres. 
\begin{figure*}
    \centering
    \includegraphics[scale=0.3]{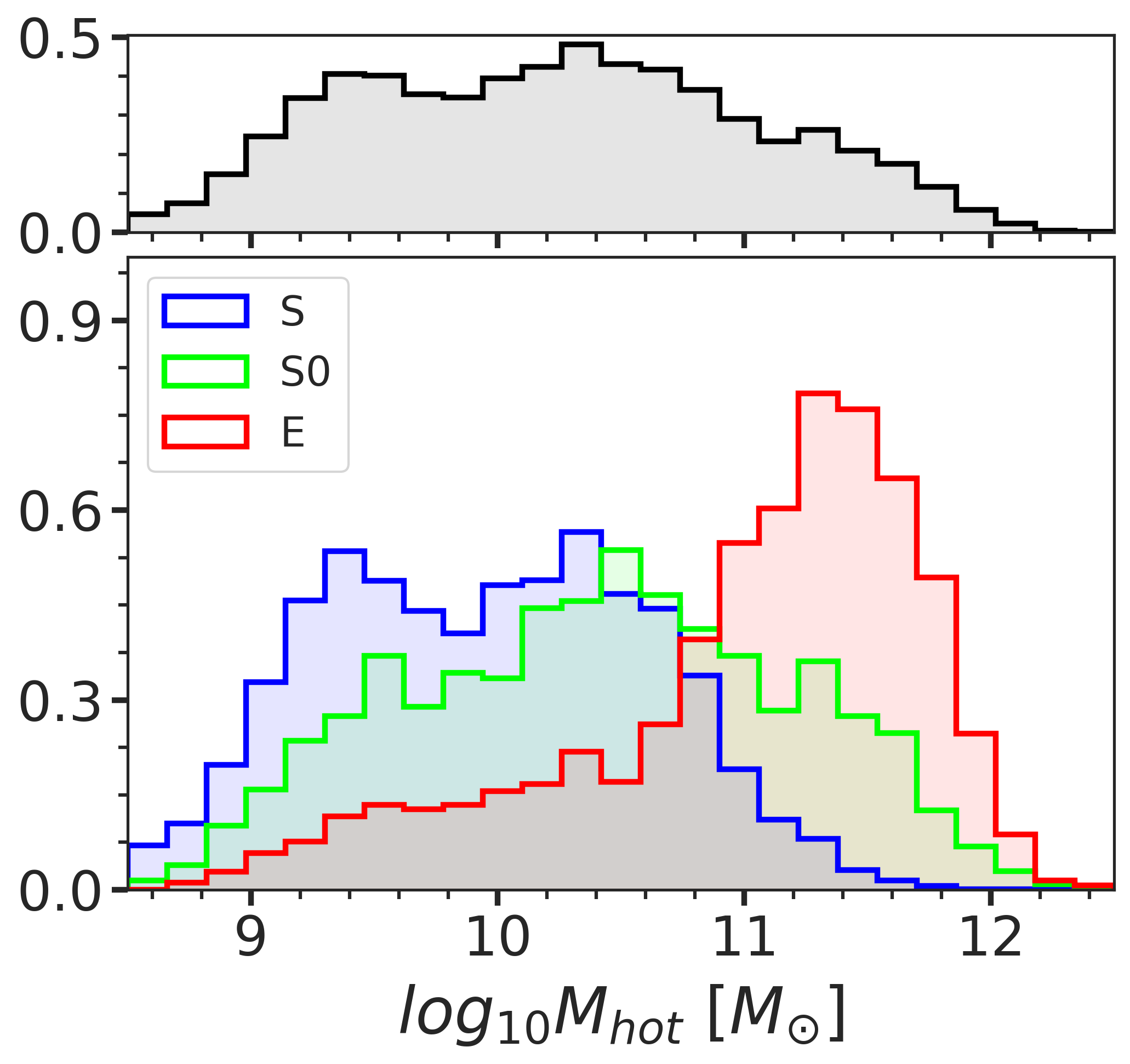}\hspace{0.01\textwidth}
    \includegraphics[scale=0.3]{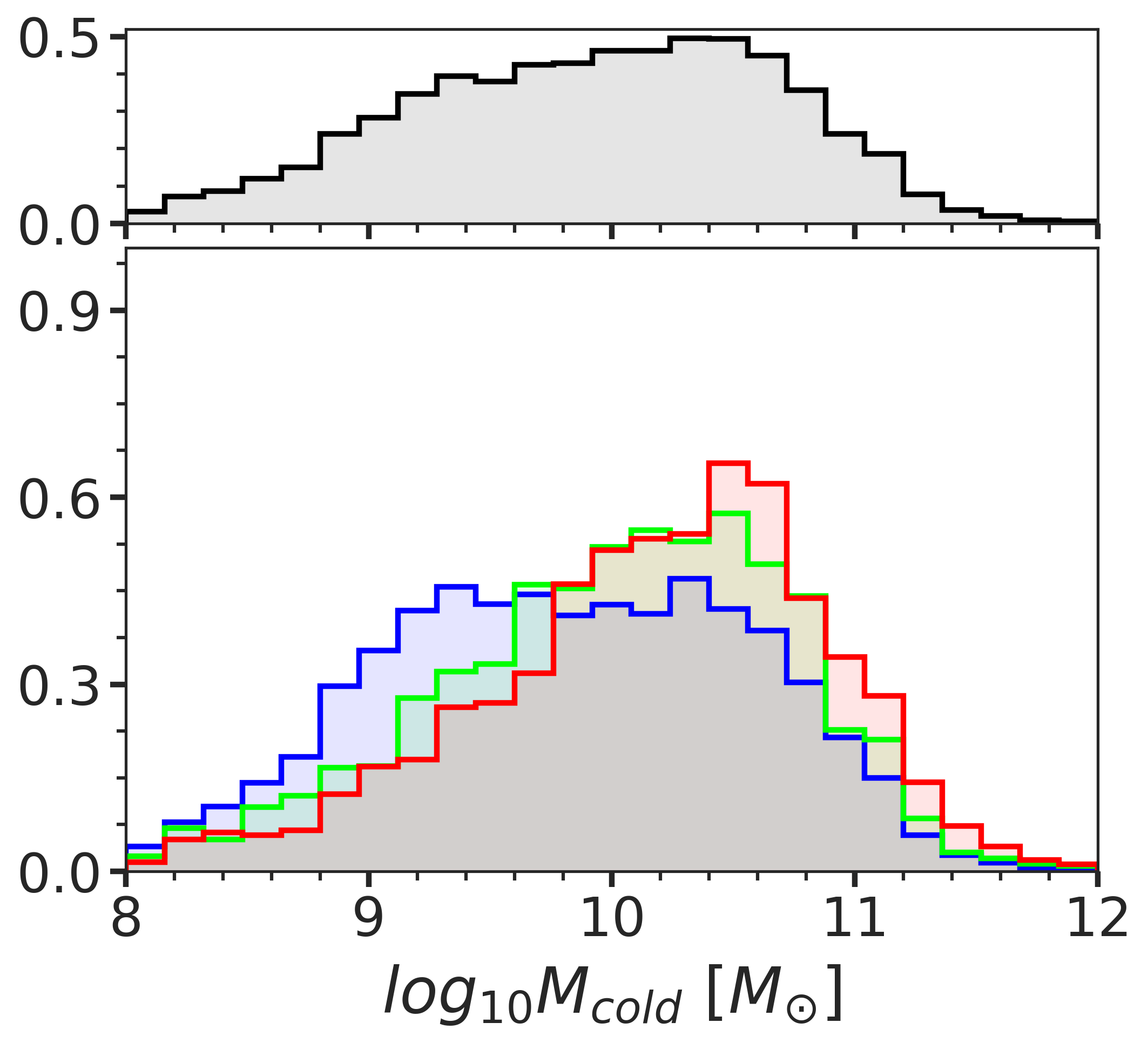}\hspace{0.01\textwidth}
    \includegraphics[scale=0.3]{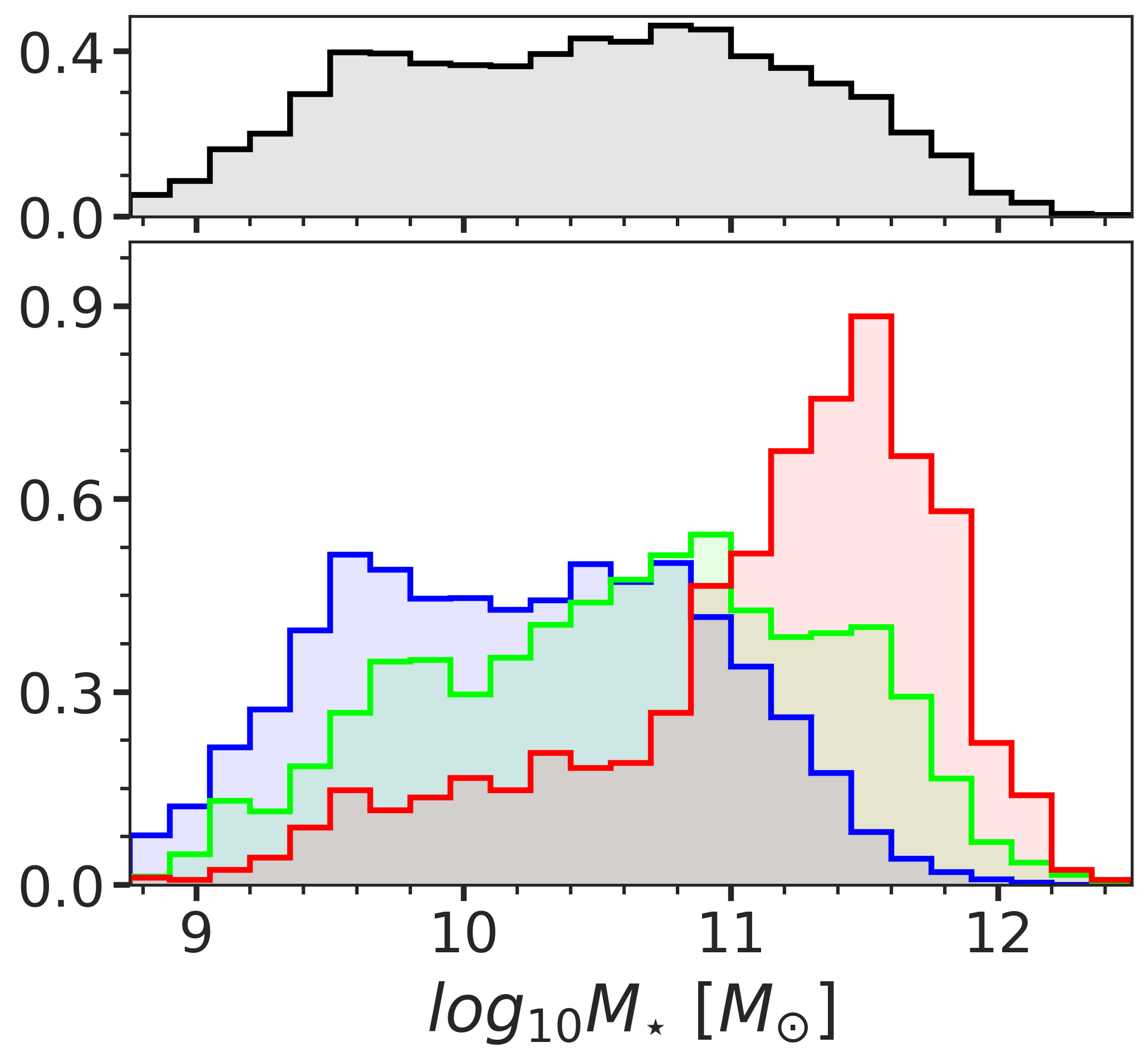}
    \caption{From left to right we show the mass distribution of the hot, cold and total stellar mass for the sample under analysis. The hot and cold mass is estimated following eqs. \ref{eq:hot_mass} and \ref{eq:cold_mass}. Each histogram is normalized to have a total area equal to one, and the distributions are colour coded according to visual morphology: blue for spirals, green for lenticulars and red for ellipticals. The panel over each picture represents the distribution of that variable for the whole population helping in understanding the relative role of the three morphological types.}
    \label{fig:Mass_distributions}
\end{figure*}

As expected, the outer disc shows a clear peak at high values of $k_2$ (i.e. $k_2\simeq0.9$) tracing, in these cases, the rotationally supported component of the galaxy. Another peak is present in the dispersion-dominated domain (i.e. $k_2\lesssim0.15$); interpreting those components as thin discs may be misleading since the majority of galaxies contributing to this are massive ellipticals dominated by dispersion. Given their not-always-spherical light distribution and the absence of ordered rotation, we model these systems as a superposition of dispersion-supported exponential "discs".

It is clear by this discussion that analysing bulges and discs, as defined by \textsc{bang}, without accounting for their kinematic state may lead to biased and incorrect conclusions; this is why we focus the following analysis on what we previously defined as "hot" and "cold" components (see Sec.~\ref{sec:Comparison with previous dynamical modelling estimates}):



\begin{equation}
    M_{\rm hot} = M_{b}+(1-k_1) M_{d,1}+(1-k_2)  M_{d,2},
    \label{eq:hot_mass}
\end{equation}
\begin{equation}
    M_{\rm cold} = k_1 M_{d,1}+k_2 M_{d,2} ,
    \label{eq:cold_mass}
\end{equation}
where $M_{\rm hot}$ and $M_{\rm cold}$ are respectively the mass of the "hot" and "cold" component.

\

Fig.~\ref{fig:Mass_distributions} shows the distribution of masses respectively for the "hot" (left) and "cold" (middle) components together with the total stellar mass (right) of the galaxy colour-coded by visual morphology (spirals in blue, lenticulars in green and ellipticals in red). Focusing on the first panel, we can notice, as expected, a clear trend between the "hot" mass and the morphology with elliptical galaxies skewed at the high-mass end of the distribution; lenticulars and spirals are, on average, less massive with a broader distribution and contributing in the whole dynamical range\footnote{As we show in appendix \ref{app:Analysis_golden_sample} even applying some quality cuts on the sample, the general trends between the different distributions are unaffected.}. 
On the top panel of the picture, we show the distribution for the whole sample demonstrating that, even though ellipticals dominate in terms of mass, their number, in the analysed sample, is less significant; we stress that these histograms are affected by the selections criteria adopted in the MaNGA survey and cannot be considered (without accounting for the proper corrections) representative of the real local galaxy population \citep[][]{Wake_2017}{}{}.

Similar trends can be seen in the $ M_{\rm cold}$ distribution. Ellipticals are still skewed at high masses peaking much closer to lenticulars. For both lenticulars and spirals the distributions span very similar ranges compared to their "hot" counterparts suggesting an increase in the rotational support (if compared to ellipticals) with a still relevant contribution from dispersion-dominated structures.

The last panel on the right shows the distribution of the total stellar mass for each galaxy mostly reflecting the behaviour of the "hot" ("cold") component in the high (low) mass bins, stressing their role and their dependency on stellar mass in the overall galaxy budget. \fr{We verified that our stellar masses are consistent with the prediction from the NASA-Sloan Atlas (NSA)\footnote{\url{http://nsatlas.org/}} settling on linear relation with a small offset of $0.04~ \rm dex$ and a scatter of $0.18~ \rm dex$.}

\begin{figure}
    \centering
    \includegraphics[scale=0.4]{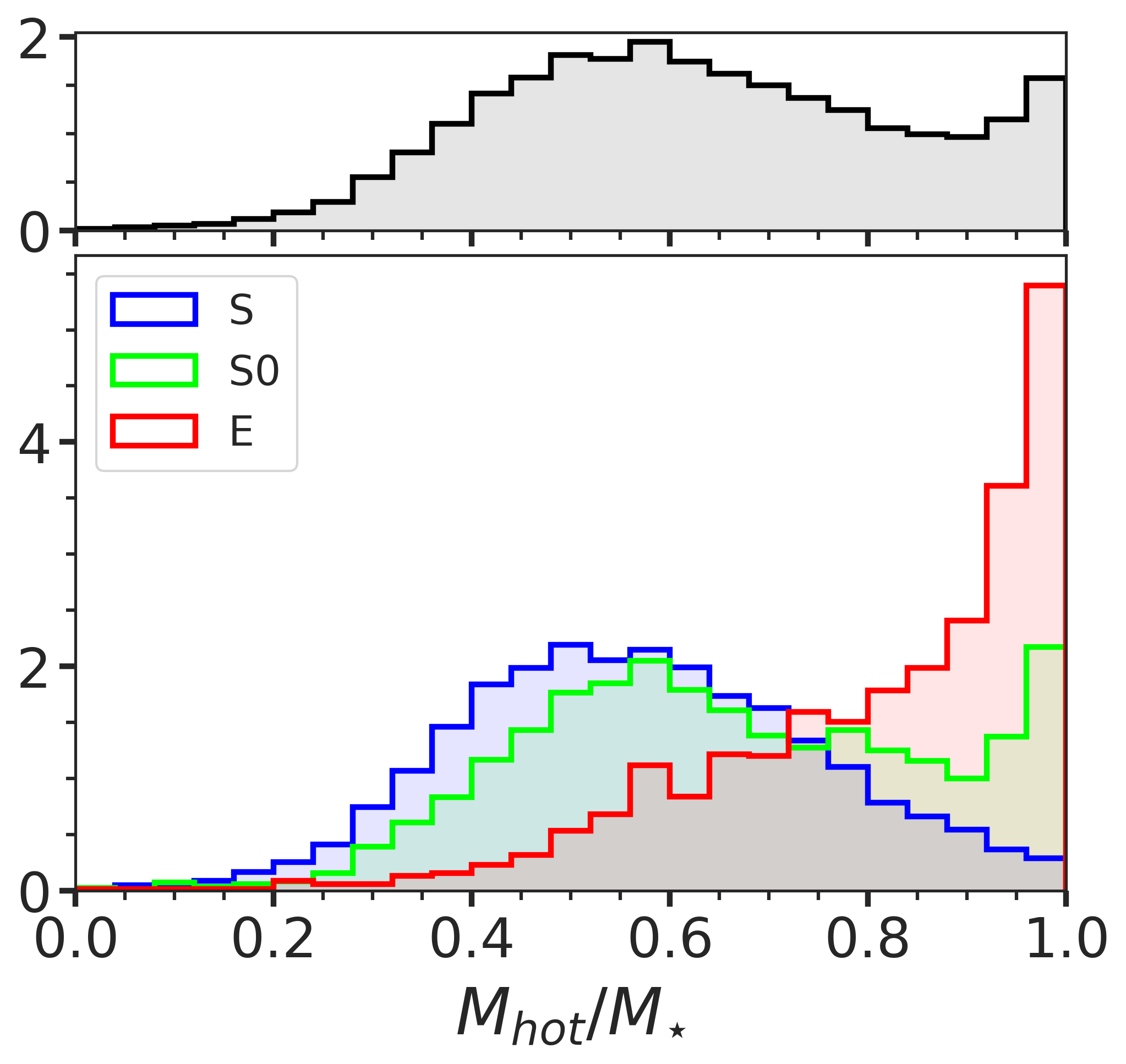}
    \caption{Distribution of the "hot" (see Eq. \ref{eq:hot_mass}) over total stellar mass fraction colour-coded according to morphology: blue for spirals, green for lenticulars and red for ellipticals. Each histogram is normalized to have a total area equal to one. The top panel shows in black and grey the distribution for the whole sample. }
    \label{fig:mass_frac_distribution}
\end{figure}

This is even clearer if we look at Fig.~\ref{fig:mass_frac_distribution} showing the fraction between the "hot" and total mass of the galaxy colour-coded in the same way as done in Fig.~\ref{fig:mass_frac_distribution}. As expected, the "hot" component largely dominates in the case of ellipticals. Interestingly, lenticulars demonstrate a bimodal distribution; a fraction of them, peaking at $M_{\rm hot}/M_{\star} \simeq 1.0$, is entirely supported by dispersion and further inspection is needed to address if these objects may have misclassified visual morphology according to the VMC-VAC \citep[][]{Vazquez_2022}{}{}. Spiral galaxies have, in terms of mass, a considerable contribution from the hot component with a broad distribution skewed towards high values of $M_{\mathrm{hot}}/M_{\star}$. Even though this could be surprising, we stress again that our definition of the "hot" component always includes a fraction of the disc mass even in cases of rotationally supported systems. We note also that a large contribution to the hot component mass of spirals comes from the two discs (second and third terms of Eq. \ref{eq:hot_mass}), possibly indicating that a simple spherical bulge cannot account for the variety of inner galactic structures (such as bars etc...) that are often present in spiral and that can be a relevant source of dispersion. The distribution of ellipticals shows a prominent tail toward small values of $M_{\mathrm{hot}}/M_{\star}$ possibly suggesting that visual morphology alone can not always be reliable while characterizing the kinematic state of a galaxy \citep[][]{Cappellari_2011,Krajnovi_2008}{}{}.

\section{Scaling relations in the whole manga sample}\label{sec:scaling_relations}

\begin{figure*}
    \centering
    \includegraphics[scale=0.42]{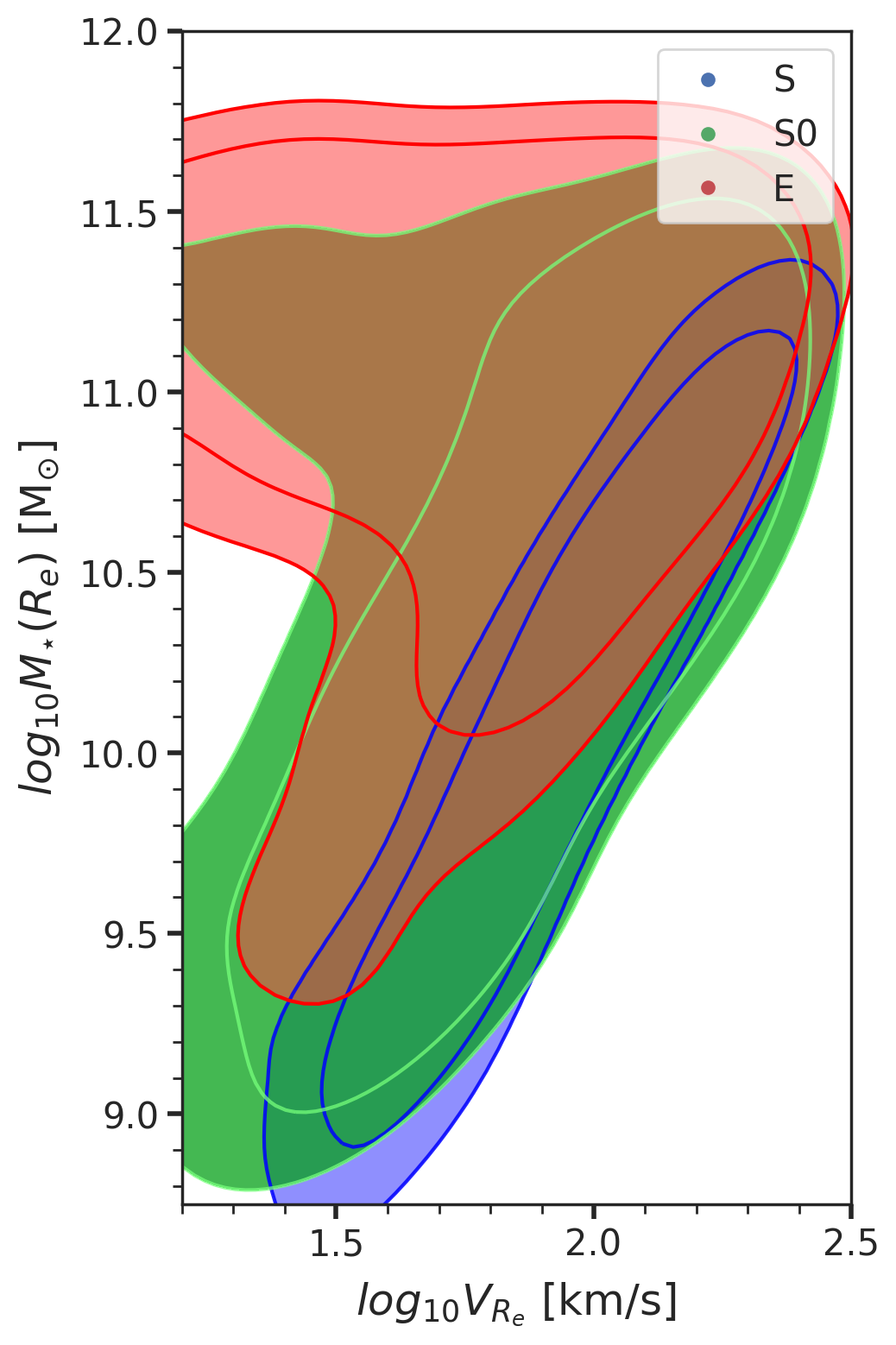}\hspace{0.01\textwidth}
    \includegraphics[scale=0.42]{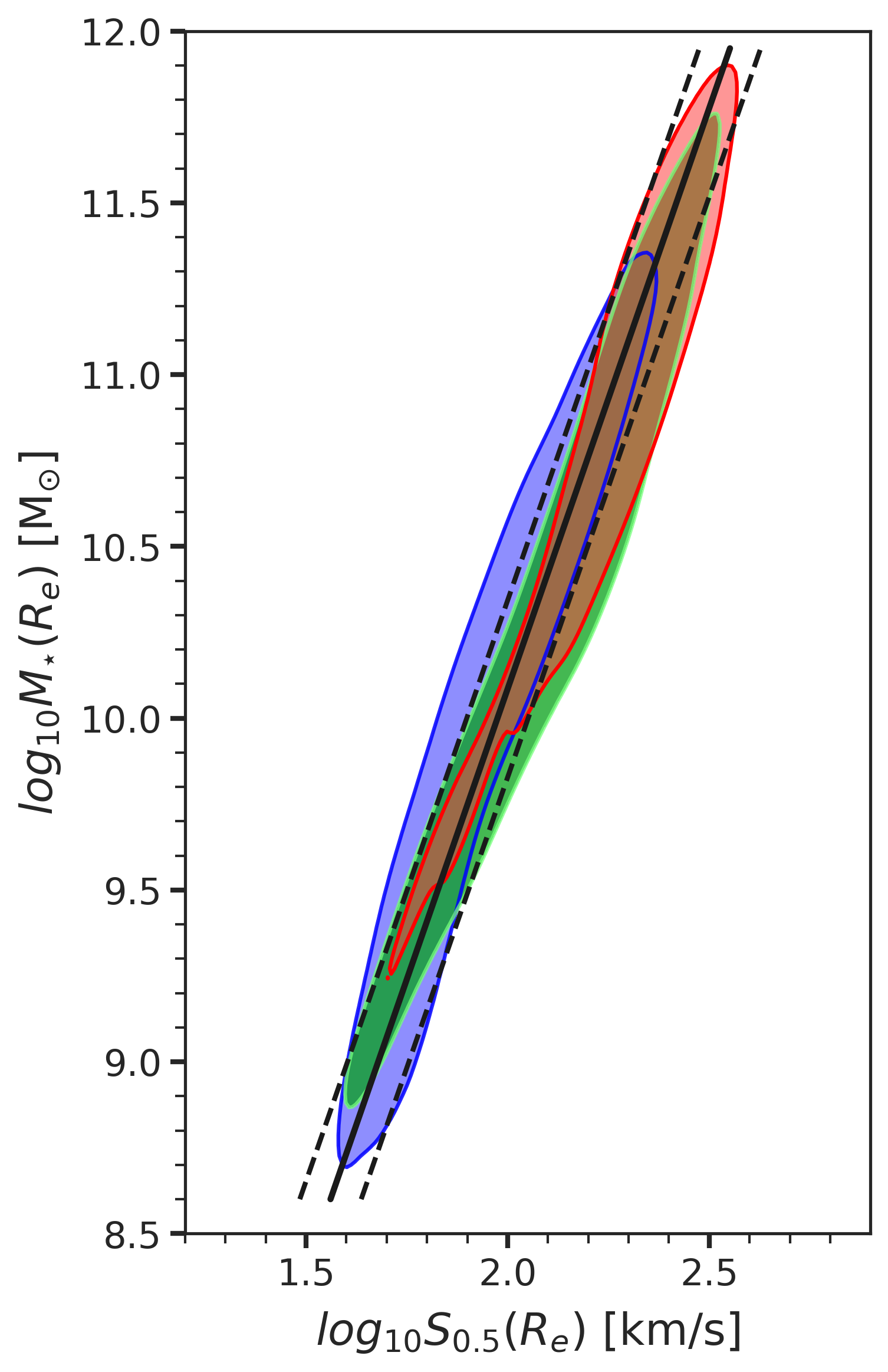}\hspace{0.01\textwidth}
    \includegraphics[scale=0.42]{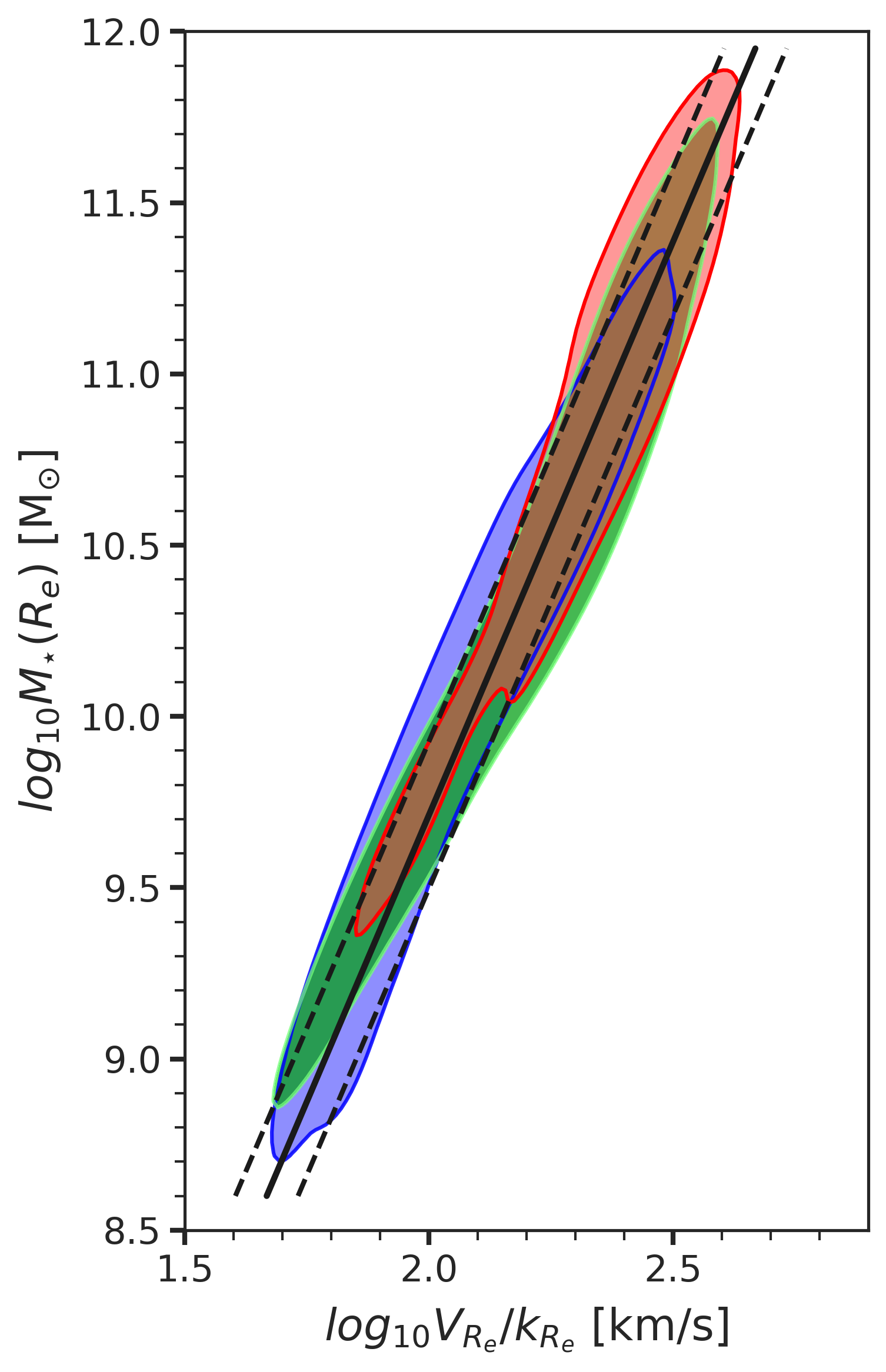}
    \caption{Scaling relations of different kinematics tracers divided according to morphology (blue for spirals, green for lenticulars and red for ellipticals). In all three plots, the y-axis is the logarithm of the stellar mass within one $ R_e$, while the horizontal axis, respectively from left to right, shows the logarithm of the velocity at one $ R_e$ ($V_{R_e}$), the logarithm of the $S_{0.5}$ parameter (see Eq. \ref{eq:S0_5}) and the logarithm of the velocity at $R_e$ corrected according to Eq. \ref{eq:arctan}. \fr{For each morphological class, the coloured area contains  80\% of the kernel density estimated probability mass, only for the first plot on the left we included a second contour at a 70\% level.} The solid black lines are the best fit linear relations while the black dotted lines represent the one$-\sigma$ confidence interval. The best-fit parameters are summarized in Tab.~\ref{tab:best_fit_parameters_linear_fit_whole}}
    \label{fig:Scaling_Relations}
\end{figure*}

Correlations between photometric and kinematic properties have been extensively studied for investigating the matter content of galaxies. As first attempts, scaling relations between the total stellar or baryonic matter with the maximum circular velocity (or a proxy of it) have been discovered for late-type galaxies \citep[LTGs, ][]{Tully_Fisher_1977}.
Similarly, in the case of ETGs, a correlation exists between stellar mass and stellar velocity dispersion \citep{Faber_Jackson_1976} as a projection of the more general fundamental plane relation \citep{Djorgovski_1987,Dressler_1987} which, at first order, can be interpreted as a consequence of the virial theorem. 

All these relations demonstrate the implicit connection, through the Poisson equation, between the distribution of mass (or density) and the stellar kinematic (or the gravitational potential). This is why, aiming at unifying ETGs and LTGs on the same scaling law, a new correlation, combining rotational and random motion, has been recently proposed \citep{Weiner_2006}.

In the following, we do not aim at any detailed study regarding scaling relations. Given their universality, we instead use them as a validating tool for our modelling scheme, probing the ability of \textsc{bang} in recovering the stellar mass and the gravitational potential, and its robustness against outliers (such as mergers, non-masked stars and type I AGNs or other contaminants).


For consistency with previous works \citep{Cortese_2014,Aquino_2018,Aquino_2020} and to show the generalization ability of our modelling scheme and its robustness against outliers \fr{together with the results on the main sample (see Sec.~\ref{sec:Preliminary analysis on the proposed decomposition}) we repeated the analysis on a carefully selected sub-sample of galaxies selected according to the same criteria adopted in \citealt{Aquino_2020}. More specifically,} in the latter case, we considered only galaxies with 
an inclination angle $25<i<75$, masking all the pixels with an error in the velocity larger than $25~\rm km/s$ ($\simeq 1/3$ of the average MaNGA spectral resolution $\sigma_{inst}\simeq 70 ~\rm km/s$), and discarding all galaxies with more than $40\%$ pixels masked within an ellipse of semi-major axis equal to the effective radius of the galaxy. As a final cut, we visually inspected all the objects in our sample excluding ongoing mergers or other contaminants (i.e. wrong redshift, central non-masked stars and type I AGNs); \fr{in the following we refer to this sample as "kinematic sample"}

We focus our analysis on the scaling relation proposed in \cite{Weiner_2006} relating the stellar rotational and dispersion support within $R_e$ to the enclosed stellar mass of the galaxy ($M_{\star}(R_e)$).

The left panel of Fig.~\ref{fig:Scaling_Relations} shows the relation between $M_{\star}(R_e)$ (y-axis) and $V_{R_e}$ color-coded by visual morphology. $M_{\star}(R_e)$ is the stellar mass within the effective radius and it is computed from the best-fit stellar mass profile as predicted by \textsc{bang}, while $V_{R_e}$ is a proxy of the rotation velocity within the effective radius estimated following \cite{Aquino_2020}. More specifically, starting from the l.o.s. velocity maps as modelled by \textsc{bang}, we considered all the pixels within an ellipse of semi-major axis equal to the effective radius of the galaxy and we estimated $V_{Re}$ by computing the difference between the $90^{th}$ and $10^{th}$ percentile (i.e. $W = (V_{90}-V_{10})/2$) of its velocity histogram and accounting for deprojection by dividing for $\sin{i}$, being $i$ the inclination angle estimated by \textsc{bang}. As expected, a linear correlation is present in the case of spirals, even though the scatter is significant and it largely increases changing morphology, with ellipticals dominating the spread in $V_{R_e}$ mainly due to the huge contribution of slow rotators.

Due to the morphology dependence of this relation, following several works in the literature, we define:
\begin{equation}
    S_{\rm K}(R_e) = \sqrt{K~V_{R_e}^2+\sigma_{R_e}^2},
    \label{eq:S0_5}
\end{equation}
where $\sigma_{R_e}$ is the average velocity dispersion within $R_e$ and the constant multiplicative factor $\rm K$ is usually set equal to 0.5 \citep{Cortese_2014,Aquino_2018,Gilhuly_2019,Barat_2019}. We estimate $\sigma_{R_e}$ starting form l.o.s. velocity dispersion best-fit model and linearly averaging over the same elliptical region as defined above.  
We estimated the errors on $M_{\star}(R_e)$, $V_{R_e}$ and $\sigma_{R_e}$ by repeating the procedure detailed above, with the model parameters associated with the 15 and 85 percentile of the posterior distribution \footnote{A proper Bayesian approach would require to evaluate our model on the whole posterior distribution estimating the best fit and its error by computing the median and the percentiles on the actual set of models instead of the parameter posteriors. For computational reasons, we refrain from following this strategy; this is why we opted for what is discussed in the text.}.   

In all cases, we fitted a line to the data points by considering errors on both the $x$ and $y$ axes. Doing that would generally require a fit with at least $N+3$ free parameters (where $N$ is the number of considered data points, see \citealt{Andreon_2013}), an impractical solution in our case. Luckily, in the case of a straight line, it is still possible to reduce the parameter space dimension back to three (slope, intercept and, optionally, intrinsic scatter) under a few simplifying assumptions (see chapter seven of \citealt{Hogg_2010} for further details).

Following \cite{Aquino_2020}, we considered the stellar mass as the independent variable fitting, with \textsc{cpnest} \citep{Del_Pozzo_CPNest}, the following functional form:

\begin{equation}
    \log_{10}{\left(\frac{S_{0.5}(R_e)}{100 ~\rm km/s}\right)} = a + b \log_{10}{\left(\frac{M_{\star}(R_e)}{10^{10.5}~\rm M_{\odot}}\right)} ,
    \label{eq:linear_relation_S05}
\end{equation}
where $a$ and $b$ are respectively the intercept and the slope of the relation and together with the intrinsic scatter $\sigma$ are the free parameters of the fit.

The $M_{\star}(R_e)-S_{0.5}(R_e)$ is presented in the middle panel \fr{(Fig.~\ref{fig:Scaling_Relations})}; the additional contribution of the dispersion component helps in reducing the scatter moving all galaxies on a similar linear relation. The best-fit results on the whole sample are presented in the first line of Tab.~\ref{tab:best_fit_parameters_linear_fit_whole}, with an exceptional agreement, in terms of the slope \footnote{A comparison of the zero point of the relation with literature results is not trivial due to possible differences in the mass estimates and the sample selection.}, with what found in \citealt{Aquino_2020}. This result is even more striking considering that we see a significant reduction (about $30 \%$ less if compared \fr{to sample and analysis of 2458 galaxies presented in} \citealt{Aquino_2020}) of the scatter in the relation, even though our sample comprises $\simeq 10,000$ galaxies, demonstrating the ability of \textsc{bang} of recovering the true gravitational potential of galaxies and its robustness against outliers, strengthening our results and motivating its blind applicability even on larger samples.  

\begin{figure}
    \centering
    \includegraphics[scale=0.5]{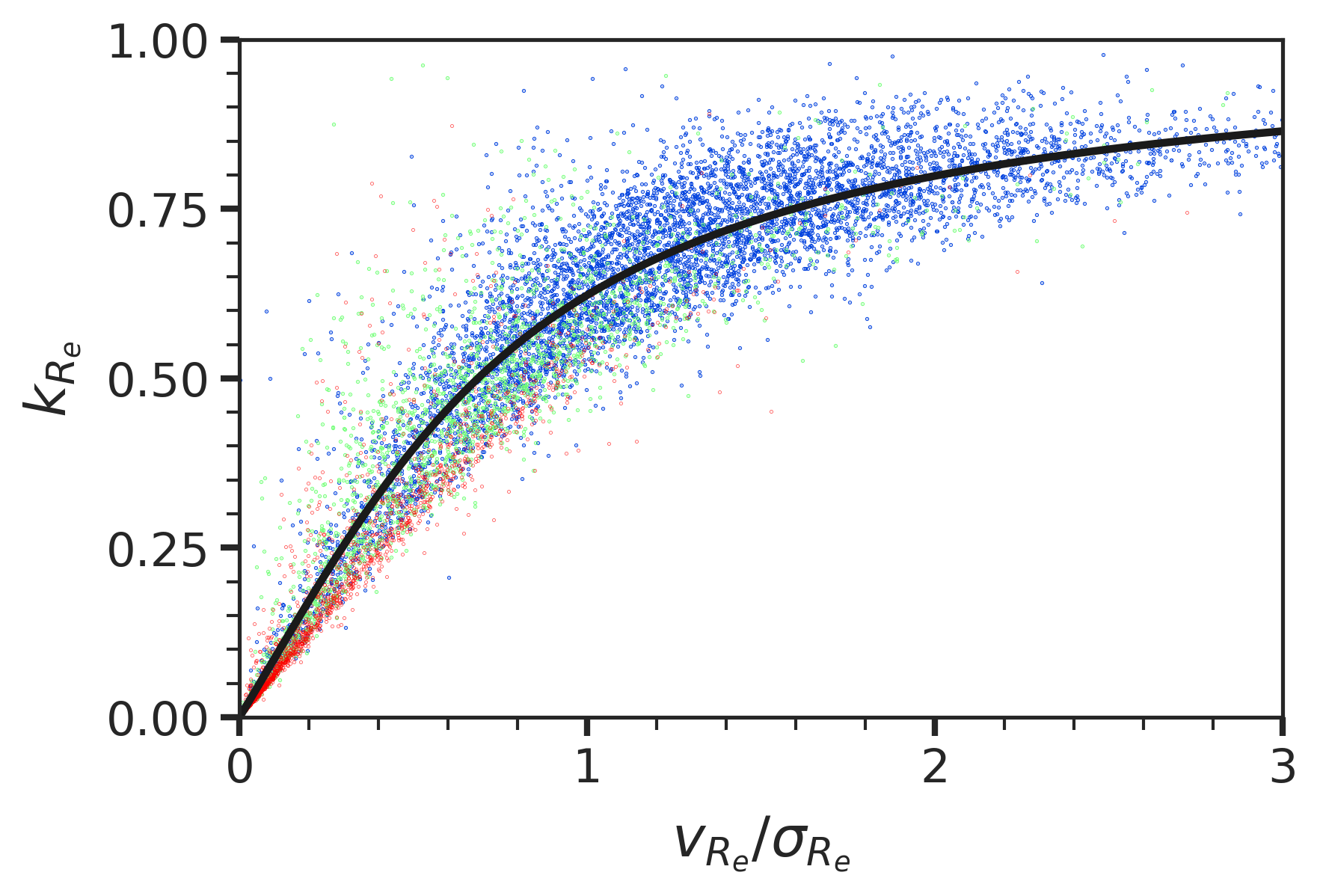}
    \caption{On the x-axis velocity at the effective radius divided by average velocity dispersion; on the y-axis the luminosity weighted average rotational support at $R_e$. Points are color coded in the same way as done in Fig.~\ref{fig:Scaling_Relations}. The black solid line is the best fit to the data of the functional form presented in Eq.~\ref{eq:arctan}.}
    \label{fig:Arctan_fit}
\end{figure}

Driven by these results, we provide in the third panel of the figure a new kinematic tracer computed by dividing $V_{R_e}$ with a proxy of the average rotational support of the galaxy ($k_{R_e}$). Similarly to what was done in \cite{Rigamonti_2022}, we compute $k_{R_e}$ for each galaxy as the brightness-weighted average of the $k_1$ and $k_2$ parameters within $R_e$. Note $k_{R_e}$ is limited between zero and one resulting in major corrections for slowly rotating galaxies. We provide through the following functional form an empirical relation for estimating $k_{R_e}$ directly from $V_{R_e}/\sigma_{R_e}$:

\begin{equation}
    k_{R_e} = 1+\frac{\mathrm  {arccot} \left[b\left(V_{R_e}/\sigma_{R_e}-x_0\right)\right]}{\mathrm {arctan}{\left(bx_{0}\right)+\pi/2}},
    \label{eq:arctan}
\end{equation}
where $b=1.47$ and $x_0=0.086$ are calibrated by fitting that profile on our estimates obtained from the MaNGA sample. Note that Eq. \ref{eq:arctan} ensures that $k_{R_e}$ is equal to zero (one) for $V_{R_e}/\sigma_{R_e}=0$ ($\infty$). Fig.~\ref{fig:Arctan_fit} shows the relation between $k_{R_e}$ and $V_{R_e}/\sigma_{R_e}$ color coded according morphology. As expected, spiral galaxies (blue dots) are supported by rotation, while ellipticals dominate the dispersion regime.

Fitting the following linear relation: 

\begin{equation}
    \log_{10}{\left(\frac{V_{R_e}/k_{R_e}}{100 ~\rm km/s}\right)} = a + b \log_{10}{\left(\frac{M_{\star}(R_e)}{10^{10.5}~\rm M_{\odot}}\right)} ,
    \label{eq:linear_relation_vRe}
\end{equation}
to the proposed correction gives the same results, in terms of the slope, as the $M_{\star}(R_e)-S_{0.5}(R_e)$ relation, reducing, even more, the scatter and possibly suggesting $V_{R_e}/k_{R_e}$ as being a better indicator of the relation between stellar mass and the underlying gravitational potential\footnote{Note that we use $k_{R_e}$ as estimated from Eq.~\ref{eq:arctan} and not directly from \textsc{bang}, thus the reduction of the scatter in the last panel of Fig.~\ref{fig:Scaling_Relations} may be also driven by that.}. Note that with the provided correction the scatter in the linear relation becomes almost consistent with the scatter in the Tully-Fisher relation for pruned samples \citep{Lelli_2019}.  

For both kinematic tracers ($S_{0.5}(R_e)$ and $V_{R_e}/k_{R_e}$) the results of a linear fit are still consistent with each other even when repeating the analysis on \fr{the "kinematic sample" defined in} the first part of Sec.~\ref{sec:scaling_relations}), with a further decrease in the scatter of the relation (Tab.~\ref{tab:best_fit_parameters_linear_fit_fiducial}). We speculate that a possible bending may be present in those relations \fr{(see. Appendix~\ref{app:Scaling_with_double_slopes})} since a fit with two lines gives a larger evidence than with a single one; this may hint at a decrease of the $\rm M_{star}/M_h$ fraction in the high stellar mass end \citep{Noordermeer_2007,Moster_2018,Behroozi_2019}. Further analysis is still required before drawing any conclusion, especially given the priors assumed on the halo.

\begin{table}
	\centering
	\caption{Summary table of the best fit parameters for the linear fit (Eqs..~ \ref{eq:linear_relation_S05}, \ref{eq:linear_relation_vRe}) of the kinematic tracers presented in the middle and left panels of Fig.~\ref{fig:Scaling_Relations}. The first column refers to the considered dependent variable (the independent variable is always $M_{\star}(R_e)$), while the other three columns are the free parameters of the fit, namely, the intercept (a), the slope (b) and the intrinsic scatter $\sigma$.}
	\label{tab:best_fit_parameters_linear_fit_whole}
	\begin{tabular}{lccc} 
	    \hline
		y-axis & a & b & $\sigma$ \\
		\hline 
		$\log_{10}{S_{0.5}(R_e)}$ & $0.13 \pm 0.01$ & $0.30 \pm 0.01$ & $0.073$\\ 
		$\log_{10}{V_{R_e}/k_{R_e}}$ & $0.24 \pm 0.01$ & $0.30 \pm 0.01$ & $0.064$\\ 
		\hline
	\end{tabular}
\end{table}

\begin{table}
	\centering
	\caption{Same as Tab.~\ref{tab:best_fit_parameters_linear_fit_whole} but for \fr{"kinematic sample"} ($\simeq 7000$ galaxies) selected according to the same criteria applied in \citealt{Aquino_2020} \fr{and defined at the beginning of  sec}.~\ref{sec:scaling_relations}. }
	\label{tab:best_fit_parameters_linear_fit_fiducial}
	\begin{tabular}{lccc} 
	    \hline
		y-axis & a & b & $\sigma$ \\
		\hline 
		$\log_{10}{S_{0.5}(R_e)}$ & $0.13 \pm 0.01$ & $0.31 \pm 0.01$ & $0.065$\\ 
		$\log_{10}{V_{R_e}/k_{R_e}}$ & $0.24 \pm 0.01$ & $0.31 \pm 0.01$ & $0.053$\\ 
		\hline
	\end{tabular}
\end{table}

\section{Summary and Conclusions}
In this work, we presented some results from the application of \textsc{bang}, a software purposely designed for bulge+discs dynamical modelling of galaxies, on the galaxies observed by the MaNGA survey. 
Being fast (GPU-based) and coupled with robust Bayesian parameter 
estimation techniques (e.g., Nested Sampling), \textsc{bang} is perfectly suited for automated applications to large galaxy data sets. We improved upon the earlier version of \textsc{bang} discussed in Paper I both in the modelling approach and, by accounting for physically motivated priors, in the parameter estimation set-up.


We successfully modelled the photometry and kinematic of $10,005$ galaxies assuming three different possible configurations (i.e., bulge+inner disc+outer disc, bulge+inner disc, inner disc+outer disc)
with two different halo prior assumptions\footnote{for a grand total of +60,000 different fits.}, estimating the structural and dynamical parameters for each galaxy of the sample (see Tab.~\ref{tab:parameters} for a summary of the free parameters in the models).

In order to test the accuracy of the methodology, we compared our results to well-known kinematic scaling relations \citep[e.g., $M_{\star}$ vs. $S_{0.5}$,][]{Weiner_2006}, finding a substantial agreement (regardless of the assumed halo prior) with previous works in the literature, for all kind of morphologies \citep[see e.g.,][]{Aquino_2020}. We also provide an empirical correction, dependent on the velocity-to-dispersion ratio, able to reduce the scatter in the stellar mass-to-velocity relation, resulting in a tighter correlation when compared to the $M_{\star}(R_e)$ vs $S_{0.5}$. These results do not assume any particular selection criteria and highlight the power of \textsc{bang} in recovering the gravitational potential of single galaxies even in an automated decomposition applied to large samples. 

We also compared stellar mass estimates for a sub-sample of 149 \fr{ETGs} galaxies analysed through Schwarzschild orbit superposition methods \citep{Jin_2020}, finding an excellent agreement. We therefore propose a new decomposition scheme that combines the visible components of our model (i.e., bulge, inner disc and outer disc) into more general, and dynamically motivated, "hot" and "cold" components. This decomposition correlates with purely orbit-based estimates, possibly helping in extending studies of the "hot" vs "cold" mass budget to larger samples.

We performed a preliminary analysis of the distribution of the structural components for different morphological types. \fr{We showed that the physical interpretation of the two exponential disc components of the model depends on their kinematical parameters $k_1$ and $k_2$. In cases where $k_1$ is small, the inner disc typically models a non-spherical central structure dominated by dispersion. Similarly, when $k_2$ is small, the outer disc (potentially in conjunction with the inner disc) models non-circular isophotes in ellipticals. The detailed analysis of the possible many real natures of the two discs is deferred to an ongoing investigation. As expected, our definition of "hot" and "cold" mass correlates with the morphological type, with ellipticals and spirals positioned at the high and low ends, respectively, of the "hot" fraction distribution.} 

Our work aimed to demonstrate the proposed methodology's ability to cope with large samples and to validate our results.
\fr{The preliminary analysis presented in Sec.~\ref{sec:Preliminary analysis on the proposed decomposition} paves the road to further developments which we will address in future works. In particular, we will exploit our parametrization $k_1$ and $k_2$ to clarify the role that the two discs have in determining the dynamic state of the internal components of galaxies. We will quantify the mass budget of each component  also in dependence on the star-formation rate of the galaxy. This will help to understand the relative role of mergers vs. coherent accretion in shaping local galaxies (Rigamonti et al., to be submitted).} 
\fr{We also plan to characterize the possible rotational support of photometric bulges, hence breaking the degeneracy between classical bulges and pseudobulges. We also aim at applying \textsc{bang} to mock simulation-based catalogues \citep[e.g., iMaNGA,][]{Nanni_2022}{}{} in order to select the parameter space region that could identify and characterise thick discs and, possibly, bars.}

\section*{Acknowledgements}
We thank the anonymous referee for their comments and suggestions that helped us to improve the quality of the paper. 

We gratefully thank Francesco Bollati, Stefano Zibetti, Matteo Fossati and Walter Del Pozzo for the many insightful and constructive discussions that improved the quality of this work. FH and MD acknowledge funding from MIUR under the grant
PRIN 2017-MB8AEZ. MD acknowledges financial support from ICSC – Centro
Nazionale di Ricerca in High-Performance Computing, Big Data and
Quantum Computing, funded by European Union – NextGeneration EU. LC acknowledges support from the Australian Research Council Future Fellowship and Discovery Project funding scheme (FT180100066,DP210100337). Parts of this research were conducted by the Australian Research Council Centre of Excellence for All Sky Astrophysics in 3 Dimensions (ASTRO 3D), through project number CE170100013.

\section*{Data Availability}

The data underlying this article will be shared on reasonable request to the corresponding author. \fr{The full database will be publicly available in a forthcoming paper (Rigamonti et al., to be submitted).}



\bibliographystyle{mnras}
\bibliography{main} 

\providecommand{\pasp}{Publications of the Astronomical Society of the
  Pacific} \providecommand{\cjaa}{Chinese Journal of Astronomy~\& Astrophysics}
  \providecommand{\mnras}{MNRAS} \providecommand{\aap}{Astronomy and
  Astrophysics} \providecommand{\apj}{ApJ} \providecommand{\apjs}{Astrophysical
  Journal, Supplement} \providecommand{\araa}{Annual Review of Astron and
  Astrophys} \providecommand{\apjl}{Astrophysical Journal Letters}
  \providecommand{\aj}{Astronomical Journal}
\begin{thebibliography}{}
\makeatletter
\relax
\def\mn@urlcharsother{\let\do\@makeother \do\$\do\&\do\#\do\^\do\_\do\%\do\~}
\def\mn@doi{\begingroup\mn@urlcharsother \@ifnextchar [ {\mn@doi@}
  {\mn@doi@[]}}
\def\mn@doi@[#1]#2{\def\@tempa{#1}\ifx\@tempa\@empty \href
  {http://dx.doi.org/#2} {doi:#2}\else \href {http://dx.doi.org/#2} {#1}\fi
  \endgroup}
\def\mn@eprint#1#2{\mn@eprint@#1:#2::\@nil}
\def\mn@eprint@arXiv#1{\href {http://arxiv.org/abs/#1} {{\tt arXiv:#1}}}
\def\mn@eprint@dblp#1{\href {http://dblp.uni-trier.de/rec/bibtex/#1.xml}
  {dblp:#1}}
\def\mn@eprint@#1:#2:#3:#4\@nil{\def\@tempa {#1}\def\@tempb {#2}\def\@tempc
  {#3}\ifx \@tempc \@empty \let \@tempc \@tempb \let \@tempb \@tempa \fi \ifx
  \@tempb \@empty \def\@tempb {arXiv}\fi \@ifundefined
  {mn@eprint@\@tempb}{\@tempb:\@tempc}{\expandafter \expandafter \csname
  mn@eprint@\@tempb\endcsname \expandafter{\@tempc}}}

\bibitem[\protect\citeauthoryear{{Abdurro'uf} et~al.,}{{Abdurro'uf}
  et~al.}{2022}]{Abdurro'uf_2022}
{Abdurro'uf} et~al., 2022, \mn@doi [\apjs] {10.3847/1538-4365/ac4414}, \href
  {https://ui.adsabs.harvard.edu/abs/2022ApJS..259...35A} {259, 35}

\bibitem[\protect\citeauthoryear{{Andreon} \& {Hurn}}{{Andreon} \&
  {Hurn}}{2013}]{Andreon_2013}
{Andreon} S.,  {Hurn} M.,  2013, \mn@doi [Statistical Analysis and Data Mining:
  The ASA Data Science Journal] {10.1002/sam.11173}, \href
  {https://ui.adsabs.harvard.edu/abs/2013SADM....6...15A} {9, 15}

\bibitem[\protect\citeauthoryear{{Aquino-Ort{\'\i}z}
  et~al.,}{{Aquino-Ort{\'\i}z} et~al.}{2018}]{Aquino_2018}
{Aquino-Ort{\'\i}z} E.,  et~al., 2018, \mn@doi [\mnras]
  {10.1093/mnras/sty1522}, \href
  {https://ui.adsabs.harvard.edu/abs/2018MNRAS.479.2133A} {479, 2133}

\bibitem[\protect\citeauthoryear{{Aquino-Ort{\'\i}z}
  et~al.,}{{Aquino-Ort{\'\i}z} et~al.}{2020}]{Aquino_2020}
{Aquino-Ort{\'\i}z} E.,  et~al., 2020, \mn@doi [\apj]
  {10.3847/1538-4357/aba94e}, \href
  {https://ui.adsabs.harvard.edu/abs/2020ApJ...900..109A} {900, 109}

\bibitem[\protect\citeauthoryear{{Barat} et~al.,}{{Barat}
  et~al.}{2019}]{Barat_2019}
{Barat} D.,  et~al., 2019, \mn@doi [\mnras] {10.1093/mnras/stz1439}, \href
  {https://ui.adsabs.harvard.edu/abs/2019MNRAS.487.2924B} {487, 2924}

\bibitem[\protect\citeauthoryear{{Behroozi}, {Wechsler}, {Hearin}  \&
  {Conroy}}{{Behroozi} et~al.}{2019}]{Behroozi_2019}
{Behroozi} P.,  {Wechsler} R.~H.,  {Hearin} A.~P.,   {Conroy} C.,  2019,
  \mn@doi [\mnras] {10.1093/mnras/stz1182}, \href
  {https://ui.adsabs.harvard.edu/abs/2019MNRAS.488.3143B} {488, 3143}

\bibitem[\protect\citeauthoryear{{Belfiore} et~al.,}{{Belfiore}
  et~al.}{2019}]{Belfiore_2019}
{Belfiore} F.,  et~al., 2019, \mn@doi [\aj] {10.3847/1538-3881/ab3e4e}, \href
  {https://ui.adsabs.harvard.edu/abs/2019AJ....158..160B} {158, 160}

\bibitem[\protect\citeauthoryear{{Bertin}, {Mellier}, {Radovich}, {Missonnier},
  {Didelon}  \& {Morin}}{{Bertin} et~al.}{2002}]{Bertin_2002}
{Bertin} E.,  {Mellier} Y.,  {Radovich} M.,  {Missonnier} G.,  {Didelon} P.,
  {Morin} B.,  2002, in {Bohlender} D.~A.,  {Durand} D.,   {Handley} T.~H.,
  eds,  Astronomical Society of the Pacific Conference Series Vol. 281,
  Astronomical Data Analysis Software and Systems XI. p.~228

\bibitem[\protect\citeauthoryear{{Blanton}, {Kazin}, {Muna}, {Weaver}  \&
  {Price-Whelan}}{{Blanton} et~al.}{2011}]{Blanton_2011}
{Blanton} M.~R.,  {Kazin} E.,  {Muna} D.,  {Weaver} B.~A.,   {Price-Whelan} A.,
   2011, \mn@doi [\aj] {10.1088/0004-6256/142/1/31}, \href
  {https://ui.adsabs.harvard.edu/abs/2011AJ....142...31B} {142, 31}

\bibitem[\protect\citeauthoryear{{Blanton} et~al.,}{{Blanton}
  et~al.}{2017}]{Blanton_2017}
{Blanton} M.~R.,  et~al., 2017, \mn@doi [\aj] {10.3847/1538-3881/aa7567}, \href
  {https://ui.adsabs.harvard.edu/abs/2017AJ....154...28B} {154, 28}

\bibitem[\protect\citeauthoryear{{Bundy} et~al.,}{{Bundy}
  et~al.}{2015}]{Bundy_2015}
{Bundy} K.,  et~al., 2015, \mn@doi [\apj] {10.1088/0004-637X/798/1/7}, \href
  {https://ui.adsabs.harvard.edu/abs/2015ApJ...798....7B} {798, 7}

\bibitem[\protect\citeauthoryear{{Cappellari} et~al.,}{{Cappellari}
  et~al.}{2011}]{Cappellari_2011}
{Cappellari} M.,  et~al., 2011, \mn@doi [\mnras]
  {10.1111/j.1365-2966.2011.18600.x}, \href
  {https://ui.adsabs.harvard.edu/abs/2011MNRAS.416.1680C} {416, 1680}

\bibitem[\protect\citeauthoryear{{Cappellari} et~al.,}{{Cappellari}
  et~al.}{2013}]{Cappellari_2013}
{Cappellari} M.,  et~al., 2013, \mn@doi [\mnras] {10.1093/mnras/stt562}, \href
  {https://ui.adsabs.harvard.edu/abs/2013MNRAS.432.1709C} {432, 1709}

\bibitem[\protect\citeauthoryear{{Cherinka} et~al.,}{{Cherinka}
  et~al.}{2019}]{Cherinka_2019}
{Cherinka} B.,  et~al., 2019, \mn@doi [\aj] {10.3847/1538-3881/ab2634}, \href
  {https://ui.adsabs.harvard.edu/abs/2019AJ....158...74C} {158, 74}

\bibitem[\protect\citeauthoryear{{Cortese} et~al.,}{{Cortese}
  et~al.}{2014}]{Cortese_2014}
{Cortese} L.,  et~al., 2014, \mn@doi [\apjl] {10.1088/2041-8205/795/2/L37},
  \href {https://ui.adsabs.harvard.edu/abs/2014ApJ...795L..37C} {795, L37}

\bibitem[\protect\citeauthoryear{{De Vaucouleurs}}{{De
  Vaucouleurs}}{1948}]{de-Vaucouleurs1948}
{De Vaucouleurs} G.,  1948, Annales d'Astrophysique, \href
  {https://ui.adsabs.harvard.edu/abs/1948AnAp...11..247D} {11, 247}

\bibitem[\protect\citeauthoryear{{Dehnen}}{{Dehnen}}{1993}]{Dehnen1993}
{Dehnen} W.,  1993, MNRAS, \href
  {https://ui.adsabs.harvard.edu/abs/1993MNRAS.265..250D} {265, 250}

\bibitem[\protect\citeauthoryear{{Djorgovski} \& {Davis}}{{Djorgovski} \&
  {Davis}}{1987}]{Djorgovski_1987}
{Djorgovski} S.,  {Davis} M.,  1987, \mn@doi [\apj] {10.1086/164948}, \href
  {https://ui.adsabs.harvard.edu/abs/1987ApJ...313...59D} {313, 59}

\bibitem[\protect\citeauthoryear{{Dressler}, {Lynden-Bell}, {Burstein},
  {Davies}, {Faber}, {Terlevich}  \& {Wegner}}{{Dressler}
  et~al.}{1987}]{Dressler_1987}
{Dressler} A.,  {Lynden-Bell} D.,  {Burstein} D.,  {Davies} R.~L.,  {Faber}
  S.~M.,  {Terlevich} R.,   {Wegner} G.,  1987, \mn@doi [\apj]
  {10.1086/164947}, \href
  {https://ui.adsabs.harvard.edu/abs/1987ApJ...313...42D} {313, 42}

\bibitem[\protect\citeauthoryear{{Drory} et~al.,}{{Drory}
  et~al.}{2015}]{Drory_2015}
{Drory} N.,  et~al., 2015, \mn@doi [\aj] {10.1088/0004-6256/149/2/77}, \href
  {https://ui.adsabs.harvard.edu/abs/2015AJ....149...77D} {149, 77}

\bibitem[\protect\citeauthoryear{{Dutton} \& {Macci{\`o}}}{{Dutton} \&
  {Macci{\`o}}}{2014}]{Dutton}
{Dutton} A.~A.,  {Macci{\`o}} A.~V.,  2014, \mn@doi [\mnras]
  {10.1093/mnras/stu742}, \href
  {https://ui.adsabs.harvard.edu/abs/2014MNRAS.441.3359D} {441, 3359}

\bibitem[\protect\citeauthoryear{{Erwin} et~al.,}{{Erwin}
  et~al.}{2015}]{Erwin_2015}
{Erwin} P.,  et~al., 2015, \mn@doi [\mnras] {10.1093/mnras/stu2376}, \href
  {https://ui.adsabs.harvard.edu/abs/2015MNRAS.446.4039E} {446, 4039}

\bibitem[\protect\citeauthoryear{{Faber} \& {Jackson}}{{Faber} \&
  {Jackson}}{1976}]{Faber_Jackson_1976}
{Faber} S.~M.,  {Jackson} R.~E.,  1976, \mn@doi [\apj] {10.1086/154215}, \href
  {https://ui.adsabs.harvard.edu/abs/1976ApJ...204..668F} {204, 668}

\bibitem[\protect\citeauthoryear{{Feng}, {Shen}, {Yuan}, {Dai}  \&
  {Masters}}{{Feng} et~al.}{2022}]{Feng_2022}
{Feng} S.,  {Shen} S.-Y.,  {Yuan} F.-T.,  {Dai} Y.~S.,   {Masters} K.~L.,
  2022, \mn@doi [\apjs] {10.3847/1538-4365/ac80f2}, \href
  {https://ui.adsabs.harvard.edu/abs/2022ApJS..262....6F} {262, 6}

\bibitem[\protect\citeauthoryear{{Freeman}}{{Freeman}}{1970}]{Freeman1970}
{Freeman} K.~C.,  1970, \mn@doi [\apj] {10.1086/150474}, \href
  {https://ui.adsabs.harvard.edu/abs/1970ApJ...160..811F} {160, 811}

\bibitem[\protect\citeauthoryear{{Gavazzi}, {Franzetti}, {Scodeggio}, {Boselli}
   \& {Pierini}}{{Gavazzi} et~al.}{2000}]{Gavazzi_2000_pV}
{Gavazzi} G.,  {Franzetti} P.,  {Scodeggio} M.,  {Boselli} A.,   {Pierini} D.,
  2000, \aap, \href {https://ui.adsabs.harvard.edu/abs/2000A&A...361..863G}
  {361, 863}

\bibitem[\protect\citeauthoryear{{Gilhuly}, {Courteau}  \&
  {S{\'a}nchez}}{{Gilhuly} et~al.}{2019}]{Gilhuly_2019}
{Gilhuly} C.,  {Courteau} S.,   {S{\'a}nchez} S.~F.,  2019, \mn@doi [\mnras]
  {10.1093/mnras/sty2792}, \href
  {https://ui.adsabs.harvard.edu/abs/2019MNRAS.482.1427G} {482, 1427}

\bibitem[\protect\citeauthoryear{{Hogg}, {Bovy}  \& {Lang}}{{Hogg}
  et~al.}{2010}]{Hogg_2010}
{Hogg} D.~W.,  {Bovy} J.,   {Lang} D.,  2010, \mn@doi [arXiv e-prints]
  {10.48550/arXiv.1008.4686}, \href
  {https://ui.adsabs.harvard.edu/abs/2010arXiv1008.4686H} {p. arXiv:1008.4686}

\bibitem[\protect\citeauthoryear{{Jethwa}, {Thater}, {Maindl}  \& {Van de
  Ven}}{{Jethwa} et~al.}{2020}]{DYNAMITE}
{Jethwa} P.,  {Thater} S.,  {Maindl} T.,   {Van de Ven} G.,  2020, {DYNAMITE:
  DYnamics, Age and Metallicity Indicators Tracing Evolution} (\mn@eprint
  {ascl} {2011.007})

\bibitem[\protect\citeauthoryear{{Jin} et~al.,}{{Jin} et~al.}{2016}]{Jin_2016}
{Jin} Y.,  et~al., 2016, \mn@doi [\mnras] {10.1093/mnras/stw2055}, \href
  {https://ui.adsabs.harvard.edu/abs/2016MNRAS.463..913J} {463, 913}

\bibitem[\protect\citeauthoryear{{Jin}, {Zhu}, {Long}, {Mao}, {Xu}, {Li}  \&
  {van de Ven}}{{Jin} et~al.}{2019}]{Jin_2019}
{Jin} Y.,  {Zhu} L.,  {Long} R.~J.,  {Mao} S.,  {Xu} D.,  {Li} H.,   {van de
  Ven} G.,  2019, \mn@doi [\mnras] {10.1093/mnras/stz1170}, \href
  {https://ui.adsabs.harvard.edu/abs/2019MNRAS.486.4753J} {486, 4753}

\bibitem[\protect\citeauthoryear{{Jin}, {Zhu}, {Long}, {Mao}, {Wang}  \& {van
  de Ven}}{{Jin} et~al.}{2020}]{Jin_2020}
{Jin} Y.,  {Zhu} L.,  {Long} R.~J.,  {Mao} S.,  {Wang} L.,   {van de Ven} G.,
  2020, \mn@doi [\mnras] {10.1093/mnras/stz3072}, \href
  {https://ui.adsabs.harvard.edu/abs/2020MNRAS.491.1690J} {491, 1690}

\bibitem[\protect\citeauthoryear{Kormendy \& Kennicutt}{Kormendy \&
  Kennicutt}{2004}]{Kormendy_2004}
Kormendy J.,  Kennicutt R.~C.,  2004, \mn@doi [Annual Review of Astronomy and
  Astrophysics] {10.1146/annurev.astro.42.053102.134024}, 42, 603–683

\bibitem[\protect\citeauthoryear{{Krajnovi{\'c}} et~al.,}{{Krajnovi{\'c}}
  et~al.}{2008}]{Krajnovi_2008}
{Krajnovi{\'c}} D.,  et~al., 2008, \mn@doi [\mnras]
  {10.1111/j.1365-2966.2008.13712.x}, \href
  {https://ui.adsabs.harvard.edu/abs/2008MNRAS.390...93K} {390, 93}

\bibitem[\protect\citeauthoryear{{Kruk} et~al.,}{{Kruk}
  et~al.}{2018}]{Kruk_2018}
{Kruk} S.~J.,  et~al., 2018, \mn@doi [\mnras] {10.1093/mnras/stx2605}, \href
  {https://ui.adsabs.harvard.edu/abs/2018MNRAS.473.4731K} {473, 4731}

\bibitem[\protect\citeauthoryear{Laurikainen, Peletier  \& Gadotti}{Laurikainen
  et~al.}{2016}]{Laurikainen_2016}
Laurikainen E.,  Peletier R.,   Gadotti D.,  2016, {Galactic Bulges}.
 Astrophysics and Space Science Library Vol. 418,
  \mn@doi{10.1007/978-3-319-19378-6, }

\bibitem[\protect\citeauthoryear{{Law} et~al.,}{{Law} et~al.}{2016}]{Law_2016}
{Law} D.~R.,  et~al., 2016, \mn@doi [\aj] {10.3847/0004-6256/152/4/83}, \href
  {https://ui.adsabs.harvard.edu/abs/2016AJ....152...83L} {152, 83}

\bibitem[\protect\citeauthoryear{{Law} et~al.,}{{Law} et~al.}{2021}]{Law_2021}
{Law} D.~R.,  et~al., 2021, \mn@doi [\aj] {10.3847/1538-3881/abcaa2}, \href
  {https://ui.adsabs.harvard.edu/abs/2021AJ....161...52L} {161, 52}

\bibitem[\protect\citeauthoryear{{Lelli}, {McGaugh}, {Schombert}, {Desmond}  \&
  {Katz}}{{Lelli} et~al.}{2019}]{Lelli_2019}
{Lelli} F.,  {McGaugh} S.~S.,  {Schombert} J.~M.,  {Desmond} H.,   {Katz} H.,
  2019, \mn@doi [\mnras] {10.1093/mnras/stz205}, \href
  {https://ui.adsabs.harvard.edu/abs/2019MNRAS.484.3267L} {484, 3267}

\bibitem[\protect\citeauthoryear{Liu, Ting  \& Zhou}{Liu
  et~al.}{2012}]{Liu_2012}
Liu F.~T.,  Ting K.,   Zhou Z.-H.,  2012, \mn@doi [ACM Transactions on
  Knowledge Discovery From Data - TKDD] {10.1145/2133360.2133363}, 6, 1

\bibitem[\protect\citeauthoryear{{Mendel}, {Simard}, {Palmer}, {Ellison}  \&
  {Patton}}{{Mendel} et~al.}{2014}]{Mendel_2014}
{Mendel} J.~T.,  {Simard} L.,  {Palmer} M.,  {Ellison} S.~L.,   {Patton} D.~R.,
   2014, \mn@doi [\apjs] {10.1088/0067-0049/210/1/3}, \href
  {https://ui.adsabs.harvard.edu/abs/2014ApJS..210....3M} {210, 3}

\bibitem[\protect\citeauthoryear{{Moster}, {Naab}  \& {White}}{{Moster}
  et~al.}{2018}]{Moster_2018}
{Moster} B.~P.,  {Naab} T.,   {White} S. D.~M.,  2018, \mn@doi [\mnras]
  {10.1093/mnras/sty655}, \href
  {https://ui.adsabs.harvard.edu/abs/2018MNRAS.477.1822M} {477, 1822}

\bibitem[\protect\citeauthoryear{{Nanni} et~al.,}{{Nanni}
  et~al.}{2022}]{Nanni_2022}
{Nanni} L.,  et~al., 2022, \mn@doi [\mnras] {10.1093/mnras/stac1531}, \href
  {https://ui.adsabs.harvard.edu/abs/2022MNRAS.515..320N} {515, 320}

\bibitem[\protect\citeauthoryear{{Navarro}, {Frenk}  \& {White}}{{Navarro}
  et~al.}{1996}]{Navarro1996}
{Navarro} J.~F.,  {Frenk} C.~S.,   {White} S. D.~M.,  1996, \mn@doi [\apj]
  {10.1086/177173}, \href
  {https://ui.adsabs.harvard.edu/abs/1996ApJ...462..563N} {462, 563}

\bibitem[\protect\citeauthoryear{{Nelson} et~al.,}{{Nelson}
  et~al.}{2019}]{TNG50_a}
{Nelson} D.,  et~al., 2019, \mn@doi [\mnras] {10.1093/mnras/stz2306}, \href
  {https://ui.adsabs.harvard.edu/abs/2019MNRAS.490.3234N} {490, 3234}

\bibitem[\protect\citeauthoryear{{Noordermeer} \& {Verheijen}}{{Noordermeer} \&
  {Verheijen}}{2007}]{Noordermeer_2007}
{Noordermeer} E.,  {Verheijen} M.~A.~W.,  2007, \mn@doi [\mnras]
  {10.1111/j.1365-2966.2007.12369.x}, \href
  {https://ui.adsabs.harvard.edu/abs/2007MNRAS.381.1463N} {381, 1463}

\bibitem[\protect\citeauthoryear{{Oh} et~al.,}{{Oh} et~al.}{2016}]{Oh_2016}
{Oh} S.,  et~al., 2016, \mn@doi [\apj] {10.3847/0004-637X/832/1/69}, \href
  {https://ui.adsabs.harvard.edu/abs/2016ApJ...832...69O} {832, 69}

\bibitem[\protect\citeauthoryear{{Pak}, {Lee}, {Oh}, {D'Eugenio}, {Colless},
  {Jeong}  \& {Jeong}}{{Pak} et~al.}{2021}]{Pak_2021}
{Pak} M.,  {Lee} J.~H.,  {Oh} S.,  {D'Eugenio} F.,  {Colless} M.,  {Jeong} H.,
   {Jeong} W.-S.,  2021, \mn@doi [\apj] {10.3847/1538-4357/ac1ba1}, \href
  {https://ui.adsabs.harvard.edu/abs/2021ApJ...921...49P} {921, 49}

\bibitem[\protect\citeauthoryear{{Peng}, {Ho}, {Impey}  \& {Rix}}{{Peng}
  et~al.}{2002}]{Peng_2002}
{Peng} C.~Y.,  {Ho} L.~C.,  {Impey} C.~D.,   {Rix} H.-W.,  2002, \mn@doi [\aj]
  {10.1086/340952}, \href
  {https://ui.adsabs.harvard.edu/abs/2002AJ....124..266P} {124, 266}

\bibitem[\protect\citeauthoryear{{Pillepich} et~al.,}{{Pillepich}
  et~al.}{2019}]{TNG50_b}
{Pillepich} A.,  et~al., 2019, \mn@doi [\mnras] {10.1093/mnras/stz2338}, \href
  {https://ui.adsabs.harvard.edu/abs/2019MNRAS.490.3196P} {490, 3196}

\bibitem[\protect\citeauthoryear{{Rigamonti}, {Dotti}, {Covino}, {Haardt},
  {Landoni}, {Del Pozzo}, {Lupi}  \& {Zibetti}}{{Rigamonti}
  et~al.}{2022a}]{Rigamonti_2022_software}
{Rigamonti} F.,  {Dotti} M.,  {Covino} S.,  {Haardt} F.,  {Landoni} M.,  {Del
  Pozzo} W.,  {Lupi} A.,   {Zibetti} S.,  2022a, {BANG: BAyesian decomposiotioN
  of Galaxies}, Astrophysics Source Code Library, record ascl:2205.022
  (\mn@eprint {ascl} {2205.022})

\bibitem[\protect\citeauthoryear{{Rigamonti}, {Dotti}, {Covino}, {Haardt},
  {Landoni}, {Del Pozzo}, {Lupi}  \& {Zibetti}}{{Rigamonti}
  et~al.}{2022b}]{Rigamonti_2022}
{Rigamonti} F.,  {Dotti} M.,  {Covino} S.,  {Haardt} F.,  {Landoni} M.,  {Del
  Pozzo} W.,  {Lupi} A.,   {Zibetti} S.,  2022b, \mn@doi [\mnras]
  {10.1093/mnras/stac1326}, \href
  {https://ui.adsabs.harvard.edu/abs/2022MNRAS.513.6111R} {513, 6111}

\bibitem[\protect\citeauthoryear{{Schwarzschild}}{{Schwarzschild}}{1979}]{Schwarzschild1979}
{Schwarzschild} M.,  1979, \mn@doi [\apj] {10.1086/157282}, \href
  {https://ui.adsabs.harvard.edu/abs/1979ApJ...232..236S} {232, 236}

\bibitem[\protect\citeauthoryear{{Simard} et~al.,}{{Simard}
  et~al.}{2002}]{simard_2002}
{Simard} L.,  et~al., 2002, \mn@doi [\apjs] {10.1086/341399}, \href
  {https://ui.adsabs.harvard.edu/abs/2002ApJS..142....1S} {142, 1}

\bibitem[\protect\citeauthoryear{{Skilling}}{{Skilling}}{2004}]{Skilling2004}
{Skilling} J.,  2004, in {Fischer} R.,  {Preuss} R.,   {Toussaint} U.~V.,  eds,
   American Institute of Physics Conference Series Vol. 735, American Institute
  of Physics Conference Series. pp 395--405

\bibitem[\protect\citeauthoryear{{Stetson}}{{Stetson}}{1987}]{Stetson_1987}
{Stetson} P.~B.,  1987, \mn@doi [\pasp] {10.1086/131977}, \href
  {https://ui.adsabs.harvard.edu/abs/1987PASP...99..191S} {99, 191}

\bibitem[\protect\citeauthoryear{{Strauss} et~al.,}{{Strauss}
  et~al.}{2002}]{Strauss_2002}
{Strauss} M.~A.,  et~al., 2002, \mn@doi [\aj] {10.1086/342343}, \href
  {https://ui.adsabs.harvard.edu/abs/2002AJ....124.1810S} {124, 1810}

\bibitem[\protect\citeauthoryear{{Tabor}, {Merrifield}, {Arag{\'o}n-Salamanca},
  {Cappellari}, {Bamford}  \& {Johnston}}{{Tabor} et~al.}{2017}]{Tabor_2017}
{Tabor} M.,  {Merrifield} M.,  {Arag{\'o}n-Salamanca} A.,  {Cappellari} M.,
  {Bamford} S.~P.,   {Johnston} E.,  2017, \mn@doi [\mnras]
  {10.1093/mnras/stw3183}, \href
  {https://ui.adsabs.harvard.edu/abs/2017MNRAS.466.2024T} {466, 2024}

\bibitem[\protect\citeauthoryear{{Tabor}, {Merrifield}, {Arag{\'o}n-Salamanca},
  {Fraser-McKelvie}, {Peterken}, {Smethurst}, {Drory}  \& {Lane}}{{Tabor}
  et~al.}{2019}]{Tabor_2019}
{Tabor} M.,  {Merrifield} M.,  {Arag{\'o}n-Salamanca} A.,  {Fraser-McKelvie}
  A.,  {Peterken} T.,  {Smethurst} R.,  {Drory} N.,   {Lane} R.~R.,  2019,
  \mn@doi [\mnras] {10.1093/mnras/stz431}, \href
  {https://ui.adsabs.harvard.edu/abs/2019MNRAS.485.1546T} {485, 1546}

\bibitem[\protect\citeauthoryear{{Thater} et~al.,}{{Thater}
  et~al.}{2022}]{Thater_2022}
{Thater} S.,  et~al., 2022, \mn@doi [\aap] {10.1051/0004-6361/202243926}, \href
  {https://ui.adsabs.harvard.edu/abs/2022A&A...667A..51T} {667, A51}

\bibitem[\protect\citeauthoryear{{Tully} \& {Fisher}}{{Tully} \&
  {Fisher}}{1977}]{Tully_Fisher_1977}
{Tully} R.~B.,  {Fisher} J.~R.,  1977, \aap, \href
  {https://ui.adsabs.harvard.edu/abs/1977A&A....54..661T} {54, 661}

\bibitem[\protect\citeauthoryear{{V{\'a}zquez-Mata} et~al.,}{{V{\'a}zquez-Mata}
  et~al.}{2022}]{Vazquez_2022}
{V{\'a}zquez-Mata} J.~A.,  et~al., 2022, \mn@doi [\mnras]
  {10.1093/mnras/stac635}, \href
  {https://ui.adsabs.harvard.edu/abs/2022MNRAS.512.2222V} {512, 2222}

\bibitem[\protect\citeauthoryear{{Veitch}, {Del Pozzo}, {Cody}  \&
  {ed1d1a8d}}{{Veitch} et~al.}{2017}]{Del_Pozzo_CPNest}
{Veitch} J.,  {Del Pozzo} W.,  {Cody}  {ed1d1a8d} 2017, {Johnveitch/Cpnest:
  Pypi Release}

\bibitem[\protect\citeauthoryear{{Wake} et~al.,}{{Wake}
  et~al.}{2017}]{Wake_2017}
{Wake} D.~A.,  et~al., 2017, \mn@doi [\aj] {10.3847/1538-3881/aa7ecc}, \href
  {https://ui.adsabs.harvard.edu/abs/2017AJ....154...86W} {154, 86}

\bibitem[\protect\citeauthoryear{{Weiner} et~al.,}{{Weiner}
  et~al.}{2006}]{Weiner_2006}
{Weiner} B.~J.,  et~al., 2006, \mn@doi [\apj] {10.1086/508921}, \href
  {https://ui.adsabs.harvard.edu/abs/2006ApJ...653.1027W} {653, 1027}

\bibitem[\protect\citeauthoryear{{Westfall} et~al.,}{{Westfall}
  et~al.}{2019}]{Westfall_2019}
{Westfall} K.~B.,  et~al., 2019, \mn@doi [\aj] {10.3847/1538-3881/ab44a2},
  \href {https://ui.adsabs.harvard.edu/abs/2019AJ....158..231W} {158, 231}

\bibitem[\protect\citeauthoryear{{White} \& {Rees}}{{White} \&
  {Rees}}{1978}]{White_1978}
{White} S.~D.~M.,  {Rees} M.~J.,  1978, \mn@doi [\mnras]
  {10.1093/mnras/183.3.341}, \href
  {https://ui.adsabs.harvard.edu/abs/1978MNRAS.183..341W} {183, 341}

\bibitem[\protect\citeauthoryear{{Yan} et~al.,}{{Yan} et~al.}{2016}]{Yan_2016}
{Yan} R.,  et~al., 2016, \mn@doi [\aj] {10.3847/0004-6256/151/1/8}, \href
  {https://ui.adsabs.harvard.edu/abs/2016AJ....151....8Y} {151, 8}

\bibitem[\protect\citeauthoryear{{Zana} et~al.,}{{Zana}
  et~al.}{2022}]{Zana_2022}
{Zana} T.,  et~al., 2022, \mn@doi [\mnras] {10.1093/mnras/stac1708}, \href
  {https://ui.adsabs.harvard.edu/abs/2022MNRAS.515.1524Z} {515, 1524}

\bibitem[\protect\citeauthoryear{{Zhu} et~al.,}{{Zhu} et~al.}{2018a}]{Zhu_2018}
{Zhu} L.,  et~al., 2018a, \mn@doi [Nature Astronomy]
  {10.1038/s41550-017-0348-1}, \href
  {https://ui.adsabs.harvard.edu/abs/2018NatAs...2..233Z} {2, 233}

\bibitem[\protect\citeauthoryear{{Zhu} et~al.,}{{Zhu}
  et~al.}{2018b}]{Zhu_2018a}
{Zhu} L.,  et~al., 2018b, \mn@doi [\mnras] {10.1093/mnras/stx2409}, \href
  {https://ui.adsabs.harvard.edu/abs/2018MNRAS.473.3000Z} {473, 3000}

\bibitem[\protect\citeauthoryear{{Zhu}, {van de Ven}, {M{\'e}ndez-Abreu}  \&
  {Obreja}}{{Zhu} et~al.}{2018c}]{Zhu_2018b}
{Zhu} L.,  {van de Ven} G.,  {M{\'e}ndez-Abreu} J.,   {Obreja} A.,  2018c,
  \mn@doi [\mnras] {10.1093/mnras/sty1521}, \href
  {https://ui.adsabs.harvard.edu/abs/2018MNRAS.479..945Z} {479, 945}

\bibitem[\protect\citeauthoryear{{Zibetti}, {Charlot}  \& {Rix}}{{Zibetti}
  et~al.}{2009}]{Zibetti_2009}
{Zibetti} S.,  {Charlot} S.,   {Rix} H.-W.,  2009, \mn@doi [\mnras]
  {10.1111/j.1365-2966.2009.15528.x}, \href
  {https://ui.adsabs.harvard.edu/abs/2009MNRAS.400.1181Z} {400, 1181}

\bibitem[\protect\citeauthoryear{{de Souza}, {Gadotti}  \& {dos Anjos}}{{de
  Souza} et~al.}{2004}]{budda_Gadotti}
{de Souza} R.~E.,  {Gadotti} D.~A.,   {dos Anjos} S.,  2004, \mn@doi [\apjs]
  {10.1086/421554}, \href
  {https://ui.adsabs.harvard.edu/abs/2004ApJS..153..411D} {153, 411}

\bibitem[\protect\citeauthoryear{{van den Bosch}, {van de Ven}, {Verolme},
  {Cappellari}  \& {de Zeeuw}}{{van den Bosch}
  et~al.}{2008}]{van_den_Bosch_2008}
{van den Bosch} R.~C.~E.,  {van de Ven} G.,  {Verolme} E.~K.,  {Cappellari} M.,
    {de Zeeuw} P.~T.,  2008, \mn@doi [\mnras]
  {10.1111/j.1365-2966.2008.12874.x}, \href
  {https://ui.adsabs.harvard.edu/abs/2008MNRAS.385..647V} {385, 647}

\makeatother
\end{thebibliography}




\appendix

\section{Scaling relation with double slopes}
\label{app:Scaling_with_double_slopes}
\begin{figure}
    \centering
    \includegraphics[scale=0.5]{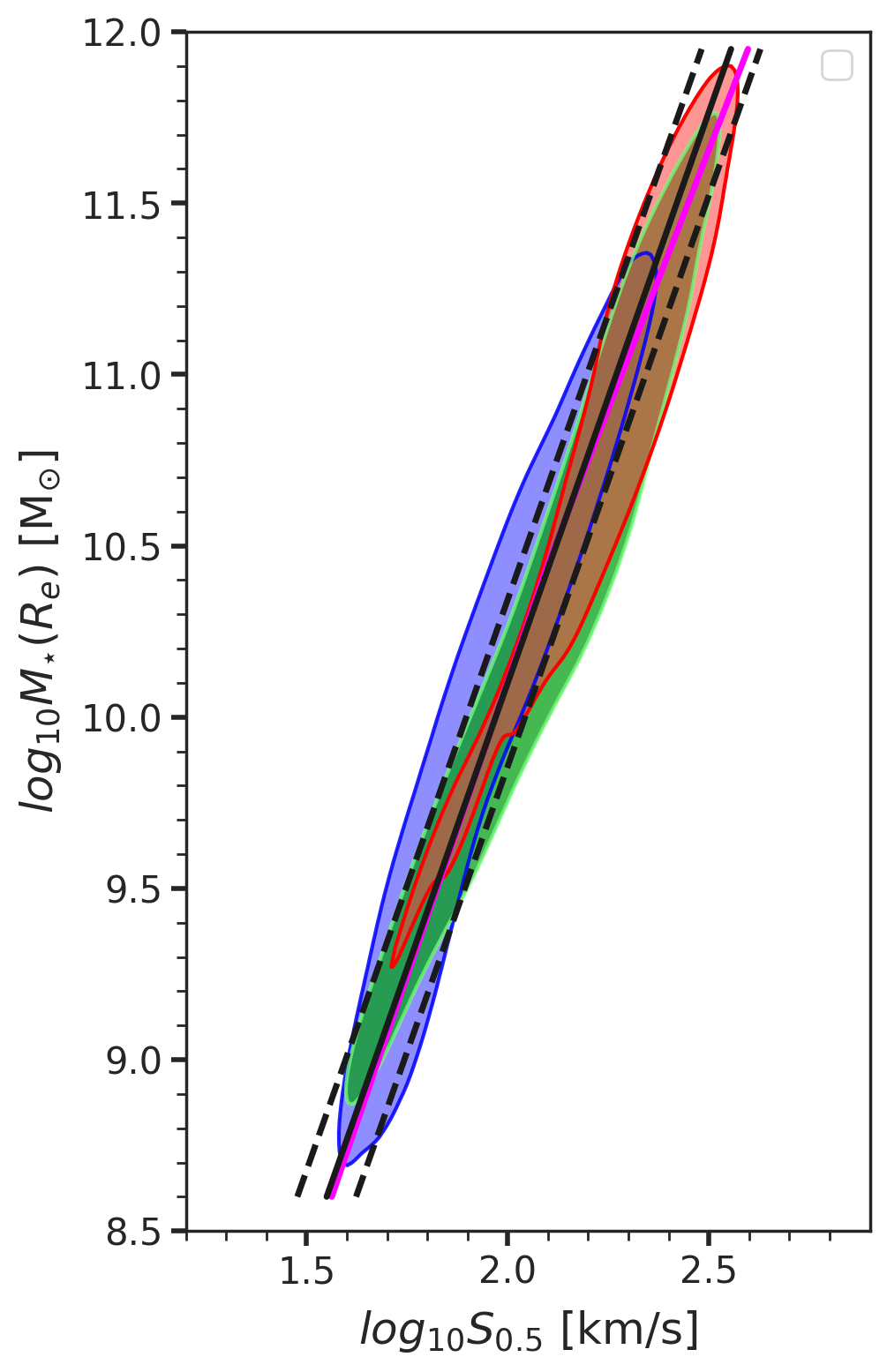}
    \caption{Same as the central panel of Fig.~\ref{fig:Scaling_Relations}. The black solid line is the best linear fit (i.e. Tab.~\ref{tab:best_fit_parameters_linear_fit_whole}) while the dashed lines represent the one-$\sigma$ confidence intervals. The fuchsia line is the result of a fit assuming two slopes (i.e. Eq.~\ref{eq:double_line}) whose best-fit parameters are presented in Tab.~\ref{tab:best_fit_parameters_bilinear_fit}.}
    \label{fig:Scaling_Relations_double_slopes}
\end{figure}

\fr{We present here the result of a fit on the $M_{\star}(R_e)-S_{0.5}(R_e)$ with two slopes (Fig.~\ref{fig:Scaling_Relations_double_slopes}). Eq.\ref{eq:double_line} represent the functional form that we used for the fit, while the best-fit parameters are reported in Tab.~\ref{tab:best_fit_parameters_bilinear_fit}. The evidence of the "two slopes" model is $\log{Z}=20,516$ against $\log{Z}=20,336$ for the single slope fit}


\begin{equation}
\log_{10}\left(\frac{S_{0.5}(R_e)}{100~\rm km/s}\right) = \begin{cases} a_1+b_1\log_{10}\left(\frac{M_{\star}}{10^{10.5}~\rm M_{\odot}}\right) & \\ \hspace{2cm} \ \ \mathrm{if} \ \ \ \ \log_{10}\left(\frac{M_{\star}}{10^{10.5}~\rm M_{\odot}}\right)\leq M_0 \\ \\ a_2+b_2\log_{10}\left(\frac{M_{\star}}{10^{10.5}~\rm M_{\odot}}\right) & \\ \hspace{2cm} \ \  \mathrm{if} \ \ \ \ \log_{10}\left(\frac{M_{\star}}{10^{10.5}~\rm M_{\odot}}\right) > M_0 ,\end{cases}
\label{eq:double_line}
\end{equation}

\fr{where the two constants $a_1$ and $a_2$ are not independent of each other due to the continuity condition of the function in $M_0$.}

\begin{table}
	\centering
	\caption{Summary table of the best-fit parameters for the fit to the  $M_{\star}(R_e)-S_{0.5}(R_e)$ with the functional form of Eq.~ \ref{eq:double_line}.}
	\label{tab:best_fit_parameters_bilinear_fit}
	\begin{tabular}{ccccc} 
	    \hline
		  $a_1$ & $a_2$ & $b_1$ & $b_2$ & $M_0$ \\
		\hline 
		 $0.12 \pm 0.01$ & $0.12 \pm 0.01$ & $0.29 \pm 0.01$ & $0.32 \pm 0.01$ & $10.57 \pm 0.01$\\ 
		\hline
	\end{tabular}
\end{table}

\section{Analysis on a galaxy subsample}
\label{app:Analysis_golden_sample}

In this section, we partially repeat the analysis outlined in Sec.~\ref{sec:Preliminary analysis on the proposed decomposition} on a galaxy subsample selected discarding all galaxies with $\log_{10}{M_{\star}}<9.5$ and  whose bulge half-mass radius is smaller (within its error) than the pixel resolution. We moreover visually inspected and discarded all the objects including ongoing mergers or other contaminants (i.e. wrong redshift, central non-masked stars and type I AGN). 

\begin{figure*}
    \centering
    \includegraphics[scale=0.3]{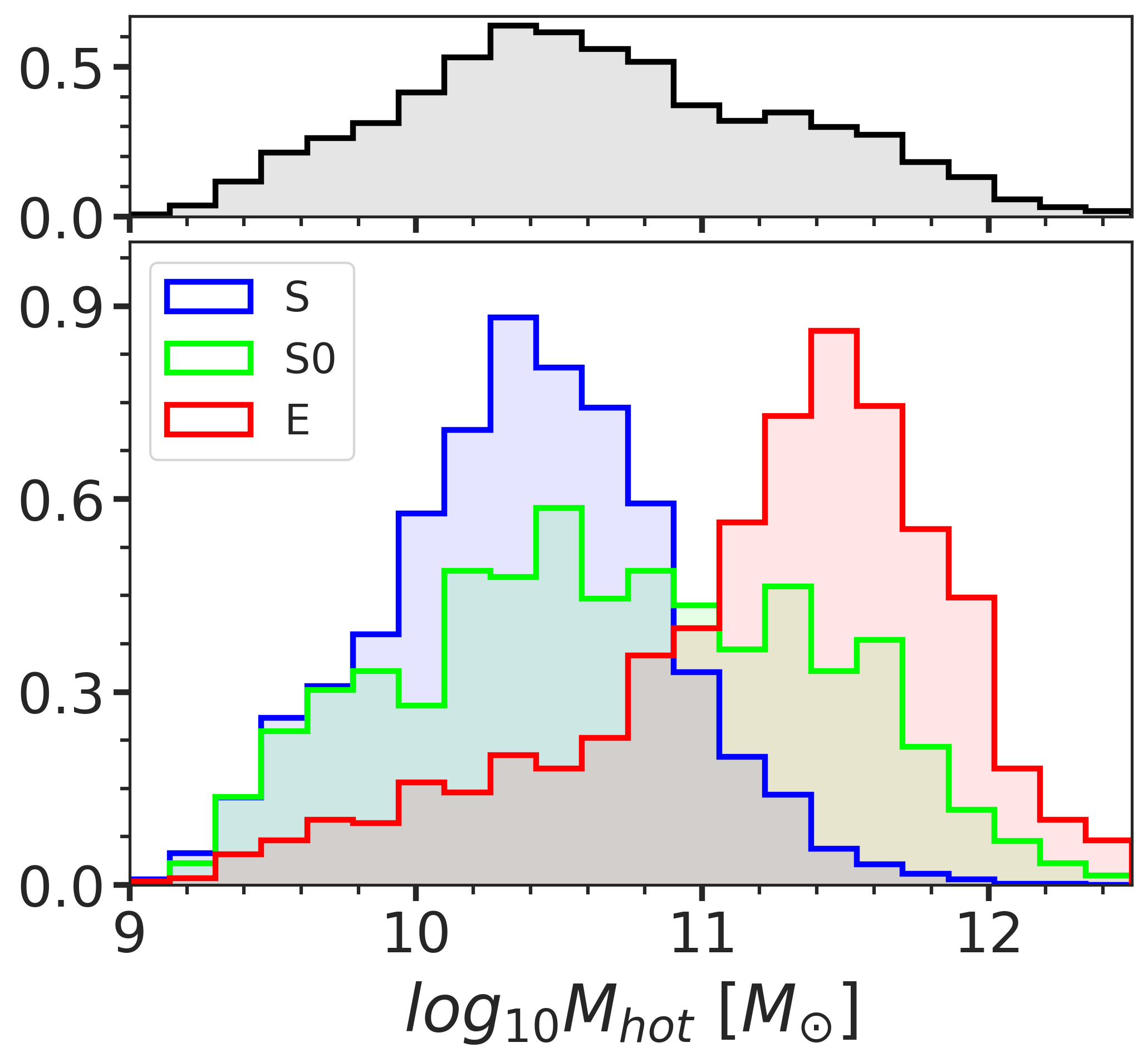}\hspace{0.01\textwidth}
    \includegraphics[scale=0.3]{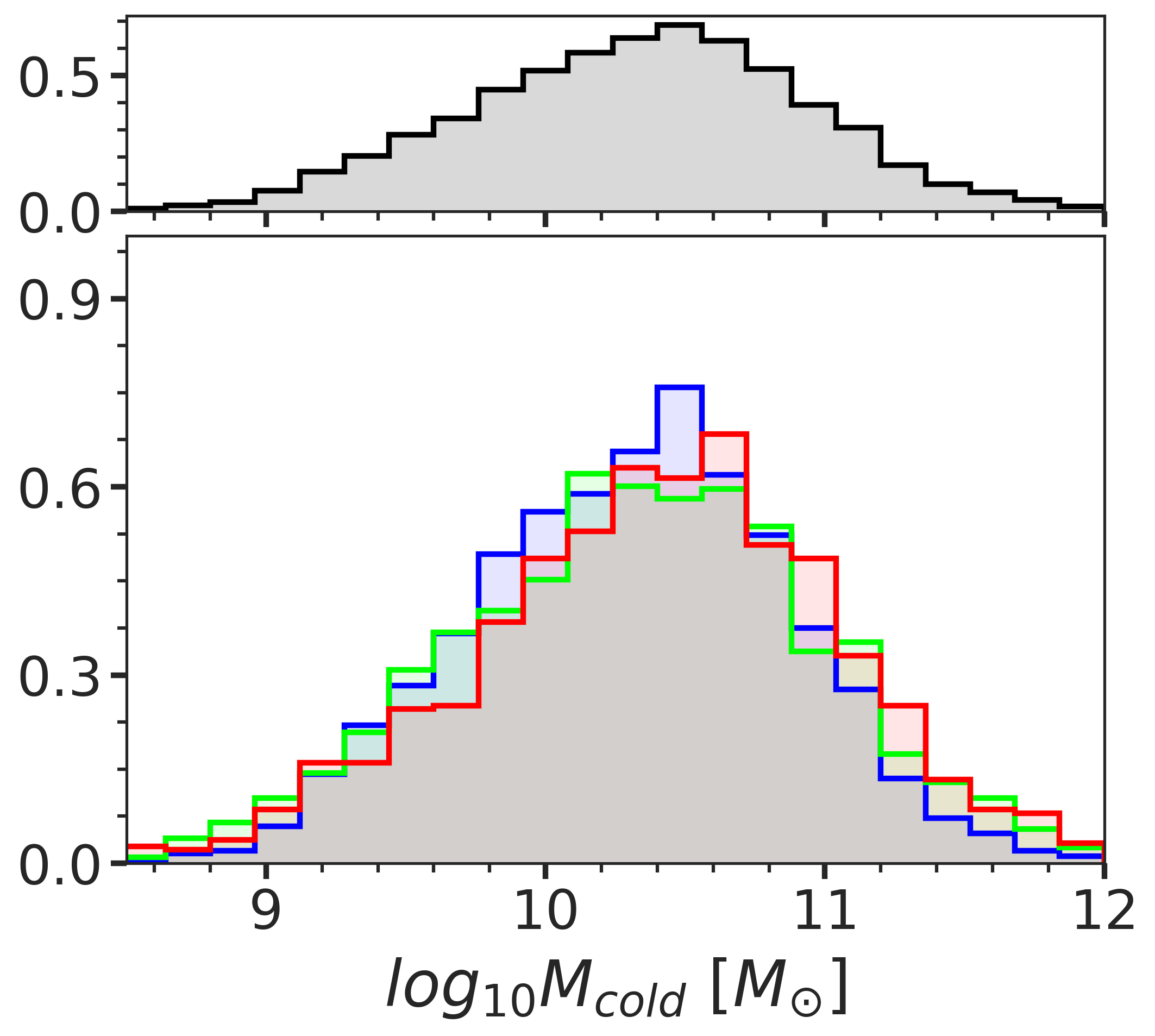}\hspace{0.01\textwidth}
    \includegraphics[scale=0.3]{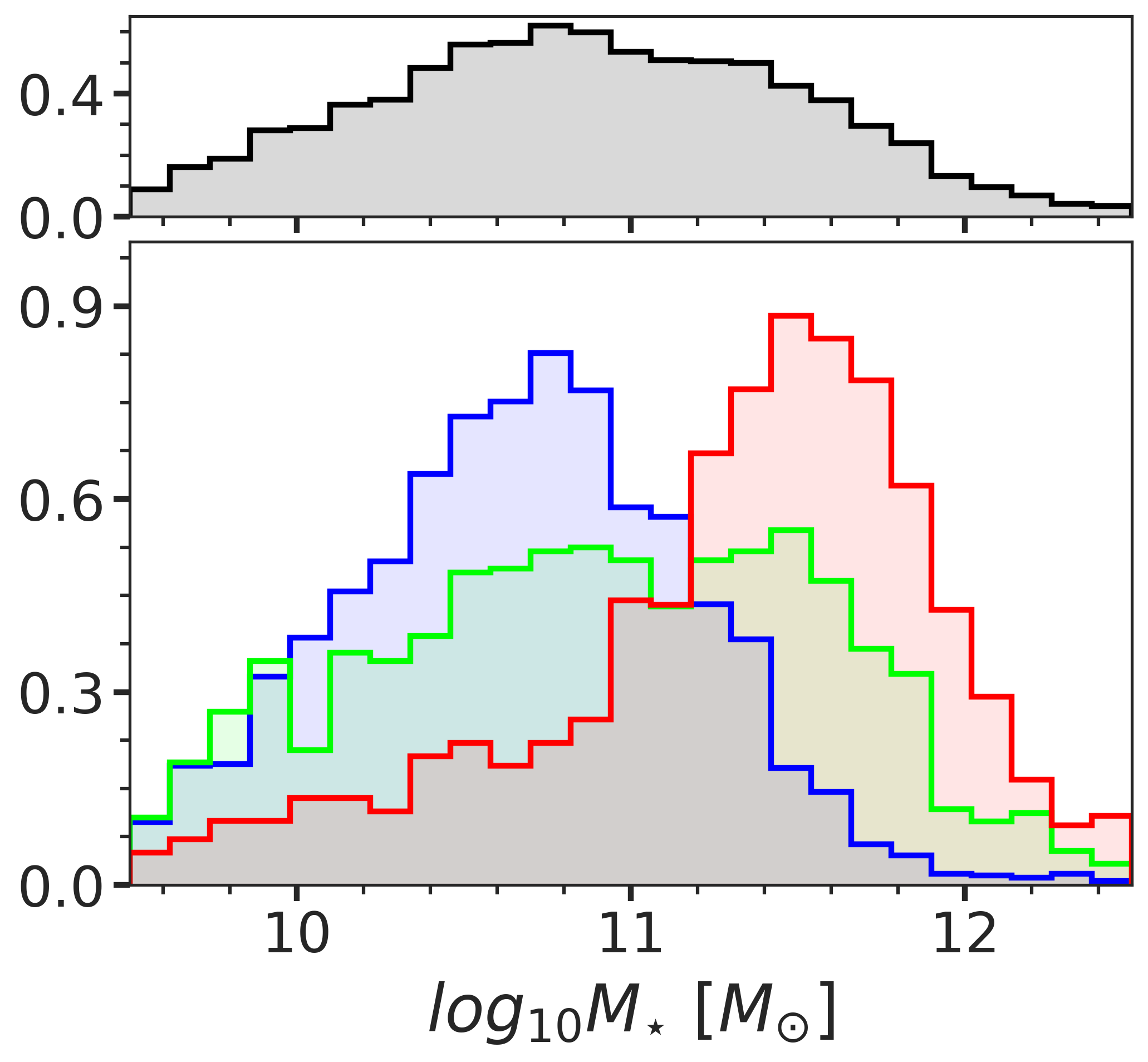}
    \caption{From left to right we show the mass distribution of the hot, cold and total stellar mass for the sample under analysis. The hot and cold mass are estimated following Eqs.. \ref{eq:hot_mass} and \ref{eq:cold_mass} for the sample selected according to the text ($\simeq 5,500$ galaxies). Each histogram is normalized to have a total area equal to one, and the distributions are colour-coded according to visual morphology: blue for spirals, green for lenticulars and red for ellipticals. The panel over each picture represents the distribution of that variable for the whole population helping in understanding the relative role of the three morphological types.}
    \label{fig:Mass_distributions_selected}
\end{figure*}

As shown in Fig.~\ref{fig:Mass_distributions_selected} the distribution of "hot", "cold" and total masses show similar trends to the one mentioned in the main body of the text (see Sec.~\ref{sec:Preliminary analysis on the proposed decomposition} with ellipticals dominating the high-mass regime with a dominant contribution of the "hot" component and spirals being colder and skewed to slightly smaller masses. The main difference from Fig.~\ref{fig:Mass_distributions} is the missing peak at small stellar masses ($\log_{10}{M_{\rm hot}} \simeq 9.4 $) in spirals mostly due to the exclusion of low mass ($\log_{10}{M_{\star}} < 9.5$) galaxies.

\section{Scaling relations in the halo-free configuration}
\label{app:Scaling_halo_fixed}

\begin{figure*}
    \centering
    \includegraphics[scale=0.42]{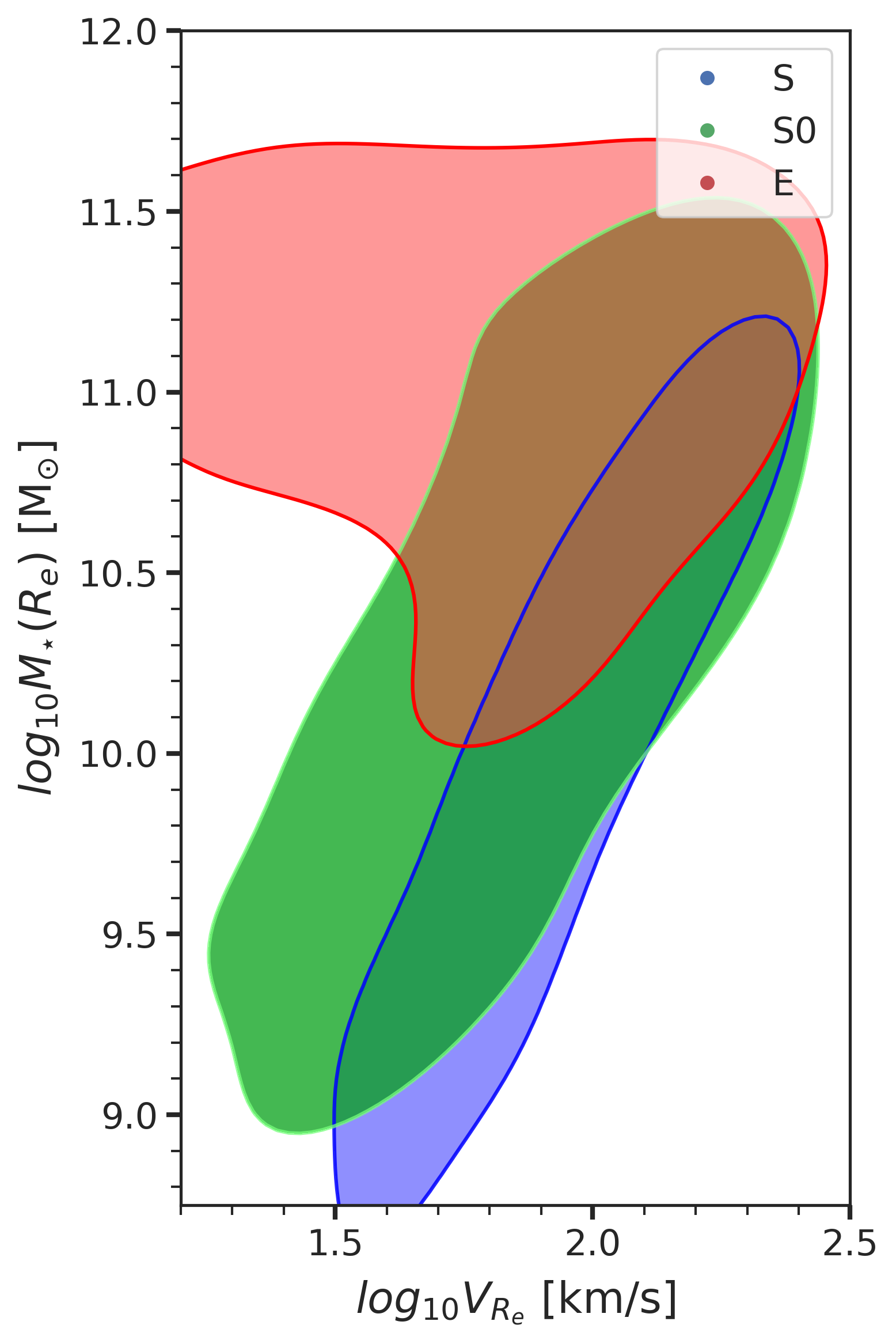}\hspace{0.01\textwidth}
    \includegraphics[scale=0.42]{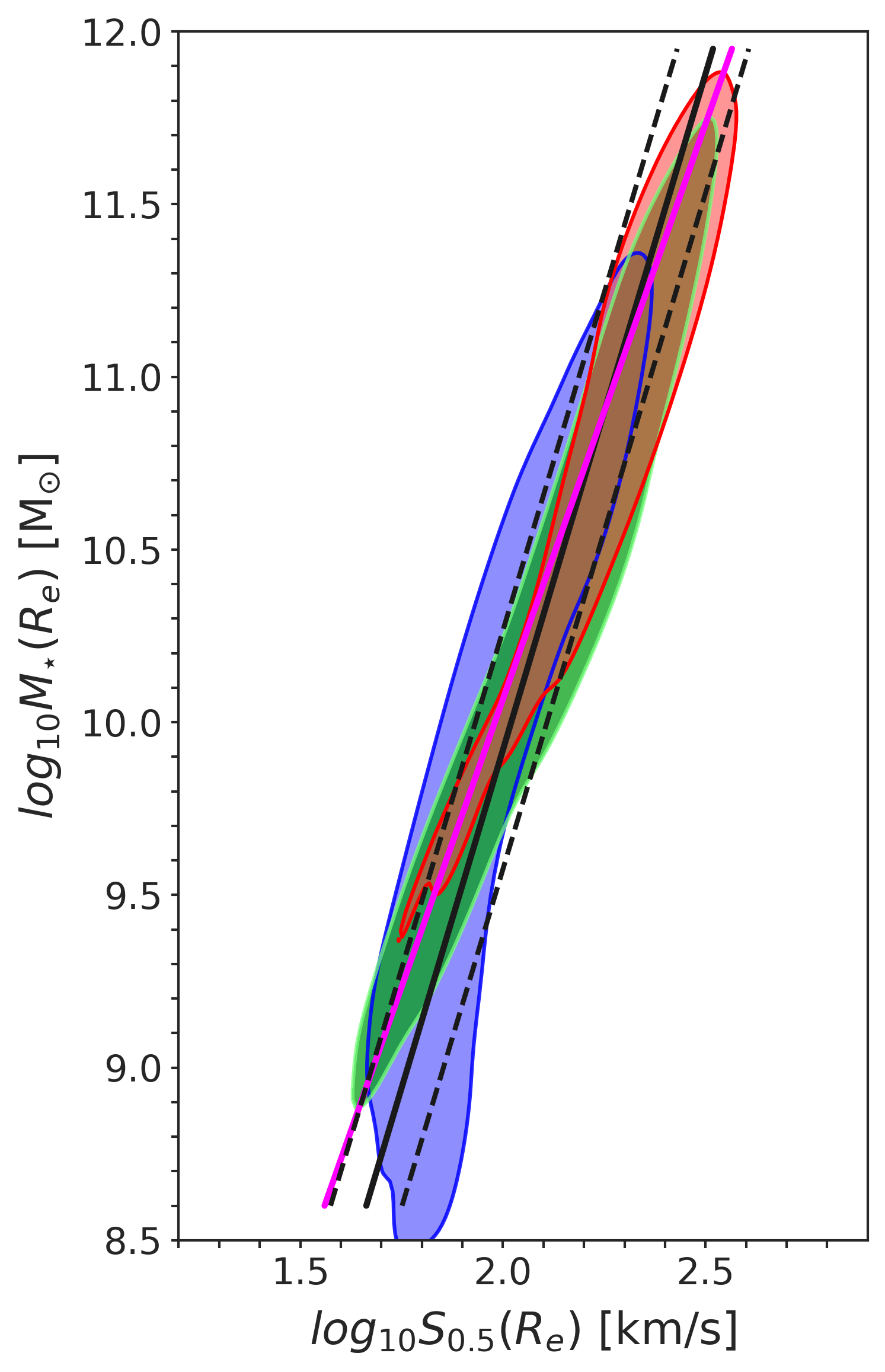}\hspace{0.01\textwidth}
    \includegraphics[scale=0.42]{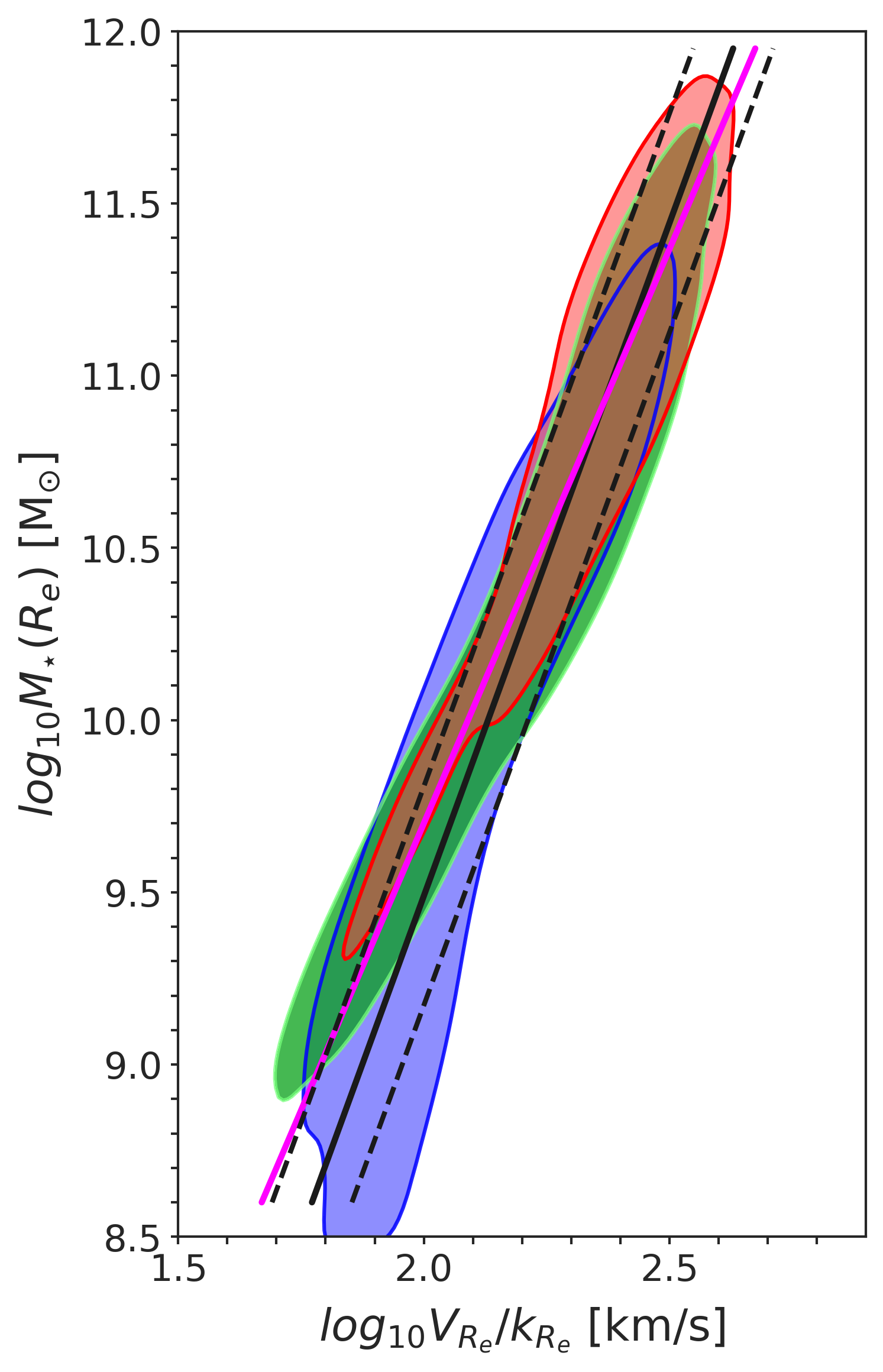}
    \caption{Scaling relations of different kinematics tracers divided according to morphology (blue for spirals, green for lenticulars and red for ellipticals) in the case of the halo-free configuration. In all three plots, the y-axis is the logarithm of the stellar mass within one $ R_e$, while the horizontal axis, respectively from left to right, shows the logarithm of the velocity at one$ R_e$ ($V_{R_e}$), the logarithm of the $S_{0.5}$ parameter (see Eq. \ref{eq:S0_5}) and the logarithm of the velocity at $R_e$ corrected according to Eq. \ref{eq:arctan}. Density contours respectively contain $75\%$, $80\%$  and $80\%$ of the kernel density estimated probability mass. The solid black lines are the best-fit linear relations while the black dotted lines represent the one $-\sigma$ confidence interval, and the fuchsia line is the best-fit relation computed from Tab.~\ref{tab:best_fit_parameters_linear_fit_whole}. The best-fit parameters are summarized in Tab.~\ref{tab:best_fit_parameters_linear_fit_whole_ten}.}
    \label{fig:Scaling_Relations_ten}
\end{figure*}

We report in this section a summary of the same analysis detailed in Sec.~\ref{sec:Comparison with previous dynamical modelling estimates} in the case of the halo-free configuration (see Sec.~\ref{sec:methodology_priors}). Tabs. \ref{tab:best_fit_parameters_linear_fit_whole_ten} and \ref{tab:best_fit_parameters_linear_fit_fiducial_ten} reports the best-fit parameters of a linear fit to the data, whose results are roughly comparable (within two $\sigma$ errors) to the one reported in the halo-fixed configuration (tabs.~\ref{tab:best_fit_parameters_linear_fit_whole} and \ref{tab:best_fit_parameters_linear_fit_fiducial}). Fig.~\ref{fig:Scaling_Relations_ten} shows the same quantities as  Fig.~\ref{fig:Scaling_Relations} for the halo-free case. Some differences can be seen, especially in the low mass regime, where, as expected, the priors on the halo parameters are more informative. The fuchsia line represents the best fit linear relation in the halo-fixed configuration which is compatible (within one$-\sigma$ scatter) with the best-fit line in the halo-free case (black solid line).

\begin{table}
	\centering
	\caption{Summary table of the best-fit parameters for the linear fit (Eqs..~ \ref{eq:linear_relation_S05}, \ref{eq:linear_relation_vRe} ) of the kinematic tracers presented in the middle and left panels of Fig.~\ref{fig:Scaling_Relations_ten}. The first column refers to the considered dependent variable (the independent variable is always $M_{\star}(R_e)$), while the other three columns are the free parameters of the fit, namely, the intercept (a), the slope (b) and the intrinsic scatter $\sigma$.}
	\label{tab:best_fit_parameters_linear_fit_whole_ten}
	\begin{tabular}{lccc} 
	    \hline
		y-axis & a & b & $\sigma$ \\
		\hline 
		$\log_{10}{S_{0.5}(R_e)}$ & $0.15 \pm 0.01$ & $0.26 \pm 0.01$ & $0.088$\\ 
		$\log_{10}{V_{R_e}/k_{R_e}}$ & $0.26 \pm 0.01$ & $0.26 \pm 0.01$ & $0.081$\\ 
		\hline
	\end{tabular}
\end{table}

\begin{table}
	\centering
	\caption{Same as Tab.~\ref{tab:best_fit_parameters_linear_fit_whole_ten} but for a sub-sample of $\simeq 7000$ galaxies selected according to the same criteria applied in \citealt{Aquino_2020} (see par.~\ref{sec:scaling_relations} for further details). }
	\label{tab:best_fit_parameters_linear_fit_fiducial_ten}
	\begin{tabular}{lccc} 
	    \hline
		y-axis & a & b & $\sigma$ \\
		\hline 
		$\log_{10}{S_{0.5}(R_e)}$ & $0.15 \pm 0.01$ & $0.27 \pm 0.01$ & $0.076$\\ 
		$\log_{10}{V_{R_e}/k_{R_e}}$ & $0.27 \pm 0.01$ & $0.27 \pm 0.01$ & $0.065$\\ 
		\hline
	\end{tabular}
\end{table}


\bsp	
\label{lastpage}
\end{document}